\documentclass[11pt]{article}

\usepackage{hyperref}
\usepackage[dvipsnames]{xcolor}
\usepackage{color,tikz}
\usepackage{mathrsfs}
\usepackage{graphicx}
\usepackage{mathtools}
\usepackage{caption,subcaption}
\usepackage{amsmath}
\usepackage{amsthm}
\usepackage{varwidth,xcolor}
\usepackage{amssymb}
\usepackage{textcomp}
\usepackage{euscript}
\usepackage{xcolor}
\usepackage{tcolorbox}
\usepackage[titletoc,title]{appendix}
\usepackage{upgreek}
\usepackage{nicefrac}
\usepackage[margin=1in]{geometry}

\usepackage{booktabs}

\makeatletter

\usepackage{float}
\usepackage[scaled]{helvet}

\def\bSig\mathbf{\Sigma}

\newcommand{\reals}{\mathbb{R}}
\newcommand{\mc}{\mathcal}
\newcommand{\m}{\mathbf}

\newcommand{\lt}{\left}
\newcommand{\rt}{\right}

\newcommand{\plsone}{\sf{PLS}1}
\newcommand{\plstwo}{\sf{PLS}2}
\newcommand{\pca}{\sf{PCA}}

\usepackage[linesnumbered,ruled,vlined]{algorithm2e}
\SetKwInput{KwInput}{Input}                
\SetKwInput{KwOutput}{Output}     
\SetKwInput{KwInit}{Initialization}   
\newcommand{\argmin}{\operatornamewithlimits{argmin}}

\usepackage{multirow}

\newtheorem{theorem}{Theorem}

\theoremstyle{plain}

\theoremstyle{definition}
\usepackage[]{graphicx}
\chardef\bslash=`\\ 

\title{Best Subset Solution Path for Linear Dimension Reduction Models using Continuous Optimization}
\author{ Benoit Liquet${}^{1,2,*}$, Sarat Moka${}^{3,*}$ and Samuel Muller${}^{1,4}$\\
\\ \small
${}^1$ School of Mathematical and Physical Sciences, Macquarie University, Sydney, Australia\\ \small
${}^2$ Laboratoire de Mathématiques et de leurs Applications, Université de Pau et des Pays de l'Adour, Pau, France.\\ \small
${}^3$ School of Mathematics and Statistics, The University of New South Wales, Sydney, Australia\\ \small
${}^4$ School of Mathematics and Statistics, University of Sydney, Sydney, Australia\\ \small
${}^*$ Corresponding author: benoit.liquet-weiland@mq.edu.au}

\date{\today}
\begin{document}
\maketitle


\begin{abstract}
The selection of best variables is a challenging problem in supervised and unsupervised learning, especially in high dimensional contexts where the number of variables is usually much larger than the number of observations. In this paper, we focus on two multivariate statistical methods: principal components analysis and partial least squares. Both approaches are popular linear dimension-reduction methods with numerous applications in several fields including in genomics, biology, environmental science, and engineering. In particular, these approaches build principal components, new variables that are combinations of all the original variables. A main drawback of principal components  is the difficulty  to interpret them when the number of variables is large. To define principal components from the most relevant variables, we propose to cast the best subset solution path method into principal component analysis and partial least square frameworks. We offer a new alternative by exploiting a continuous optimization algorithm  for best subset  solution path. Empirical studies show the efficacy of our approach for providing the best subset solution path. The usage of our algorithm is further exposed through the analysis of  two real datasets. The first dataset is analyzed using the principle component analysis while the analysis of the second dataset is based on partial least square framework.
\end{abstract}


\section{Introduction}
\label{s:intro}

The selection of best variables is a challenging task, particularly in a high dimensional context where the number of variables $p$ is usually much larger than the number of observations $n$.  Analysing each variable separately is time consuming, while describing the results using graphs and numerical measures  may not sufficiently aid in drawing conclusions as either too many features are   visualized or the summary information may be inconclusive. A solution to circumvent this problem is to use multivariate statistical methods such as principal components analysis (PCA) and partial least squares (PLS), which are well established linear dimension-reduction methods for analysing data resulting from observations with a large number of variables. 
In PCA and PLS, a few number of new variables are constructed, which are linear combinations of the original variables.  These linear combinations are called components, scores. The relation between these new variables and the original ones is characterized by the weights involved in the linear combinations. In PCA, the weights are defined in such a way that the variance of each component is maximal, under the constraint that the score variables are orthogonal (see, e.g. \cite{Jolliffe2005}). As a result, PCA offers a low-dimensional representation of the variables that attempts to capture the most important information from the data. In many applications, only a few components are required to recover a large proportion of the overall multidimensional variability present in the original dataset, thereby performing a dimension-reduction  while most of the information is preserved. 

While PCA tackles the analysis of a single dataset,  {\it Projection to Latent Structures} models focus on multiple sets of data, each comprising a large number of variables measured on the same statistical units. Projection to Latent Structures was first introduced by \cite{Wold1966} 
under the name {\it Partial Least Squares} (PLS) in the context of regression models to deal with high collinearity of the predictors where the number of variables is larger than the number of observations. PLS methods offer a wide range of multivariate supervised and unsupervised statistical techniques on multiple blocks of data. In PLS, algorithms also construct new variables that are linear combinations of the original variables. Here, these new components are obtained by maximizing a covariance criterion for capturing the relationships between the sets of data. A recent review of PLS modelling for two blocks of data is provided by \cite{de2019pls} where both asymmetric and  symmetric situations are presented. The asymmetric situation deals with the case where one block of predictors is used to explain the other block while the symmetric situation corresponds to the case where the two blocks are interchangeable.

PCA and PLS are now extremely popular linear dimension-reduction techniques with numerous applications; see, for example, \cite{mehmood2016diversity}) in genomics, \cite{Ji2011} and \cite{wang2020partial} in neuroimaging, \cite{tu2011new} in biology, \cite{khatri2021review} in environment science, and \cite{chen2022hybrid} in engineering.

However, in the case of large number of variables, the main drawback of these algorithms remains the difficulty to interpret the new linear combinations obtained from the large number of original variables. This difficulty has been addressed by proposing sparse modelling techniques for constructing new components using a small number of the original variables \cite{Shen2008,Chun2010}.   The sparsity into the new components can be achieved via lasso penalization \cite{Shen2008,Witten2009,LeCao2008a,Chun2010,Wang2016}. These lasso penalization based techniques have the potential to improve interpretability and to get better estimators, especially for the analysis of large datasets \cite{Lin2014b,LeCao2011,Liquet2016,broc2021penalized}. 

In this paper, we present a new and more suitable approach for identifying components based on the most relevant variables. In particular, we present the challenge of defining sparse components that appear on the so-called {\em{best subset solution path}}, which contains for a given size $k$ that model that is a best subset of $k$ variables for constructing the components.  This terminology of best subset solution (BSS) path follows the terminology in \cite{muller2010} and \cite{hui2017joint}. The BSS path approach aims to collect $p$ models of varying subset size $k$ that solve the PCA model and, respectively, the PLS model. That is the goal in finding the BSS path is to attempt to recover for each subset size $k=1,\ldots,p$, the best subset of size $k$ that is obtained through an exhaustive search approach. Note that the best subset selection problem in the traditional sense is a different problem, as it aims to find a single best model from as many as $2^p$ possible models with a desirable optimality property, such as having highest accuracy.

The problem of finding a single best subset has been intensively studied in the case of linear regression  \cite{HockingLeslie1967,BKM16}, which is often one of the subsets that appear on the best subset solution path. 
Recently proposed methods \cite{hastie2020best,YM20,HazimehMazumder2020} offer a solution beyond the exhaustive search using  the Furnival Wilson algorithm \cite{leaps2000,tarr2018mplot}. An exhaustive search using the Furnival Wilson algorithm is not practical when the number of variables $p$ is larger than 30 to 40 (depending on the computational power available). Our approach for BSS for PCA and PLS models is based on the continuous optimization algorithm recently developed by \cite{COMBSS22} for BSS in linear regression.
More specifically, we frame the BSS for PCA and PLS models as continuous optimization algorithms which can  take advantage of standard continuous optimization methods, such as gradient descent, to visit a large set of subsets. We refer to the proposed method as ``best'' subset solution path approaches because this is what is often achieved numerically where approaches first reduce the vast model space to $p$ subsets on the best subset solution path.

The rest of the paper is organized as follows: Section 2 briefly reviews PCA and PLS models, and their sparse versions. In Section 3, we cast BSS into PCA and PLS models. Section 4 presents the main algorithm and gives more details on its implementation. A simulation study is presented in Section 5 where we highlight the ability of our algorithms to recover best subsets in PCA and PLS models. Section 6 presents applications of our algorithm for two different real datasets. Finally, Section 7 completes the paper with some concluding remarks. We present additional empirical research and additional theoretical results in the supplementary material. All numerical results of our simulation study are reproducible,  R code that is made available at
\url{https://github.com/benoit-liquet/BSS-PCA-PLS}.

\section{Sparse PCA and PLS}
\label{s:model}

In this section, we briefly review  sparse PCA (sPCA) as presented in \cite{Shen2008} and sparse PLS (sPLS) as proposed in \cite{LeCao2008a}. The approaches are based on singular value decomposition (SVD), where sparsity is achieved using lasso type-penalties. Further, these are iterative algorithms based on deflation in each iteration, where deflation removes the information contained in the previous components. We adopt the standard deflation used by \cite{Witten2009} and \cite{LeCao2008a}. 
We first detail the procedures to define the first component for the PCA framework and the first pair of components for the PLS framework. Subsequent components for PCA or subsequent pair of components for PLS repeat the  same procedure on the {\em deflated} data matrices. We remark that the type of deflation procedure used determines the mode of PLS.

\subsection{Notation}

Let $X\in\operatorname{mat}(n,p)$ and $Y\in\operatorname{mat}(n,q)$ be two data matrices, both consisting of $n$ observations of $p$ and $q$ variables, respectively. Without any loss of generality, we assume these matrices are column centered. 
When $q=1$, the observed centered response vector of size $n$ is denoted by $\m y$. 
We use $\langle\, \cdot\, , \, \cdot\, \rangle$ to denote the inner product between two vectors of the same dimension. 
The Frobenius norm of an $n\times p$ matrix $A$ is $\|A\|_F= \sqrt{\operatorname{trace}(A^{\top}A)}$. The soft thresholding function  is $g^{\text {soft }}(x, \tau)=\textrm{sign}(x)(|x|-\tau)_{+}, \text {where }(a)_{+}=\max (a, 0)$.

\subsection{PCA and sPCA}

The first PCA component of $X$ is obtained by solving,
\begin{equation}
\max_{\m u \in \reals^p,\,\, \|\m u \| = 1} \textrm{var}( X \m u)= \max_{\m u \in \reals^p,\,\, \|\m u \| = 1}\frac{1}{n}\m u^\top X^\top X \m u, \label{optPCA}
\end{equation}
where $\textrm{var}(\cdot)$ is the sampling variance operator. Application of Lagrange multiplier techniques shows that an optimal solution  $\m u_1  \in \reals^p$, called the loading vector (also called weight vector), is the eigenvector associated with the largest eigenvalue of the sample covariance of the data $S=\frac{1}{n} X^\top X$.
In practice, this loading vector  $\m u_1  \in \reals^p$  is usually obtained by computing the truncated SVD of $X$ which gives the best rank-one approximation matrix $\tilde{X}=\delta_1\m v_1\m u_1^\top$ with respect to the Frobenious norm. The vectors $\m v_1 \in \reals^n$ and $\m u_1 \in \reals^p$ are respectively the first left singular vector and the first right singular vector of $X$, associated with the largest singular value $\delta_1$.

In order to introduce some sparsity into the loading vector $\m u$, \cite{Shen2008} proposed to solve the problem
\begin{equation}
\min_{\m v  \in \reals^n \| \m v \| = 1,  \m u \in \reals^p} \| X - \m v \m u^\top\|^2_F+ P_{\tau}(\m u), \label{PCAlasso}
\end{equation}
where $P_{\tau}(\m u)=\sum_{j=1}^pp_{\tau}(|u_j|)$ is a penalty function and $p_{\tau}(\cdot)$ is a non-negative function parameterized by $\tau\geq 0$. Let $(\m u^*,\m v^*)$ be the solution of  \eqref{PCAlasso}. The sparse loading of unit length is then  $\tilde{\m u}=\m u^*/\| \m u^*\|$. A  soft thresholding penalty $p_{\tau}(|\theta|)=2\tau|\theta|$ has been implemented in the R package \texttt{mixOmics} \cite{mixOmics}. Note that without any penalty term, this procedure matches with the non-sparse PCA where $\tilde{\m u} =\m u_1$. The subsequent sparse components are defined sequentially using \eqref{PCAlasso} on residual matrices obtained through the deflation step presented in Table 1 in the supplementary information. 

\subsection{PLS and sPLS}
The first component pair for the PLS model for the data matrices $X$ and $Y$ is obtained by solving, 
\begin{equation}
\max_{\m u \in \reals^p,\,\, \|\m u \| = 1, \m v \in \reals^q,\,\, \|\m v \| = 1} \textrm{cov}( X \m u, Y\m v)= \max_{\m u \in \reals^p,\,\, \|\m u \| = 1, \m v \in \reals^q,\,\, \|\m v \| = 1} \frac{\langle X \m u,\,  Y \m v \rangle}{n}. \label{PLS-g}
\end{equation}
where $\textrm{cov}(\cdot,\cdot)$ is the sampling covariance operator. An efficient way to solve this optimization problem is to exploit the SVD of the matrix $M=(X^\top Y)/n$ of rank $r\leq \text{min}(p,q)$: 
\begin{equation}
M=U\Delta V=\sum_{k=1}^r\delta_k\m u_{k} \m v_{k}^\top, \label{Msvd}
\end{equation}
 where  $\m u_{k}$ and $\m v_{k}$ are the $k$-th left and right singular vector of $M$ associated with the $k$-th singular value. The first left singular vector $\m u_1$ and the first right singular vector $\m v_1$ are the solution of \eqref{PLS-g}.
The subsequent component pairs  are obtained in a similar manner using deflated versions of $X$ and $Y$ to ensure the appropriate orthogonal constraint  depending the mode of the PLS used. Some of the popular deflation techniques are presented in Table 1 in the supplementary material. 

Sparsity in the weight vectors $\m u$ and $\m v$ can be introduced by solving  \cite[see,][]{de2019pls}
\begin{equation}
\textrm{maximize}\ \ \ \  \m u^{\top} M \m v-P_{\tau_1}(\m u)-P_{\tau_2}(\m v ) \textrm{    subject to } \|\m u\|_2 \leq 1,\|\m v\|_2 \leq 1,
\end{equation}
where $P_{\tau_1}(\cdot)$ and $P_{\tau_2}(\cdot)$ are convex penalty functions parameterized by tuning parameters $\tau_1$ and $\tau_2$. The first pair of sparse weight vectors  $(\m  u, \m v)$ can be found by iteratively calculating
\begin{equation}
\begin{aligned}
&\tilde{\m u} \leftarrow \underset{\tilde{\m u} \in \mathbb{R}^p}{\operatorname{argmin}}\left\{\left\|M-\tilde{\m u} \m v^\top\right\|_F^2+P_{\tau_1}(\tilde{\m u})\right\}, \\
&\tilde{\m v} \leftarrow \underset{\tilde{\m v} \in \mathbb{R}^q}{\operatorname{argmin}}\left\{\left\|M^{\top}-\tilde{\m v} \m u^\top\right\|_F^2+P_{\tau_2}(\tilde{\m v})\right\},\label{PenaltyPLS}
\end{aligned} 
\end{equation}
and then use scaling $\m u=\tilde{\m u}/\left\|\tilde{\m u}\right\|_2$  and  $\m v=\tilde{\m v}/\left\|\tilde{\m v}\right\|_2$.  

An $\ell_1$-norm penalty has been adopted by \cite{LeCao2008a} and \cite{Chung2010}: 
\[
P_{\tau_1}(\tilde{\boldsymbol{u}})=\sum_{i=1}^p 2 \tau_1\left|\tilde{u}_i\right| \quad \text { and } \quad P_{\tau_2}(\tilde{\boldsymbol{v}})=\sum_{j=1}^q 2 \tau_2\left|\tilde{v}_j\right|.
\]
These $\ell_1$-norm penalties have the advantage to provide a closed form solution of \eqref{PenaltyPLS} given by
 $\tilde{\m u}=g^{\text {soft }}\left(M \m v, \tau_1\right), \quad \tilde{\m v}=g^{\text {soft }}\left(M^\top \m u, \tau_2\right),
 $
 where $g^{\text {soft }}(\cdot,\tau)$ is the soft thresholding function applied element-wise.
Similar to the non-sparse PLS, the subsequent component pairs are obtained using the procedure above using the deflated versions of $X$ and $Y$ (see Table 1 in the supplementary material).

\section{Best Subset Solution Path for PCA and PLS models}

In this paper, we refer to the `best subset solution' (BSS) path. The BSS path contains $p$ models of varying subset size $k$ that solve the PCA model (as defined by equation (1)) and, respectively, the PLS model (as defined by equation (3)). That is the goal in finding the BSS path is to attempt to recover for each subset size, the subset obtained through an exhaustive search approach.

We first present the problem of the BSS path for the PLS model with univariate response, which is the simpler optimization problem to solve. This particular PLS model is known as PLS1. Then, we move on to present the BSS path for the multivariate case of PLS, called PLS2. We further show that the BSS path for PCA can be  easily derived from the BSS path for PLS2.

\subsection{Best Subset Solution Path for  PLS with Univariate Response}

We now consider the BSS path framework for constructing the first component score. When $q=1$, finding the optimal solution $\m u^*  \in \reals^p$  of \eqref{PLS-g} is given by
\begin{equation}
\m u^* = \frac{X^\top \m y}{\|X^\top \m y\|}. \label{loadingPLS1}
\end{equation}
Now suppose we want to introduce sparsity, in the sense that the new optimization problem is
\begin{align}
\max_{\m u_{[\m s]} \in \reals^k,\,\, \|\m u_{[\m s]} \| = 1} \frac{\langle X_{\m [\m s]} \m u_{[\m s]},\, \m y \rangle}{n}, \quad \text{subject to}\quad \m s \in \{0, 1\}^p, \,\, \sum_{j=1}^p s_j  \leq k, \label{eqn:pls_sparse}
\end{align}
where $X_{\m [\m s]}$ is the matrix constructed from $X$ by removing all its columns with indices $j$ where $s_j = 0$, $k$ is the sparsity parameter that represents the subset size, and $|\m s|$ denotes the number of ones in the binary vector $\m s$. Observe that for any fixed binary vector $\m s$, the optimal solution of \eqref{eqn:pls_sparse} is $
\m u^*_{\m [\m s]} = X_{\m [\m s]}^\top \m y /\left(\|X_{\m [\m s]}^\top \m y\|\right).$

Thus, the optimization problem \eqref{eqn:pls_sparse} can be expressed as 
\begin{align*}
\max_{\m s \in \{0, 1\}^p} \frac{\langle  X_{\m [\m s]} \m u^*_{\m [\m s]},\, \m y \rangle}{n}, \quad \text{subject to}\quad |\m s| \leq k,
\end{align*}
Since,
$\langle X_{\m [\m s]} \m u^*_{\m [\m s]},\, \m y \rangle \,=\, \lt(\m u^*_{\m [\m s]}\rt)^\top X_{\m [\m s]}^\top \m y
\,=\, \frac{\| X_{\m [\m s]}^\top \m y\|^2}{\| X_{\m [\m s]}^\top \m y\|} \,=\, \| X_{\m [\m s]}^\top \m y\|,$
we can express \eqref{eqn:pls_sparse} as
\begin{equation}
\min_{\m s \in \{0, 1\}^p} \lt[ - \frac{\| X_{\m [\m s]}^\top \m y\|}{n}\rt], \quad \text{subject to}\quad\sum_{j=1}^p s_j  \leq k. \label{PLS1S}
\end{equation}
This problem defines the best subset solution path for PLS1. However, solving this problem is NP-hard, and hence, we consider, by exploiting the idea of \cite{COMBSS22}, a Boolean relaxation of \eqref{PLS1S} is given by
\begin{align}
\min_{\m t \in [0, 1]^p} \lt[ -\frac{\|X_{\m t}^\top \m y\|}{n}\rt], \quad \text{subject to}\quad \sum_{j=1}^pt_j \leq k, \label{PLS1t}
\end{align}
where $\m t =(t_1,\ldots,t_p)^\top$, with each $t_j\in[0,1]$, and $X_{\m t}$ is obtained from $X$ by multiplying its $j$-th column with $t_j$ for every $j = 1, \dots, p$. Since minimizing $ - \|X_{\m t}^\top \m y\|$ is equivalent to minimizing $ - \|X_{\m t}^\top \m y\|^2$, to simplify the gradient expression later, we rewrite \eqref{PLS1t}
as
\begin{align}
\min_{\m t \in [0, 1]^p} \lt[ -\frac{\|X_{\m t}^\top \m y\|^2}{n^2}\rt], \quad \text{subject to}\quad \sum_{j=1}^pt_j \leq k. \label{PLS1t-2}
\end{align}
From Theorem~\ref{thm:equi-s-t} (i),  it turns out the solution of the Boolean relaxation \eqref{PLS1t-2} is indeed the exact solution obtained by \eqref{PLS1S}.

\begin{figure}[h] 
    \includegraphics[scale=0.45]{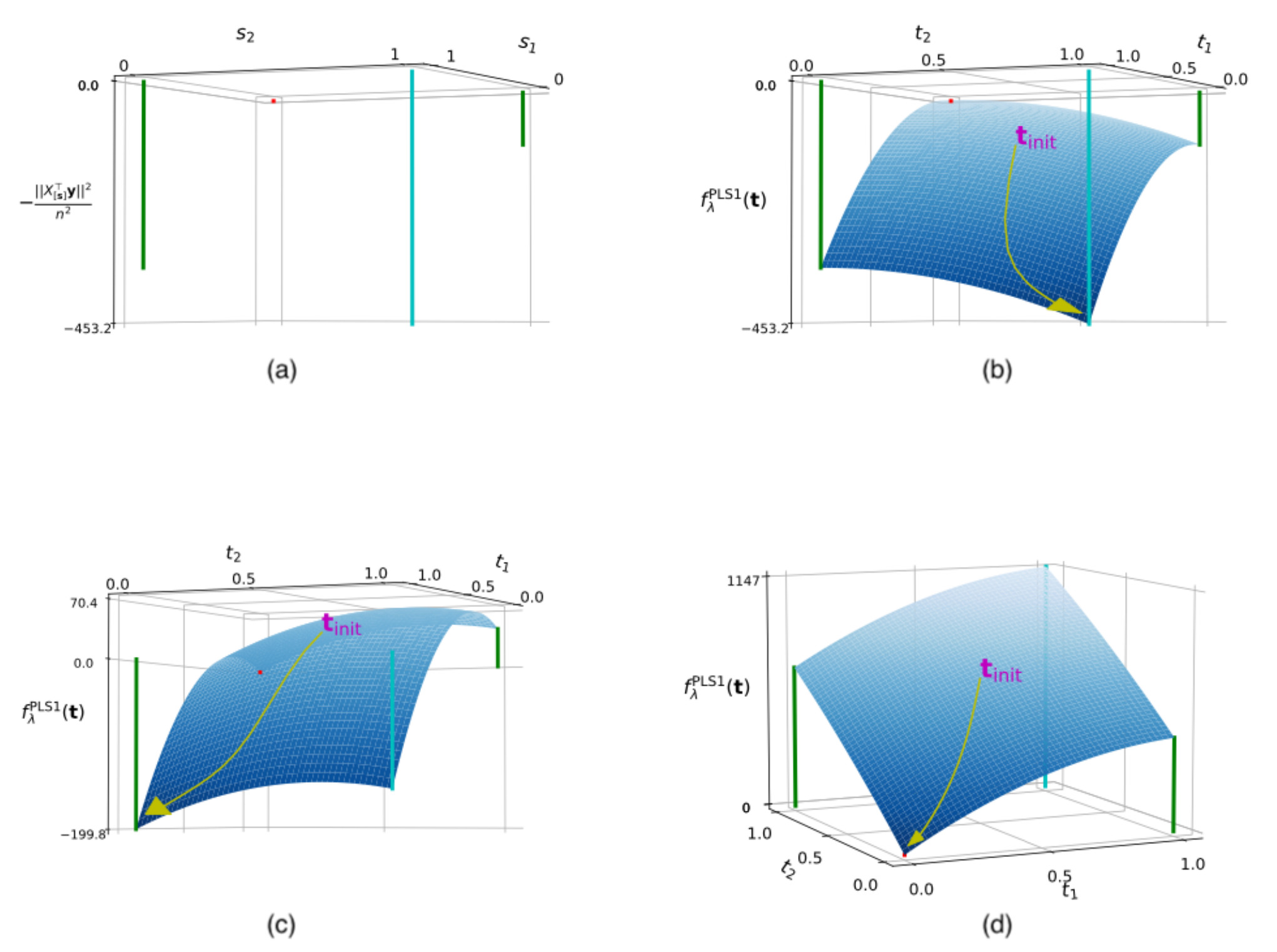}
         \caption{Illustration of the workings of our continuous optimization method using basic gradient descent for an example data with $p = 2$. Plot (a) shows the objective function of the PLS model with univariate response at binary points $\m s \in \{0,1\}^2$. Observe that the best subsets correspond to $k =0$, $k =1$, and $k =2$ are $(1,1)^\top$, $(0, 1)^\top$, and $(0,0)^\top$, respectively.
      Plots (b) - (d) show the objective function of our optimization method \eqref{PLS1} for different values of the parameter $\lambda$. In each of these three plots, the curve (in yellow) shows the execution of basic gradient descent algorithm that, starting at the initial point $\m t_{\mathsf{init}} = (0.5, 0.5)^\top$, converges towards the best subsets of sizes $0, 1$, and $2$. }
  \label{fig:exact-vs-combss}
\end{figure}

Note that the optimization problem in \eqref{PLS1S} is defined using $X_{[\m s]}$ constructed by removing columns from the design matrix $X$ (and hence $X_{[\m s]}$ and $X$ are of different sizes) while $X_{\m t}$ in optimization problem in \eqref{PLS1t} is constructed by multiplying the $j$-th column of $X$ by $t_j$ for every $j$. Thus, both $X_{\m t}$ and $X$ are of the same size. This construction allows us to define our new estimator of the weight vector $\m u_{\m t}$ for all $\m t \in [0, 1]^p$ while guaranteeing that 
 \[
\|X_{\m t}^\top \m y\|=\|X_{\m [\m s]}^\top \m y\|, \quad \textrm{for} \, \, \m t=\m s,
\]
at the corner points $\m s$ of the hypercube $[0, 1]^p$. This construction also
guarantees that the new objective function $-\frac{\|X_{\m t}^\top \m y\|^2}{n^2}$ is smooth over the hypercube as illustrated in Figure~\ref{fig:exact-vs-combss}.

Finally, instead of solving \eqref{PLS1t-2}, we consider
\begin{align}
\label{eqn:f-lambda}
f_{\lambda}^{\plsone}(\m t) = - \frac{\|X_{\m t}^\top \m y\|^2}{n^2} + \lambda  \sum_{j=1}^pt_j,
\end{align}
and solve 
\begin{equation}
\min_{\m t \in [0, 1]^p} f_{\lambda}^{\plsone}(\m t), \label{PLS1}
\end{equation}
using a continuous optimization method, such as basic gradient descent or Adam (as shown in the example of Figure~\ref{fig:exact-vs-combss}).  Theorem~\ref{thm:equi-s-t}   shows that for each $k$ there exists a value of $\lambda$ such that an optimal solution of the box constrained optimization \eqref{PLS1t-2} provides an exact solution of the best subset solution path problem \eqref{PLS1}. A proof of the theorem is presented in Appendix A in the supplementary material. 

\begin{theorem}
\label{thm:equi-s-t}
We have the following equivalence between the optimization problems \eqref{PLS1S}, \eqref{PLS1t-2}, and \eqref{PLS1}.
\begin{itemize}
    \item[] (i) The optimal solutions of the minimization problems defined by \eqref{PLS1S} and \eqref{PLS1t-2}  are identical. 
    \item[] (ii) For every $k =1, \dots, p$,  there exists $\lambda$ such that an optimal solution of \eqref{PLS1t-2} is an optimal solution of \eqref{PLS1}. 
\end{itemize}
\end{theorem}

To execute a gradient descent algorithm to solve \eqref{PLS1}, we use the gradient expression given by
\begin{equation}
\nabla f_{\lambda}^{\plsone}(\m t) = \lambda \mathbf{1} - \frac{2}{n^2}\lt(\m t \odot X^\top \m y \odot X^\top \m y \rt), \label{gradPLS1}
\end{equation}
where $\mathbf{1}$ is is a vector of all ones and $\odot$ is the element-wise product operator.

\textbf{Remark 3.2} From Theorem~\eqref{thm:equi-s-t}, we note that our target continuous optimization problem \eqref{PLS1} provides a solution to the exact best subset solution path problem \eqref{PLS1S}. However, we encounter few challenges in solving \eqref{PLS1}  using a gradient descent algorithm. Lemma A1 in the supplementary material shows that our objective function $f_{\lambda}^{\plsone}(\m t)$ is concave on the hypercube $[0,1]^p$. Thus, depending on the initial point, the converging point of the continuous optimization can be suboptimal. 
This can be particularly an issue if the maximum point $\m t_{\max}$ of $f_{\lambda}^{\plsone}(\m t)$, as a function on $\reals_+^p$, lies within the hypercube.
Note that $\m t_{\max}$ is obtained by equating the gradient \eqref{gradPLS1} to zero. That is, with $ \m z = (X^\top \m y)/n$ (which is independent of $\m t$),
\[
\m t_{\max} = \frac{\lambda \m 1}{2 (\m z \odot \m z)},
\]
where the division is element-wise. Notice that $\m t_{\max}$ can be inside or outside the hypercube $[0, 1]^p$ depending on $\lambda$. The larger the $\lambda$ value, the farther away is $\m t_{\max}$ from the hypercube.  In particular,  $\m t_{\max} \in [0, 1]^p$ if and only if,
\begin{align}
\label{eqn:lambda_bdd}
\lambda \leq 2 \min_{j = 1, \dots, p} z_j^2.    
\end{align}
Since we are interested in sparse solutions, which are achieved when $\lambda$ is larger, the $\m t_{\max}$ usually stays outside the hypercube, allowing our algorithm to depend less on the initial point. In fact, in our simulations, we select a grid of $\lambda$ values over $[0, \lambda_{\max}]$, where $\lambda_{\max} = \sum_{j = 1}^p z_j^2$ which can be much larger than the upper bound in \eqref{eqn:lambda_bdd} and corresponds to the empty model. Desired sparse models are achieved for large values of $\lambda$ on the grid. 

\textbf{Remark 3.3} Exploiting the continuity of the new objective function enables gradient descent algorithms to explore a huge space of models while converging in a few iterations towards identifying the best subset. By increasing the value of $\lambda$, we can increase the sparity of the solution of the optimization problem \eqref{PLS1t}, because the penalty $\lambda \sum_{j=1}^pt_j$ encourages sparsity in $\m t$ (see Figure~\ref{fig:exact-vs-combss}).  Note that even though there is a mapping between the sparsity of the solution and the value of $\lambda$, since in practice we obtain solutions over a grid of $\lambda$, there is a chance of not seeing models corresponds to some values of $k$. To overcome this issue and to reduce the reliance on the $\lambda$ parameter, in Section~\ref{s:algorithm}, we provide an improved version of the algorithm so that a single run of the algorithm provides a list of subsets of desired sizes $k$. In this case, $\lambda$ can be viewed as an exploration parameter rather than a sparsity parameter.

\subsection{Best Subset Solution Path for  PLS with Multivariate Response}

Recall that, for a design matrix $X \in \reals^{n \times p}$ and a multivariate response matrix $Y \in \reals^{n \times q}$, the solution of \eqref{PLS-g} is given by the pair $(\m u_1, \m v_1)$ being the first left and right singular vector of $M=(X^\top Y)/n$ associated with the largest singular value $\delta_1$. Further, it is well known that
\begin{equation} \label{PLS-max}
\max_{\m u \in \reals^p,\,\, \m v \in \reals^q \|\m u \| = 1, \|\m v \| = 1} \frac{\langle X \m u,\,  Y \m v \rangle}{n}=\delta_1.
\end{equation}
 Note that the square of this largest singular value is the largest eigenvalue of the symmetric matrices $M^\top M$ and $M M^\top$. 
Indeed,
$M^\top M= \sum_{k=1}^r\delta^2_k\m v_{k}\m v_{k}^\top$, and $M M^\top = \sum_{k=1}^r\delta^2_k\m u_{k}\m u_{k}^\top.$

Consequently, 
$
\delta_1^2=\m v_1^\top M^\top M \m v_1=\m u_1^\top M M^\top \m u_1.
$
Note that the largest eigenvalue $\eta^*$ of any symmetric matrix $A$ (of size $p\times p$) can be 
 attained by exploiting the \textit{power method} which is described by the recurrence relation,
$ \omega^{(\ell+1)} = A \m \omega^{(\ell)}/\left(\left\|A \m \omega^{(\ell)}\right\|\right),$
with $\omega^{(0)}\in \Re^p$ a random unit vector. The sequence of eigenvalues $\eta^{(\ell+1)}=(\m \omega^{(\ell+1)})^\top A \m \omega^{(\ell+1)}$ converges to the largest eigenvalue of $A$. That is, 
$\eta^{(\ell+1)} \longrightarrow \eta^*$ as $\ell \to \infty$.  By choosing $A=MM^\top$ or $A=M^\top M$ the power method enables us to get $\delta_1^2$.
Similar to the BSS path for PLS1, we introduce sparsity into the $X$ matrix through the matrix $X_{\m t}$ as described earlier (see Section 3.1). According to \eqref{PLS-max}, we get
\begin{equation} \label{PLS-with-t}
\delta_{\m t}=\max_{\m u \in \reals^p,\,\, \m v \in \reals^q \|\m u \| = 1, \|\m v \| = 1} \frac{\langle X_{\m t} \m u,\,  Y \m v \rangle}{n}=\sqrt{\m v_{\m t}^\top M_{\m t}^\top M_{\m t} \m v_{\m t}} = \sqrt{\m u_{\m t}^\top M_{\m t} M_{\m t}^\top \m u_{\m t}},
\end{equation}
where 
$M_{\m t}= (X_{\m t}^\top Y)/n$, and $\m v_{\m t}$ and $\m u_{\m t}$ are respectively the first left and the first right singular vectors of $M_{\m t}$ associated to the largest singular value $\delta_{\m t}$. In other words, $\m v_{\m t}$ is the eigenvector associated with the highest eigenvalue, $\delta^2_{\m t}$, of  $M_{\m t}^\top M_{\m t}$ and $\m u_{\m t}$ is the eigenvector associated with the highest eigenvalue, again $\delta^2_{\m t}$, of  $M_{\m t} M_{\m t}^\top$.

Define,
\begin{align}
\label{eqn:flambda_pls2}
f_{\lambda}^{\plstwo}(\m t) = -\delta^2_{\m t} + \lambda  \sum_{j=1}^pt_j.
\end{align}
Then, our aim is to solve the following relaxation optimization problem:
\begin{equation}
\max_{\m t \in [0, 1]^p} \left(\delta^2_{\m t} - \lambda  \sum_{j=1}^pt_j\right) \ \ \ \ \ \textrm{or, equivalently,} \ \ \ \ \ \ \ \ \ \ \min_{\m t \in [0, 1]^p} f_{\lambda}^{\plstwo}(\m t) \label{optPLS2}
\end{equation}
Towards this, we need the gradient $\nabla \delta^2_{\m t} = \lt(\partial \delta^2_{\m t}/\partial t_1, \dots, \partial \delta^2_{\m t}/\partial t_p \rt)$. Each term of this gradient can be expressed as either
\begin{align}
\frac{\partial \delta^2_{\m t}}{\partial t_j} &= \frac{\partial \m v_{\m t}^\top M_{\m t}^\top M_{\m t} \m v_{\m t}}{\partial t_j}, \label{eqn:der_v}
\end{align}
or, 
\begin{align}
\frac{\partial \delta^2_{\m t}}{\partial t_j}&= \frac{\partial \m u_{\m t}^\top M_{\m t} M_{\m t}^\top \m u_{\m t}}{\partial t_j}. \label{eqn:der_u}
\end{align}
Due to computational reasons, whether we use \eqref{eqn:der_v} or \eqref{eqn:der_u} depends on whether $q < p$ or $q > p$, respectively. Suppose, we consider \eqref{eqn:der_v}. Then,
\begin{align}
\frac{\partial \delta^2_{\m t}}{\partial t_j} &= \frac{\partial \m v_{\m t}^\top M_{\m t}^\top M_{\m t} \m v_{\m t}}{\partial t_j}
 = \m v_{\m t}^\top\frac{\partial  M_{\m t}^\top M_{\m t}}{\partial t_j} \m v_{\m t}+ 2\left(\frac{\partial \m v_{\m t}}{\partial t_j}\right)^\top  M_{\m t}^\top  M_{\m t} \m v_{\m t}.\label{eqn:delta2_dervative} 
\end{align}
Recall that $\delta_{\m t}^2$ is an eigenvalue of $M_{\m t}^\top  M_{\m t}$ with the corresponding unit length eigenvector being $\m v_{\m t}$. Thus, 
$M_{\m t}^\top  M_{\m t} \m v_{\m t} = \delta_{\m t}^2 \m v_{\m t}.$

Also, since $\m v_{\m t}^{\top} \m v_{\m t} = \|\m v_{\m t}\|^2 = 1$, for every $j$, we get
$0 = \frac{\partial \m v_{\m t}^\top \m v_{\m t}}{\partial t_j} = 2 \lt(\frac{\partial \m v_{\m t}}{\partial t_j}\rt)^\top \m v_{\m t}.
$
Thus, the second term on the right hand side of \eqref{eqn:delta2_dervative} is equal to $0$ because
$
\left(\frac{\partial v_{\m t}}{\partial t_j}\right)^\top  M_{\m t}^\top  M_{\m t} \m v_{\m t} = \delta_{\m t}^2\left(\frac{\partial \m v_{\m t}}{\partial t_j}\right)^\top  \m v_{\m t} = 0.
$
Therefore, 
$
\frac{\partial \delta^2_{\m t}}{\partial t_j} =  \m v_{\m t}^\top\frac{\partial  M_{\m t}^\top M_{\m t}}{\partial t_j} \m v_{\m t}.
$
Now note that 
\[
\frac{\partial  M_{\m t}^\top M_{\m t}}{\partial t_j} = \frac{\partial  M^\top T_{\m t}^2 M}{\partial t_j} = M^\top \frac{\partial   T_{\m t}^2}{\partial t_j} M = 2 t_j M^\top E_j M,
\]
where $E_j$ is a $p\times p$-dimensional matrix with the $j$th diagonal element being $1$ while every other element is $0$. Thus, we get,
\[
\frac{\partial \delta^2_{\m t}}{\partial t_j} = 2t_j (M \m v_{\m t})^\top E_j (M \m v_{\m t}).
\]
Hence, 
\begin{align}
\nabla \delta^2_{\m t} = 2 \lt(\m t \odot (M \m v_{\m t}) \odot (M \m v_{\m t}) \rt).
\label{eqn:delta2_dervative_final_v} 
\end{align}
Similarly, we can also show that 
\begin{align}
\nabla \delta^2_{\m t} = 2 \lt(\m u_{\m t} \odot \lt(M M^\top (\m t \odot \m u_{\m t}) \rt) \rt).
\label{eqn:delta2_dervative_final_u} 
\end{align}

\textbf{Remark 3.4} \normalfont Note that ${\m v}_{\m t}$ or ${\m u}_{\m t}$ are obtained using the power method, which has computational complexity for finding the eigenvector of square matrix $A$ depending on the dimension of $A$. In particular, the smaller the dimension of $A$, the faster the power method. Since ${\m v}_{\m t}$ is obtained using $A = M_{\m t}^\top M_{\m t}$ and ${\m u}_{\m t}$ is obtained using $A = M_{\m t} M_{\m t}^\top$, it is reasonable to use \eqref{eqn:delta2_dervative_final_v} when $q < p$ and to use \eqref{eqn:delta2_dervative_final_u} when $q \geq p$.

\subsection{Best Subset Solution Path for  PCA}

Recall that in PCA, the optimal solution $\m u^*  \in \reals^p$ of \eqref{optPCA}  is given by the eigenvector associated to the largest eigenvalue of the sample covariance of the data $S=(X^\top X)/n$.  Hence, the sampling variance of the first component score is equal to the largest eigenvalue $\delta^{*}=\textrm{var}( X \m u^*)$. 

Similar to the BSS path for PLS1 and PLS2, we introduce sparsity in PCA by defining the optimization problem given by
\begin{align}
 \max_{\m u \in \reals^p,\,\, \|\m u \| = 1}\frac{1}{n}\m u^\top X_{\m s}^\top X_{\m s} \m u,
 \quad \text{subject to}\quad \m s \in \{0, 1\}^p, \,\, |\m s| \leq k. \label{eqn:pca_sparse}
\end{align}
Observe that for any fixed binary vector $\m s$, the optimal solution of \eqref{eqn:pca_sparse} is the eigenvector associated with the largest eigenvalue of the sample covariance of the data $S_{\m s}=(X_{\m s}^\top X_{\m s})/n$.
Thus, the optimization problem \eqref{eqn:pca_sparse} can be expressed as
\begin{align*}
\min_{\m s \in \{0, 1\}^p} -\frac{1}{n}\m u^\top X_{\m s}^\top X_{\m s} \m u, \quad \text{subject to}\quad |\m s| \leq k.
\end{align*}
This problem defines the best subset solution path for PCA. 

We again use $X_{\m t}$ to write a continuous relaxation of \eqref{eqn:pca_sparse}. In particular, we consider 
\begin{equation} 
\delta_{\m t}=\max_{\m u \in \reals^p,\ \|\m u \| = 1} \textrm{var} (X_{\m t}\m u)=\frac{1}{n}\m u_{\m t}^\top X_{\m t}^\top  X_{\m t} \m u_{\m t}, \label{PCA-with-t}
\end{equation}
where $\m u_{\m t}$ is the eigenvector associated to the largest eigenvalue $\delta_{\m t}$ of the matrix $S_{\m t} = X_{\m t}^\top  X_{\m t}/n$. Thus, by taking $f^{\pca}_{\lambda}(\m t)=-\delta_{\m t}+ \lambda \sum_{j=1}^p t_j$, our goal is to solve

\begin{equation}
\min_{\m t \in [0, 1]^p} f^{\pca}_{\lambda}(\m t). \label{PCA}
\end{equation} 
Towards this, we use the gradient expression given by
$
\nabla f^{\pca}_{\lambda}(\m t) = \lambda \mathbf{1}_p - \nabla \delta_{\m t}.$
By observing the similarity with the PLS2 framework, especially \eqref{PLS-with-t} and \eqref{PCA-with-t},  the gradient vector 
$\nabla \delta_{\m t}$ is obtained using \eqref{eqn:delta2_dervative_final_u} by substituting $MM^\top$ with $X^\top  X/n$.

\section{Implementation}
\label{s:algorithm}
Building on \cite{COMBSS22}, we reformulate the box constrained problems \eqref{PLS1}, \eqref{optPLS2} and \eqref{PCA} into an equivalent unconstrained problem by considering $\m t = \m t(\m r)$ given by 
\begin{align}
\label{eqn:Def_tw}
t_j(r_j) = 1 - \exp(-r_j^2),\quad j = 1, \dots, p.
\end{align}
Then we rewrite \eqref{PLS1}, \eqref{optPLS2}, or \eqref{PCA} as,
\begin{align}
\label{eqn:uccbss1}
\min_{\m r \in \reals^p} f_{\lambda}\lt(\m t(\m r)\rt),
\end{align}
where $f_{\lambda}$ is either $f_{\lambda}^{\plsone}$, $f_{\lambda}^{\plstwo}$, or $f_{\lambda}^{\pca}$ depending on whether the model is PLS1, PLS2, or PCA, respectively. 
The unconstrained optimization problem \eqref{eqn:uccbss1} is equivalent to the box constrained problem (\eqref{PLS1}, \eqref{optPLS2}, or \eqref{PCA}), because for any $a, b \in \reals$,
${1 - \exp(-a^2) < 1 - \exp(-b^2)}$ if and only if $a^2 < b^2.$
Thus, by defining, $g_{\lambda}(\m r) = f_{\lambda}\lt(\m t(\m r)\rt),$
we solve,
\begin{align}
\label{eqn:uccbss}
\min_{\m r \in \reals^p} g_{\lambda}(\m r),
\end{align}
via a continuous optimization method. Note that the gradient expression of the objective function of linear regression in \cite{COMBSS22} is complicated and requires linear equation solvers like conjugate gradient descent in its implementation of the algorithm. On the other hand, the objective functions $f_{\lambda}(\m t)$ in this paper have simpler gradient expressions, making the algorithm faster and easy to implement.

\subsection{Algorithm for the first component score}

Algorithm~\ref{alg:COMBSS} presents pseudo-code of our continuous optimization method.
Step~\ref{step:GD} of Algorithm~\ref{alg:COMBSS} calls a gradient descent method of choice to minimize the unconstrained objective function $g_{\lambda}(\m r)$ using the gradient $\nabla g_{\lambda}(\m r)$ with $\m r^{(0)}$ as the initial point. The gradient descent algorithm terminates when a predefined termination condition is satisfied to return $\m r_{\mathsf{path}} = ( \m r^{(0)}, \m r^{(1)}, \dots )$, the sequence of all the points $\m r$ visited during its execution. A most common robust termination condition is to stop when the change in $\m r$ is significantly small over a fixed number of consecutive iterations. In Step~\ref{step:w-to-t}, each $\m r^{(l)}$ is mapped to a point $\m{t}^{(l)}$ on the hypercube $[0, 1]^p$ via the map \eqref{eqn:Def_tw} to obtain the sequence  $\m t_{\mathsf{path}} = ( \m t^{(0)}, \m t^{(1)}, \dots )$. 

Steps~\ref{step:initiate_empty} to \ref{step:collection_of_models} collect several subsets for each size  $k = 1,\ldots,K$ using the points in $\m t_{\mathsf{path}}$. In particular, we start with an empty set $\mc M_k$ for each $k$. We then take each point $\m t$ in $\m t_{\mathsf{path}}$ and sort the elements of $\m t$ in descending order. Suppose $j_1,  \dots, j_K$ are the indices of the first $K$ largest elements of $\m t$ in descending order. Then, we take $\m s_k \in \{0, 1\}^p$ to be a binary vector with ones only at positions $j_1, \dots, j_k$ and add $\m s_k$ to $\mc M_k$. Finally, at Step~\ref{step:model_selection}, for each $k$, we select a best subset $s^*_k$ among all the subsets in the set $\mc M_k$.

\begin{algorithm}
\DontPrintSemicolon
 \KwInput{Data: $X$ for PCA; $(X, Y)$ for PLS
 
  \hspace{1.3cm} Tuning parameter $\lambda$
  
  \hspace{1.3cm} The initial point $\m r^{(0)}$
  
   \hspace{1.3cm} Largest subset size $K$
   
  }
  
  \KwOutput{A list of $K$ subsets of sizes from $1$ to $K$}

$\m r_{\mathsf{path}} \leftarrow \mathsf{GradientDescent}\lt(\m r^{(0)}, g_{\lambda}, \nabla g_{\lambda}\rt)$ \label{step:GD}
 
Obtain $\m t_{\mathsf{path}}$  from $\m r_{\mathsf{path}}$ using the map $\m t \leftarrow \m 1 - \exp(- \m r \odot \m r)$ \label{step:w-to-t}

 $\mc M_k \leftarrow \{ \}$ for each $k \leq K$ \label{step:initiate_empty}
 
\For{each $\m t = (t_1, \dots, t_p)^\top$ in $\m t_{\mathsf{path}}$}{
Let $t_{j_1}, t_{j_2}, \dots, t_{j_K}$ be the $K$ largest elements of $\m t$ in the descending order

           \For{$k = 1$ to $K$}{
                       Take $\m s_k \in \{0,1\}^p$ with non-zero elements only at $j_1, \dots, j_k$
                        
                       $\mc M_k \leftarrow \mc M_k \cup \{\m s_k\}$
           }
 } \label{step:collection_of_models}
\For{$k = 1$ to $k = K$}{
        $\m s^*_k \leftarrow \argmin_{{\m s} \in \mc M_k} f_0(\m s)$ where $f_0(\m s)$ is the objective function with zero penalty \label{step:model_selection}
}
\Return ${\mc M = \{\m s^*_1, \dots, \m s^*_K\}}$ 
 \label{step:end2}
\caption{Best Subset Solution Path}
\label{alg:COMBSS}
\end{algorithm}

In practice, we call the algorithm for several values of $\lambda$ and each value of $\lambda$ provides one subset for every $k = 1, \dots, K$.  For instance, if we use $100$ values of $\lambda$, we get $100$ subsets of size $k$, for every $k$. At the end, for each $k$, we select the best subset among all the $100$ options. Therefore, the final solution for each $k$ depends on the values we select for $\lambda$. One simple approach is to preselect a grid of values for $\lambda$ before using the algorithm. However, since the surface of the objective function $f_{\lambda}(\m t)$ is data dependent, it is more meaningful to select the values for $\lambda$ in a data dependent manner so that the surface of  $f_{\lambda}(\m t)$ is explored well by the algorithm. Below we describe one such data dependent approach for selecting a grid of values for $\lambda$.

\subsection{Dynamic grid of $\lambda$ values}
Suppose we want to call Algorithm~\ref{alg:COMBSS}  for a grid of at most $L$ values for $\lambda$. For each $\lambda$, the algorithm converges to a point ${\m t} \in [0, 1]^p$ where some of the $t_j$'s are very close to $0$ indicating that the corresponding columns of $X$ are insignificant for that $\lambda$.  We can create a subset from this terminal $\m t$ by mapping all the insignificantly small values to $0$ and others to $1$ using a threshold parameter $\rho\in (0, 1)$. That is, we have a subset, say $\m s \in \{0, 1\}^p$, obtained by $
s_j = I(t_j > \rho)$, $j = 1, \dots, p.$
Let $k_\lambda = |\m s|$, the size of the terminal subset. 

Now to create a dynamic grid of at most $L$ values, we take $\lambda_{\mathsf{max}}$ to be the largest eigenvalue of $M^\top M$ for the PLS model and take it to be the largest eigenvalue of $X^\top X/n$ for the PCA model. In either case, the terminal subset obtained for $\lambda_{\mathsf{max}}$ is empty, that is, $\m s$ is an all zero vector. Then the dynamic grid is constructed as follows:
\begin{itemize}
\item[] {\em Step 1}: For $\ell = 1, 2, \dots$, call Algorithm~\ref{alg:COMBSS} with $\lambda = \lambda_{\mathsf{max}}/2^{\ell}$ until either $\ell = L$ or $k_\lambda \geq K$. Let $\ell'$ be the final value of $\ell$, that is, $\ell'$ is the number of times Algorithm~\ref{alg:COMBSS} is called so far. Also, let $\Lambda = \{\lambda_{\mathsf{max}}/2^{\ell} : \ell = 0, 1, \dots, \ell'\}$. If $\ell' < L$, go to Step 2; otherwise, terminate the procedure.
\item[] {\em Step 2}: Suppose that the sequence $\lambda_1 < \cdots < \lambda_{|\Lambda |}$ are the elements of $\Lambda$ in the ascending order, $|\Lambda|$ denotes the number of elements in $\Lambda$. Moving from left to the right on the sequence, if $k_{\lambda_\ell} > k_{\lambda_{\ell+1}} + 1$ for some $\ell = 1, \dots, |\Lambda|$,  call Algorithm~\ref{alg:COMBSS} with $\lambda = (\lambda_\ell + \lambda_{\ell+1})/2$ and add this $\lambda$ to $\Lambda$. Terminate the procedure either if the number of times Algorithm~\ref{alg:COMBSS} is called in this step is $L - \ell'$ or there is no $\ell$ such that $k_{\lambda_\ell} > k_{\lambda_{\ell+1}} + 1$; otherwise, repeat Step 2.
\end{itemize}

\subsection{Subsequent component score}
The output of the BSS path algorithm is $K$ subsets (one for each subset size $k = 1, \dots, K$) and so $K$ different sparse scores which are linear combinations of the variables included in the potential subset. The subsequent score (for PCA) and the pair of scores (for PLS) are obtained using the same algorithm on the respective deflated matrices (with the same dimension as the original matrices), i.e., after removing the information contained in the previous component or pairs of scores. Then, the subsequent sparse component is more likely to be constructed using variables that are different from the variables used for constructing the previous components. The construction of the deflated matrices are provided in Table 1 in the supplementary material. One can think of exploring all the $K$ potential deflation matrices by using the BSS path algorithm on each of them to get a best subset for the second component for PCA (or, the second pair of components for PLS), and repeat this for the further subsequent  components. However, this strategy can be computationally expensive. To reduce the complexity, we suggest in practice to pick only one subset from all the $K$ best subsets to create the first sparse component before each deflation step using a specific strategy. Some useful strategies are described at the end of this section.

For the PCA model, we propose two ad-hoc approaches based on the percentage of variance explained (PEV). One can choose the size of the best subset for the first component by monitoring the percentage decrease in the PEV compared to a non-sparse PCA. This strategy is  used in \cite{zou2006sparse}. The second strategy corresponds to the ad-hoc approach proposed in \cite{Shen2008} which is based on the cumulative PEV (CPEV). In particular, we select the smallest best subset whose CPEV  is within, say, $10\%$ of the CPEV of the largest best subset.
Note that the definition of CPEV  from \cite{Shen2008} is adjusted to take into account the non-orthogonality between sparse components. This strategy is illustrated with an application in Section~\ref{sec:real-PCA}.

For the PLS model in a regression mode, one can select the best subset before deflation (number of variables to keep for constructing the score) using the best prediction accuracy such as Mean Absolute Error (MAE), Mean Squared Error (MSE), or $R^2$ (square of the correlation between prediction and observed outcome). In Appendix C in the supplementary material, we provide explicit formulae to express the PLS model in terms of the original variables in a regression setting. For the PLS model in the canonical mode, the best subset before deflation can be chosen using the absolute correlation between the pair of scores. Typically, these measurements are obtained using $v$-fold cross-validation. Note that all these proposed ad-hoc approaches are guidelines for selecting a best subset before each deflation step. However, selection of an appropriate approach is based on the domain knowledge of the study.

Finally, we want to stress a possible side effect of working with deflated matrices. The first score $\boldsymbol{\xi}_1=X \boldsymbol{u}_1$ is built as a sparse linear combination (with weights in $\boldsymbol{u}_1$) of the original variables. The second score $\boldsymbol{\xi}_2=X_1 \boldsymbol{u}_2$ is built as a sparse linear combination (with weights in $\boldsymbol{u}_2$) of the original variables that have not been already explained by the first score variables. More generally, the $h$-th score variable, $\boldsymbol{\xi}_h=X_{h-1} \boldsymbol{u}_h$, is built as a sparse linear combination of the original variables, from which we extract (by projection) the information not already brought by the previous score variables. However, it is possible to calculate the adjusted weights ${w}_h$ such that $\boldsymbol{\xi}_h=\boldsymbol{X} \boldsymbol{w}_h$.
These weights allow for direct interpretation of the selected variables in the PLS model. Note that although $\boldsymbol{w}_h$ and $\boldsymbol{z}_h$ allow for direct interpretation of the selected variables, the sparsity is enforced on $\boldsymbol{u}_h$. So if $\boldsymbol{u}_h$ is sparse, this does not necessarily mean that the adjusted weights $\boldsymbol{w}_h$ will be sparse. We provide in Appendix C in the supplementary material information on how to estimate the adjusted weights and we refer to \cite{de2019pls} for more details.

\section{Simulation Study}

In this section, we first focus on the efficacy of the proposed approach in retrieving potential best subsets of given sizes in constructing the first component for the PLS model. More precisely, we focus on the capacity of our method for providing the optimal subset for the first component (i.e., solution of \eqref{eqn:pls_sparse}) for any subset size $k$. We also provide out-of-sample prediction measures such as the mean square error in prediction (MSEP) and subset selection accuracy  through the evaluation of  sensitivity, specificity and F1-score for retrieving the support recovery of the true simulated signal.  Then, we present a simulation for a model with two components.  Prediction power and variable selection is investigated in the  univariate response case and compared to a lasso model \cite{Tibshirani1994}. Finally, we present an experiment for the PCA model where we show the ability of constructing sparse components without a loss of variance explained.
 
We compare our approach to the sparse PLS and sparse PCA methods (denoted respectively as sPLS and sPCA) as offered in the package \texttt{mixOmics}. For the PCA, we also compare our method to the sparse PCA method presented in 
\cite{zou2006sparse} as available in \texttt{elasticnet} and denoted SPCA. All numerical results of our simulation study are reproducible using R code that is made available at \url{https://github.com/benoit-liquet/BSS-PCA-PLS}.

\subsection{Simulation design}
We use a model similar to the model used in \cite{sutton2018sparse}. In particular, we consider the latent PLS underlying model with multivariate responses given by
\begin{equation}
X=T C^\top+E_X, \quad Y=SD^\top+E_Y, \label{simumodel}
\end{equation}
where $T=(\boldsymbol{\xi}_1,\dots,\boldsymbol{\xi}_H) \in \mathbb{R}^{n\times H}$ collects the latent variables whose elements are independently generated from the uniform distribution $U(-1,3)$. 
The rows of the residual matrix $E_X$ (respectively, $E_Y$ ) are generated from a mean-zero multivariate normal distribution with covariance matrix $\Sigma_X=\sigma I_p$ (respectively, $\Sigma_Y=\sigma I_q$ ). The regression coefficients in $C\in \mathbb{R}^{p\times H}$  enable us to specify the `true' (i.e., active) X-variables linked to the response Y-variables.  In the regression setting, we use the inner relationship $S=TB$, and so  the X-score variables are simulated and used to construct $S$ (see Appendix~C in the supplementary material for more details). The response $Y$ is simulated with $q=10$ variables. We arbitrarily set the elements of the matrix $BD^\top\in \mathbb{R}^{H\times Q}$ with elements independently generated from the uniform distribution $U(0.5,10)$. 

We first consider the case of a single component, that is, $H=1$. Then, $C=\left(c_1, \ldots, c_p\right)^\top$ is a $p$-dimensional vector  with $c_j \neq 0$ if the corresponding variables $X^j$ ($j$th column of $X$) are true variables (i.e., associated to one of the latent variable $\xi_1$) and $c_j = 0$ otherwise. 

In this simulation study, we investigate the effect of the noise (through $\sigma$ parameter), the effect of the sample size ($n$),  effect of the true signal (through sparsity parameter $\gamma$) and the effect of the dimension $p$ of the data matrix $X$ on the efficacy of our algorithm. For a small dataset $X$, we use an exhaustive method to find the exact (``optimal") solution of the best subset for any subset size ranging from $1$ to $p$. Here,  by ``optimal" solution we mean a solution of the problem as stated in  \eqref{eqn:pls_sparse}. We assess our method in retrieving the exact (``optimal") best subset for each subset size.

\subsection{Effect of the noise}
Here, we investigate the effect of the noise level on the performance of our approach in finding the best subsets.
We take $p=15$, $q=10$, $n=100$ and the sparsity of the model $\gamma=5$, meaning that only $p-\gamma=10$ variables from the $X$ data matrix are associated to the multivariate response~$Y$. We set  $C =(0,0,0,0,0,1,-1,1,-1,1,-1,1,-1,1,-1)^\top$. We use $4$ standard deviation values $\sigma \in \{1.5, 3,6, 8\}$ for the noise and this corresponds to an estimated signal-to-noise of 2.6, 0.8, 0.3 and 0.2.
For each level of noise, the BSS path for PLS (BSS-PLS) consists most of  the time of the ``optimal" best subset for every subset size, while sparse PLS selects  the ``optimal" best subset  relatively less frequent (see Table~\ref{tabnoise}) which is expected as sparse PLS is not designed/optimized to find the ``optimal subset''.

Table 2 in the supplementary material presents out-of-sample prediction trough the MSEP for each subset size using a new test set of size $n/2$ to mimic the situation $2/3$ of the data for training and $1/3$ of the data for testing. In this case,  our BSS path method is slightly better than the sparse PLS method. In terms of support recovery, both methods perform similarly in terms of specificity, sensitivity and F1 score (see Table 3 and 4 in the supplementary material).

\begin{table}[h]
\centering
\caption{Number of times  BSS-PLS and sparse PLS retrieve the true best subset for different subset sizes  over 100 runs for varying noise levels. Here, $p=15$, $q=10$, $n=100$, and $\gamma=5$.} \label{tabnoise}
\begin{tabular}{ccccc|cccc}
  \hline
& \multicolumn{4}{c}{BSS-PLS} &  \multicolumn{4}{c}{Sparse PLS}\\
\cmidrule{2-5}\cmidrule{6-9}
Subset size &$\sigma =  1.5$ & $\sigma =  3$ & $\sigma =  6$& $\sigma =  8$&$\sigma =  1.5$ & $\sigma =  3$ & $\sigma =  6$& $\sigma =  8$\\
  \hline
 \hline
1 & 100 & 100 & 100 & 100 & 100 & 97 & 95 & 85 \\ 
  2 & 99 & 96 & 96 & 98 & 99 & 95 & 92 & 82 \\ 
  3 & 100 & 100 & 97 & 96 & 100 & 97 & 89 & 81 \\ 
  4 & 99 & 98 & 98 & 95 & 97 & 96 & 85 & 81 \\ 
  5 & 99 & 100 & 95 & 93 & 98 & 98 & 83 & 87 \\ 
  6 & 100 & 98 & 96 & 96 & 100 & 97 & 89 & 75 \\ 
  7 & 99 & 99 & 95 & 99 & 99 & 99 & 89 & 82 \\ 
  8 & 100 & 100 & 98 & 98 & 100 & 99 & 90 & 79 \\ 
  9 & 100 & 99 & 97 & 97 & 99 & 99 & 93 & 75 \\ 
  10 & 100 & 100 & 99 & 99 & 100 & 100 & 96 & 80 \\ 
  11 & 100 & 100 & 100 & 100 & 100 & 100 & 94 & 86 \\ 
  12 & 100 & 100 & 100 & 100 & 100 & 98 & 95 & 95 \\ 
  13 & 100 & 100 & 100 & 100 & 100 & 99 & 94 & 95 \\ 
  14 & 100 & 100 & 100 & 100 & 100 & 100 & 99 & 94 \\  \hline
 \end{tabular}
\end{table}

\subsection{Effect of the sample size}
We investigate the performance of our method when the sample
size $n$ is increasing, by varying $n$ over $\{100,200,500\}$. This simulation corresponds to the situation where $p=15$, $q=10$, $\sigma=6$, and the sparsity of the model is set to be $5$, similar as in the previous simulation setting. Results are presented in Table 5 in the supplementary material. We observe that for every $n$,  BSS-PLS retrieves most of  the time the ``optimal" best subset for every subset size. The sparse PLS selects  the ``optimal" best subset  relatively less frequent. However, the performance of the sparse PLS seems to improve as the sample size increases. 

Table 6 in the supplementary material presents out-of-sample prediction trough the MSEP for each subset size. Our BSS path method is again slightly better than the sparse PLS method. In terms of the support recovery, the two methods have similar performance in terms of specificity, sensitivity and F1-score (see Table 7 and 8 in the supplementary material).

\subsection{Effect of the sparsity}
We investigate the performance of our method when the sparsity of the generated model is varied. In particular, we take the  sparsity $\gamma \in\{3,7,9,11\}$.
This simulation corresponds to the situation when $p=15$, $q=10$, $\sigma=5$, and $n=100$. Results are presented in Table 9 in the supplementary material. For every sparsity level $\gamma$ of the true generated model,  BSS-PLS enables us to retrieve most of the time the ``optimal" best subset of any subset size. In this simulation setting, sparse PLS selects the exact best subset less frequently.

Table 10 in the supplementary material presents out-of-sample prediction trough the MSEP for each subset size. Our BSS path method exhibits slightly better performance than the sparse PLS method. In terms of the support recovery, both methods perform similarly in terms of specificity, sensitivity and F1-score (see Tables 11 and 12 in the supplementary material).


\subsection{Effect of the dimension ${p}$}
\begin{figure}
\centerline{
 \includegraphics[scale=0.47]{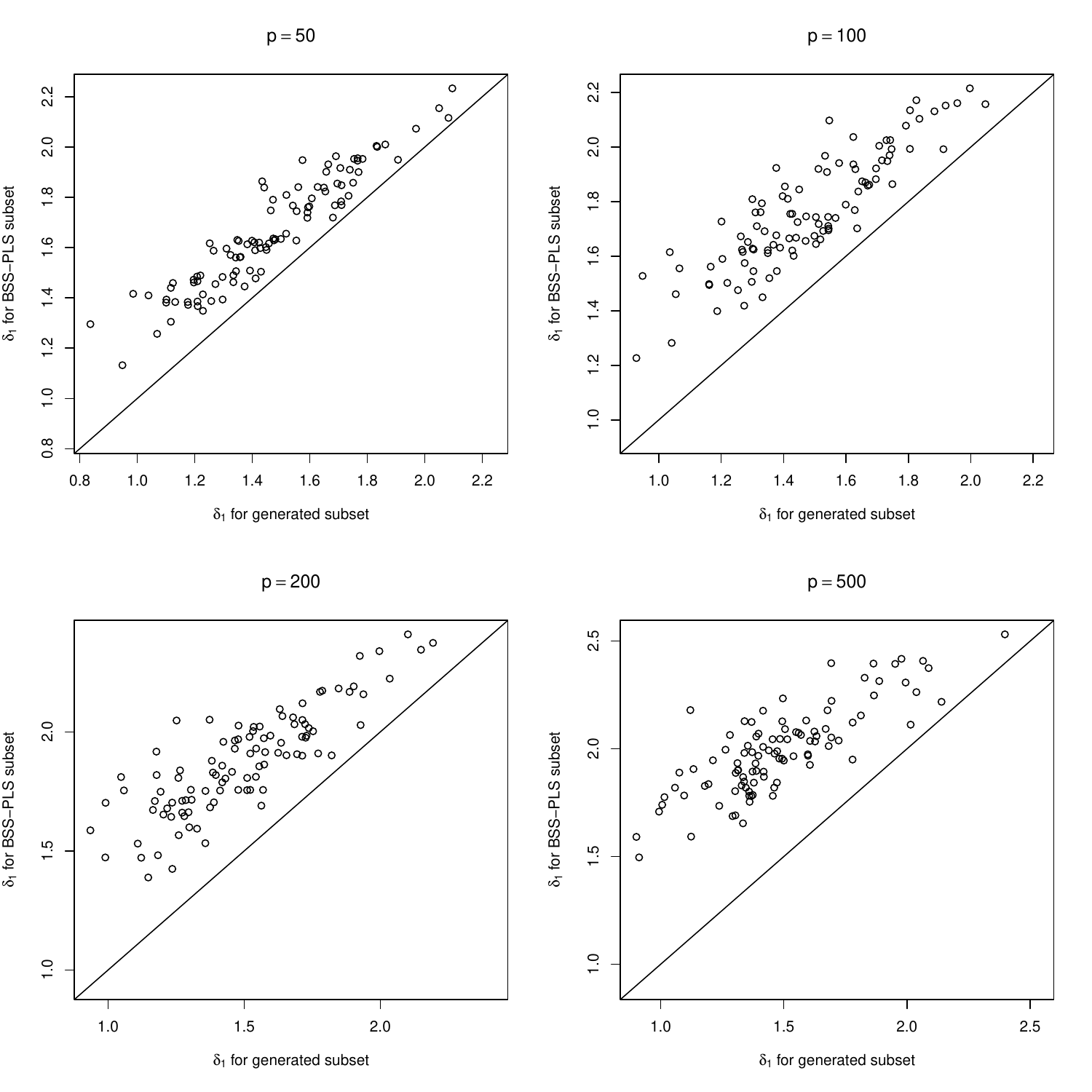}}
 \caption{Ability of BSS-PLS to propose a better subset  than the one used to generate the model (over $50$ replications and different values of $p$). Each dot represents the value of $\delta_1$ defined in \eqref{PLS-max} obtained from the subset of size $10$ corresponds to BSS-PLS and the true generated subset.}\label{crit}
\end{figure}

We investigate the performance of our method by varying the dimension $p$ over $\{50,100,200,500\}$. In this simulation, we take $q=10$, $\sigma=5$, $\gamma=p-10$, and $n=100$. Note that  the exact best subsets for this set-up are unknown as $p \geq 50$, since it is computationally impractical to conduct an exhaustive search over all the subsets of sizes $1$ to $p$. Thus, to assess the performance of our method in retrieving a {\textit{competing}} best subset, since the generated true (active) subset size is $10$, we compare  the ``best" subset  obtained from BSS-PLS for subset size $10$ to the true generated subset. For this comparison, we use the PLS optimization criterion defined in \eqref{PLS-max}. Figure~\ref{crit} plots these results with $100$ replications for every dimension $p$ of $X$ mentioned above.

For every $p \in \{50,100,200,500\}$, the value of the criterion \eqref{PLS-max} is higher for the subset given by the BSS-PLS than the corresponding criterion value for the true generated subset. This indicates that the BSS-PLS provides a better subset than the true generated subset.  
Indeed, empirically, the ``best'' subset selection solves
$\max_{\m s \in \{0, 1\}^p} \langle  X_{\m [\m s]} \m u^*_{\m [\m s]},\, Y \m v \rangle$ subject to $|\m s| \leq k$
and this is the data driven optimal subset. This is different to the data generating subset, that is the empirically best subset is not optimized for support recovery but for giving the highest values of  $\langle  X_{\m [\m s]} \m u^*_{\m [\m s]},\, Y \m v \rangle$.  Thus, even if the data has been simulated, say with  $k=10$ active variables, empirically it is not guaranteed that this subset of true active variables will reach the highest values of $\langle  X_{\m [\m s]} \m u^*_{\m [\m s]},\, Y \m v \rangle$, and depending on the signal-to-noise ratio and other factors, either overfitting or underfitting can occur. In this simulation setting, we also compare to the sparse PLS in terms of MSEP on a test set and on the support recovery (see Table 13 in the supplementary materiel). MSEP is slightly better for BSS-PLS but gives similar performance in terms of sensitivity, specificity and F1-score. Regarding the running time, on average over the 100 runs, BSS-PLS is obtained respectively in 2, 3.3, 5.8, 12.7 seconds for $p=50,100,200$ and $500$. The sPLS method implemented in \texttt{mixOmics} is faster and takes respectively 0.1, 0.3, 0.5 and 2 seconds for $p=50,100,200$ and $500$.

\subsection{PLS model with 2 components}

\begin{figure}[h]
 \includegraphics[scale=0.55]{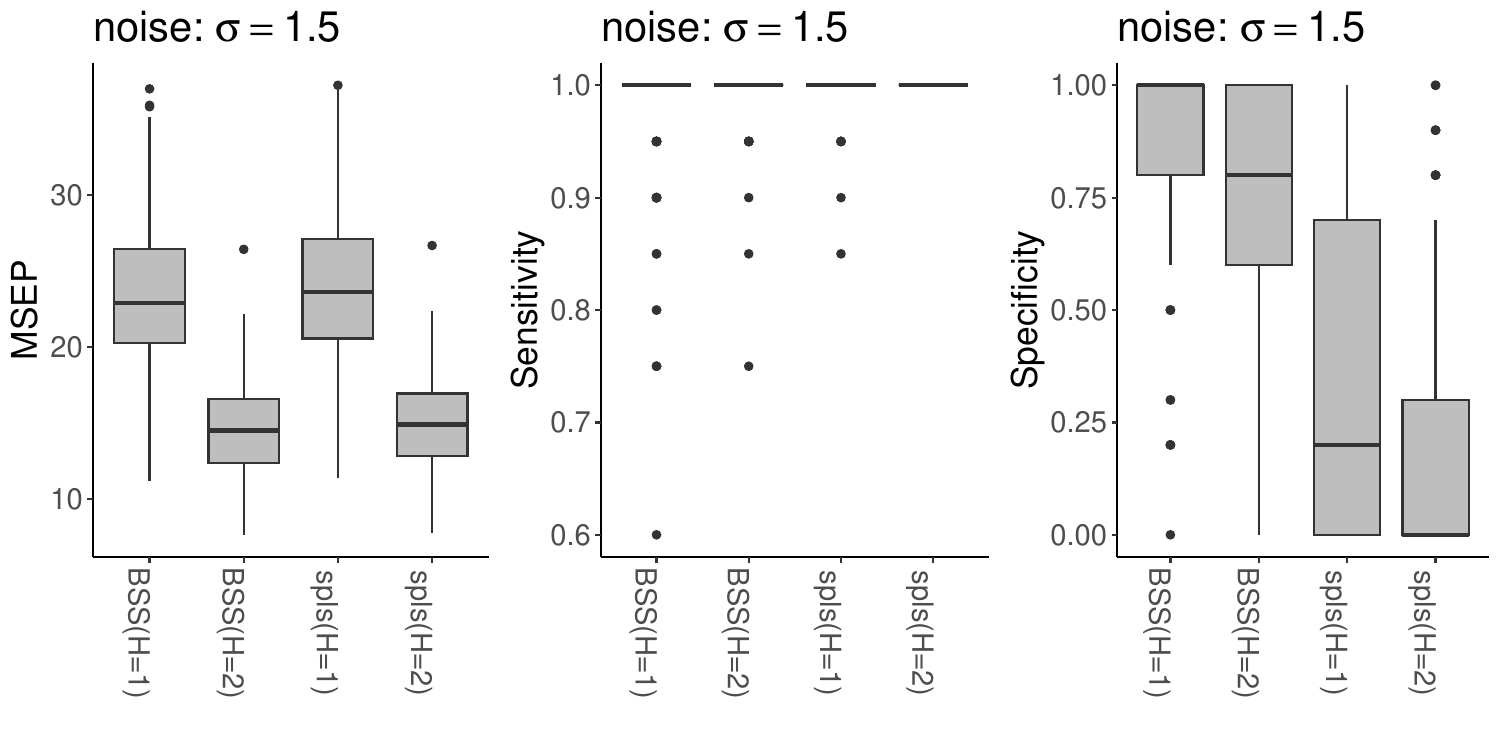}\\
 \includegraphics[scale=0.55]{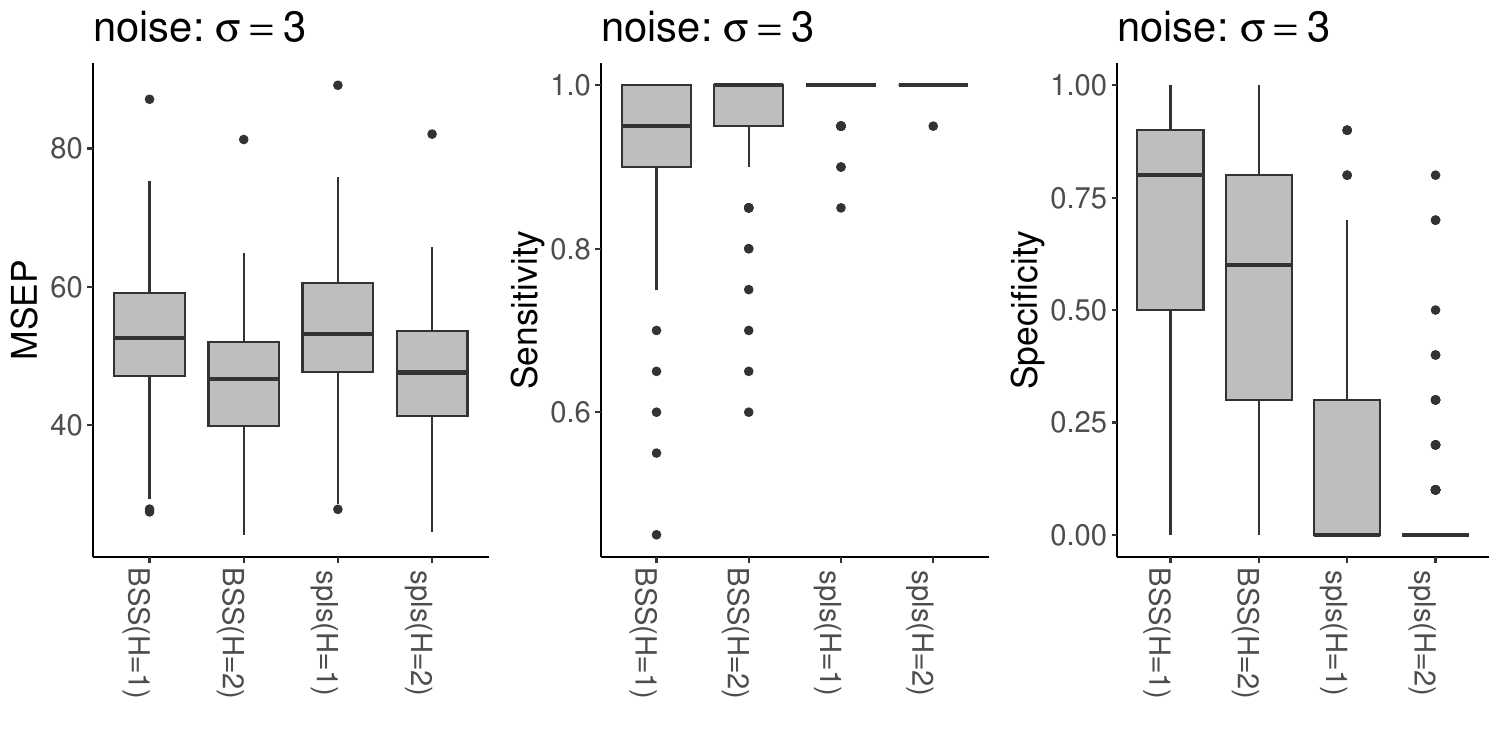}
 \caption{MSEP, sensitivity and specificity results for a PLS model with 1 and 2 components and two different noise level.} \label{2component}
\end{figure}

In this simulation setting, we simulate the data from model \eqref{simumodel} with $H=2$ components, $p=30$, and $q=10$. The two columns of the matrix $C\in \mathbb{R}^{p\times 2}$ are set to  $C_1 =(1,-1,1,-1,1,-1,1,-1,1,-1,\boldsymbol{0}^\top_{20})^\top$ and  $C_2 =(\boldsymbol{0}^\top_{10},1,-1.5,1,-1.5,1,-1.5,1,-1.5,1,-1.5,\boldsymbol{0}^\top_{10})^\top$, where $\boldsymbol{0}_r$ denotes the $r$-vector with all entries equal to zero. We use three standard deviation values $\sigma \in \{1.5, 3,6\}$.  We first run BSS-PLS with one component. As a result we end up with a list of possible subsets, one for each size, for constructing the first component. We pick the one which gives the smallest  MSEP on a test set. Then, we run the BSS-PLS on the deflated matrices (see section 4.3) and then end up with  a list of subsets for constructing the second component. We pick the one corresponding to the smallest MSEP.  Note that the sPLS model from \texttt{mixOmics} package is also implemented using deflated matrices. Results for the cases $\sigma= 1.5$ and $\sigma=3$ are presented in Figure \ref{2component}. 

As expected, a model with $H=2$ components is performing better in terms of prediction (MSEP) than a one component model. BSS-PLS is slightly better than sPLS in term of MSEP. Regarding the support recovery, we stress that a model with two components includes a subset of variables selected to construct the first component and a subset of variables to construct the second component. However, the subset of variables selected for constructing the second X-component is a subset from the deflated $X$ matrix (i.e., after removing the information contained in the previous score) and not from the original matrix. In this simulation setting, the second component has a better sensitivity but inferior specificity. Overall, BSS-PLS has better performance in terms of support recovery than the sPLS model.

\subsection{PLS model with univariate response}

We present here the case with a univariate response variable. We use the same setting as in \cite{Chun2010}.  We consider the case when $n>p$ ($n=400$ and $p=40$) and the case when  $p>n$ ($n=40$ and $p=80$). We vary the sparsity of the model by varying the number of spurious variables: $\gamma=10$ and $30$ when $n>p$ and  $\gamma=20$ and $40$ when $n<p$. Hidden variables $H_1, H_2$ and $H_3$ are from $N(0,25\boldsymbol{I}_n)$  and the columns of the matrix $X$ are generated by $X_i=H_i+\epsilon_i$ for $n_{j-1}+1\leq i\leq n_j$, where $j=1,2,3$, $(n_0,n_1,n_2,n_3)=(0,(p-q)/2,p-q,p)$ and $\epsilon_1,\ldots,\epsilon_p$ are drawn independently from $N(0,\boldsymbol{I}_n)$. The response $Y$ is generated by $3H_1-4H_2+f$, where $f$ is normallly distributed with mean 0 and variance to match a signal-to-noise ratio that is around 3 and 6.  In this simulation, we use a one component model for BSS-PLS and sparse PLS. We compare their performances to a lasso model as implemented in the \texttt{glmnet} R package. From the list of ``best'' subsets from BSS-PLS we pick the one which has the smallest MSEP on a test set for constructing the PLS model.  The same strategy is applied for sPLS and we choose the tuning parameter of the lasso model using the MSEP criterion over a grid of 50 tuning parameter values. Results are presented in Table 14 in the supplementary material.
The three models give similar results in terms of MSEP.  However, for the model selection accuracy, BSS-PLS and sPLS show good performance, whereas the lasso exhibits poor performance by missing relevant variables. Overall BSS-PLS performs better than the other methods considered. When $n < p$, the lasso fails to identify important variables, whereas BSS-PLS and sPLS regression succeeds. This is because the actual number of variables that makes up a  component score can exceed $n$.

\subsection{Numerical experiment on PCA}

For this synthetic example we use the same data generating process and setting used in \cite{Shen2008}. 
In this situation, the ability of sparse PCA procedures is applied to data whose covariance matrix actually has sparse eigenvectors. We consider a covariance matrix with two specified leading sparse eigenvectors. We consider a data matrix $X\in \mathbb{R}^{n\times p}$ with $p=10$ and each row of $X$  generated as $X\sim N\left(0, \boldsymbol{\Sigma}_1\right)$. Let
$$
\widetilde{\mathbf{u}}_1=(1,1,1,1,0,0,0,0,0.9,0.9)^T, \quad \widetilde{\mathbf{u}}_2=(0,0,0,0,1,1,1,1,-0.3,0.3)^T .
$$
The first two eigenvectors of $\Sigma_1$ are then chosen to be
$$
\begin{aligned}
& \mathbf{u}_1=\tilde{\mathbf{u}}_1 /\left\|\widetilde{\mathbf{u}}_1\right\|=(0.422,0.422,0.422,0.422,0,0,0,0,0.380,0.380)^T, \\
& \mathbf{u}_2=\tilde{\mathbf{u}}_2 /\left\|\widetilde{\mathbf{u}}_2\right\|=(0,0,0,0,0.489,0.489,0.489,0.489,-0.147,0.147)^T,
\end{aligned}
$$
both of which have a degree of sparsity of 4 . The 10 eigenvalues of $\Sigma_1$ are, respectively, 200, 100, $50,50,6,5,4,3,2$ and 1 (see \cite{Shen2008} for more details of the data generation). The first two eigenvectors explain about $70 \%$ of the total variance.

We simulate 100 data sets of size $n=30$, 100 and 300, respectively, with the covariance matrix $\Sigma_1$.  For each simulated data set, the first two sparse loading vectors are obtained from BSS-PCA, sPCA and SPCA (proposed in \cite{zou2006sparse}).

 To facilitate comparison we use the true degree of sparsity for each model, meaning that the first two components are based on 6 variables each. Table \ref{PCAsimultation} reports the percentages of correctly/incorrectly identified zero loadings for the loading vectors. All considered methods appear to perform reasonably well and give comparable results even though SPCA is less powerful for the second loading vector. We also report the percentage of variance explained using sparse components compared to a non-sparse PCA (noted PCA). Results show the ability to construct sparse components by keeping most of the information of the data.

\begin{table}[ht]
\centering
\caption{Comparison of PCA and sparse PCA methods: percentage of variance explained, percentages of correctly/incorrectly identified zero loadings} \label{PCAsimultation}
\begin{tabular}{lccccccc}
  \hline

& \multicolumn{3}{l}{$\boldsymbol{u}_1$} &&  \multicolumn{3}{l}{$\boldsymbol{u}_2$}\\
\cmidrule{2-4}\cmidrule{6-8}
 & $\%$ variance & Correct& Incorrect && $\%$ variance & correct& Incorrect \\ 
Method& explained & ($\%$) & ($\%$) && explained & ($\%$)&  ($\%$)\\ 
  \hline
& $n=50$ &  &  && &  &  \\ 

PCA & 0.48 &  &  && 0.72 &  &  \\ 
  BSS-PCA & 0.47 & 1.00 & 0.00 && 0.71 & 0.90 & 0.10 \\ 
  sPCA & 0.47 & 0.99 & 0.01 && 0.71 & 0.88 & 0.12 \\ 
  SPCA & 0.46 & 0.95 & 0.05 && 0.69 & 0.85 & 0.15 \\ 
 & $n=100$ &  &&  & &  &  \\ 
  PCA& 0.48 &  &&  & 0.72 &  &  \\ 
  BSS-PCA & 0.48 & 1.00 & 0.00& & 0.71 & 0.93 & 0.07 \\ 
  sPCA & 0.48 & 1.00 & 0.00 && 0.71 & 0.93 & 0.07 \\ 
  SPCA & 0.47 & 0.98 & 0.02 && 0.70 & 0.89 & 0.11 \\ 
  & $n=300$ &  &  & & & &  \\ 
  PCA& 0.48 &  &&  & 0.71 &  &  \\ 
  BSS-PCA & 0.47 & 1.00 & 0.00 && 0.71 & 0.98 & 0.02 \\ 
  sPCA & 0.47 & 1.00 & 0.00 & &0.71 & 0.98 & 0.02 \\ 
  SPCA & 0.47 & 1.00 & 0.00 && 0.71 & 0.95 & 0.05 \\ 

   \hline
\end{tabular}
\end{table}

\section{Case Studies}

In this section, we illustrate the usage of our method on two datasets: \texttt{multidrug} and \texttt{Hopx}. The \texttt{multidrug} dataset is analyzed through a PCA model while the \texttt{Hopx} dataset is analyzed through a PLS model. Two vignettes for running the case studies using BSS path for PCA and PLS models are detailed in \url{https://github.com/benoit-liquet/BSS-PCA-PLS}.

\subsection{Illustration of Best Subset Solution Path for PCA}
\label{sec:real-PCA}
The dataset \texttt{multidrug}  contains the expression of 48 known human ABC transporters with patterns of drug activity in 60 diverse cancer cell lines (the NCI-60) used by the National Cancer Institute to screen for anticancer activity. This dataset is available from the \texttt{mixOmics} package. We desire to provide a best subset of variables which reproduces the general characteristics of the observations in a best possible way. 

We first run a full PCA to decide the number of components to retain in the model. According to the scree plot (see Figure~1 in the supplementary material), we choose to investigate a model with 3 components which explained 29.9\% of the total variation of the data. 

Next, we run our algorithm with a budget of 50 different values of $\lambda$ to explore best subsets for the first component.  The results of best subset solution obtained for each subset size from 1 to 48 are presented in Table 15 in the supplementary material.  
We use  a drop of 10\% of the CPEV to select a best subset among all the $48$ best subsets given by BSS-PCA.
Figure~4 shows the CPEV as a function of the sparsity ($p - $ size of the subset) and the blue vertical line indicates the largest value of the sparsity where the CPEV does not exceed a drop of 10\%. The best subset of size 20 is selected for the deflation step. 

Then, we again use BSS-PCA to perform best subset solution path for the second component. 
Results are presented in Table 16 in the supplementary material and the CPEV plot for selecting the size of the best subset for the second component is given in Figure 4 in the supplementary material.

\begin{figure}[h]
\centerline{
 \includegraphics[scale=0.41]{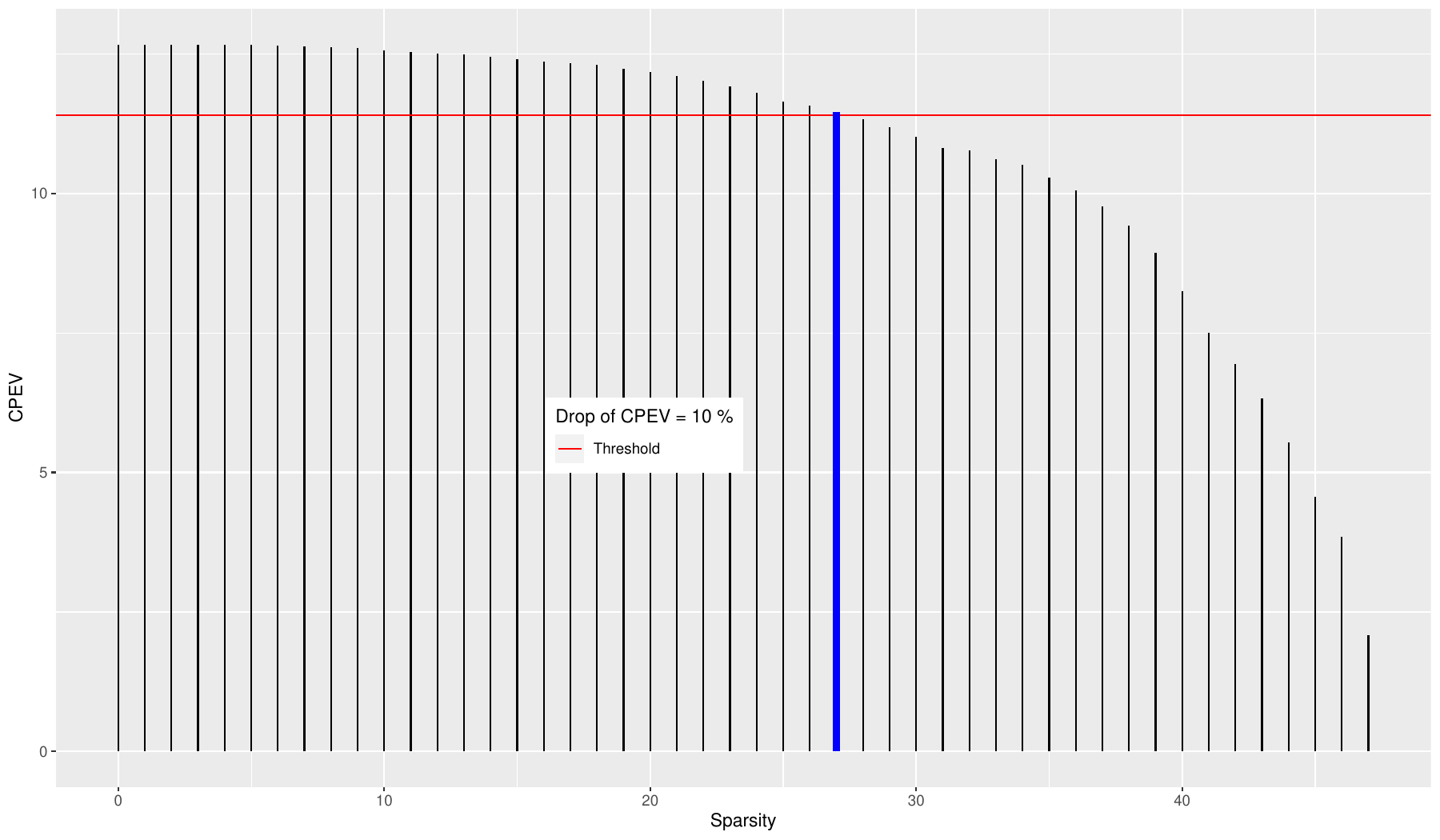}}
 \caption{CPEV as a function of the sparsity ($p$ - size of the subset) for the first component. Blue vertical line indicates the largest value of the sparsity such CPEV does not exceed a drop of 10\%} \label{drop-CPEV1}
\end{figure}

In a similar manner, we  obtain a best subset for component 3 as well (see results in Table 17 and Figure 5 in the supplementary material). 

We observe that the BSS-PCA provides components $1$, $2$, and $3$ with $20$, $12$ and $4$ variables, respectively, with a CPEV equal to  $23.5 \%$.  In this example we remark that the number of variables for constructing the components decreases with the number of components. The first two sparse components capture most of the information (CPEV of 19.12\%) compared to a CPEV of  22\% for a non-sparse PCA with two components. Then, the third component according to the CPEV strategy requires  only 4 variables and adds little information compared to the first two components (CPEV 23.5\%). 

\begin{figure}[h!]
\centerline{
 \includegraphics[scale=0.45]{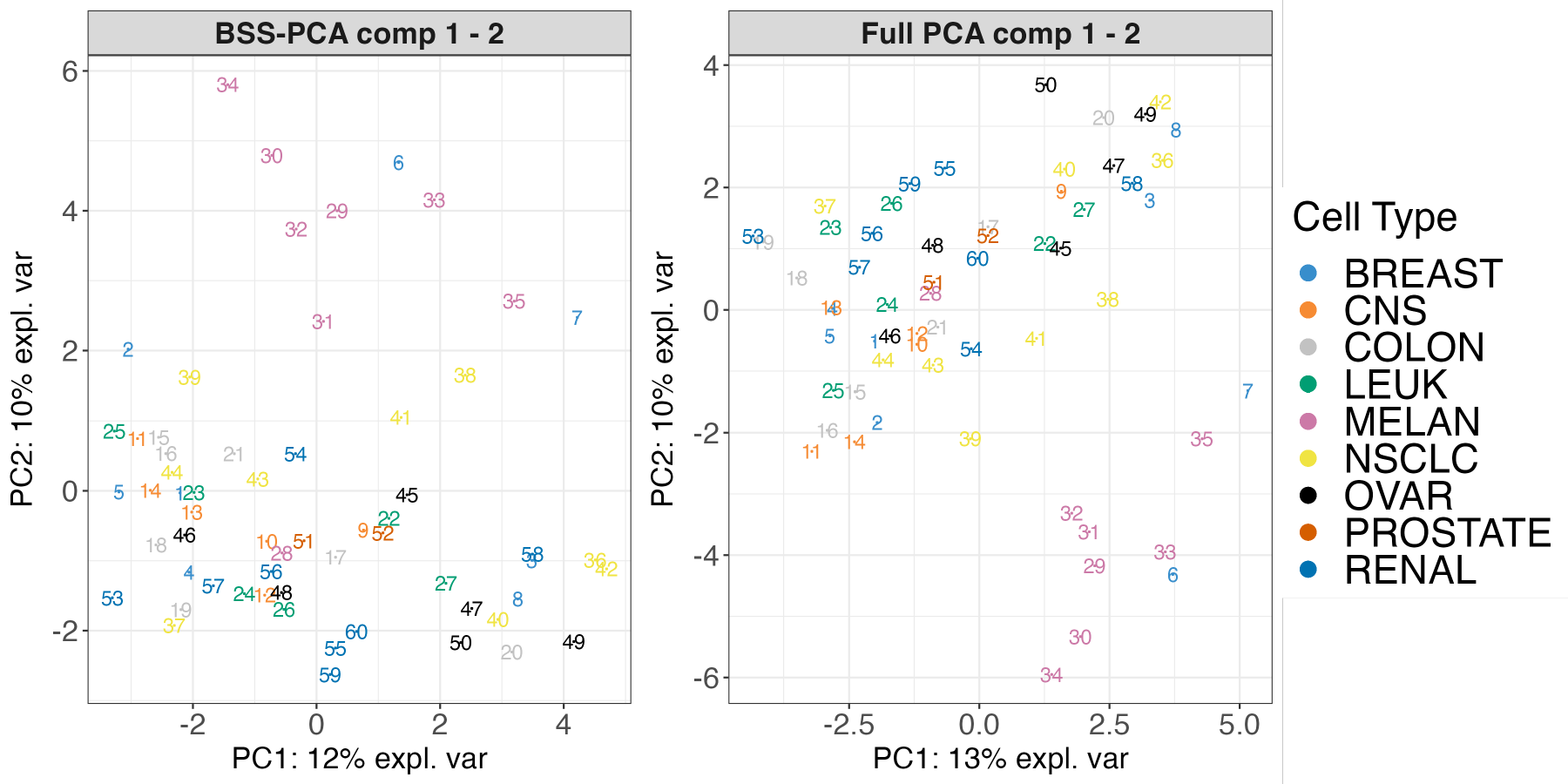}}
 \caption{Sample projected onto the first two components of the BSS-PCA (left panel) and onto the ones from the full-PCA (right panel)} \label{sample_plot}
\end{figure}

Note that one can increase the CPEV by choosing a larger subset in each component. For example, a subset of 18 variables for constructing component 3 will reach a CPEV to $25.8\%$. The left panel of Figure~\ref{sample_plot} displays the samples projected onto the first two components using $20$ and $12$ variables, respectively,  while the right panel displays the samples projected onto the first two components of the non-sparse PCA. The samples are colored according to their cancer type. The sample plot from BSS-PCA is similar to that of the non-sparse PCA, meaning that only a little information is lost. In both representations,  component 2 shows a separation of the melanoma samples. The correlation plot of the BSS-PCA, presented in Figure 2 in the supplementary material, identifies a group of transporters (ABCA9, ABCB5, ABCC5, and ABCD1) which are highly positively correlated to component 2 and  thus contributes to the explanation of the variation in the melanoma samples. Similar results have been shown in \cite{le2021multivariate} using the sparse PCA method.


\subsection{Illustration of Best Subset Solution Path for the PLS2 model}

We illustrate the usage of our approach in the context of genetic regulation. In expression Quantitative Trait Loci (eQTL) analysis, in order to discover the genetic causes of variation in the expression (i.e., transcription) of genes, gene expression data are treated as a quantitative phenotype while genotype data (SNPs) are used as predictors. 
Here, we use a dataset from a larger study (\cite{heinig2010trans}) from which we selected the Hopx genes, as in \cite{Pedretto}. This dataset has been also analyzed by \cite{liquet2016r2guess}, who used a Bayesian model to identify a parsimonious set of predictors that explains the joint variability of gene expression in four tissues (adrenal gland, fat, heart, and kidney) and by \cite{liquet2017bayesian} using sparse group 	Bayesian multivariate regression model for a similar purpose.	

The \texttt{Hopx} dataset consists of $770$ SNPs from $29$ inbred rats as a predictor matrix ($n = 29$, $p = 770$), and the $29$ measured expression levels in the four tissues as the outcome ($q = 4$). A full description of the dataset is provided in \cite{Pedretto} and it is available from the R package \texttt{R2GUESS} \cite{R2GUESS}.

We decide to explore BSS-PLS2 for only one component as the $Q^2$ criterion proposed by \cite{tenenhaus1998}, which measures the predictive power of the components, is not improved by increasing the number of components (see Figure 6 and 7 in the supplementary material).

\begin{figure}
\centerline{
 \includegraphics[scale=0.4]{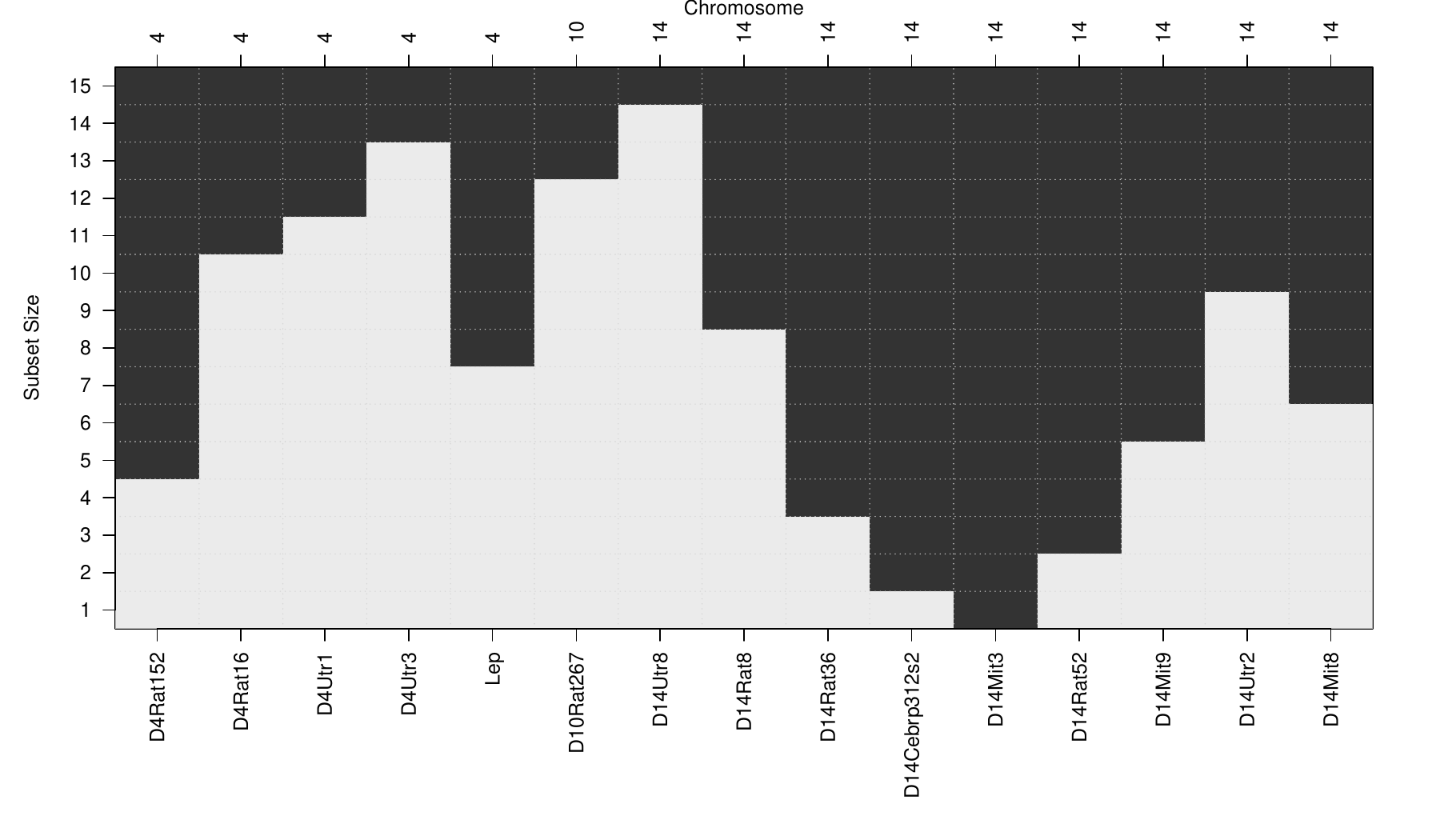}}
 \caption{Best Subset Solution Path for PLS2 for subset size ranging from 1 to 15} \label{bestmodelPLS2}
\end{figure}

We perform BSS-PLS2 using a dynamic grid of $50$ $\lambda$ values with initial $\m t_{\mathsf{init}} = 0.5\times\m 1_p$. The results of the best subset solution for PLS2 with subset size ranging from $1$ to $15$ are given in figure~\ref{bestmodelPLS2}. The full list of best subsets for every subset size from $1$ to $770$ are given in the vignette for BSS-PLS available at \href{https://github.com/benoit-liquet/BSS-PCA-PLS/}{https://github.com/benoit-liquet/BSS-PCA-PLS/}. 
	
The SNP \texttt{D14Mit3} is included in all the best subsets. This SNP has been previously identified by \cite{liquet2016r2guess}, as the most associated with the four levels of expression,  and has also been selected by the sparse group Bayesian model proposed by \cite{liquet2017bayesian}. Also, all the SNPs correspond to the subset of size $4$ obtained by BSS-PLS2 (which are \texttt{D14Mit3}, \texttt{D14Rat36}, \texttt{D14Cebrp312s2}, \texttt{D14Rat52}) have been selected by the  sparse group Bayesian model of  \cite{liquet2017bayesian}.  A clustered image map is provided in Figure 8 in the supplementary material for presenting the similarity values between the SNPs from the subset of size 15 obtained by BSS-PLS2 and the four tissues. From this clustered image maps, we identify a cluster of $10$ SNPs that are highly positively correlated to the 4 tissues and 5 SNPs that are highly negatively correlated. Note that nine of the ten highly correlated SNPs belong to chromosome 14 and the remaining one belongs to chromosome 10. All the  highly negatively correlated SNPs belong to the chromosome 4. In our modeling, the group structure of the predictors (i.e., grouping of SNPs across chromosomes) is not taken into account.

\section{Concluding Remarks}

In conclusion, our work in this paper develops a simple unconstrained continuous optimization approach for addressing the best subset solution path problem within the framework of the PLS and PCA models, naming these methods BSS-PLS and BSS-PCA respectively. The effectiveness of our method is demonstrated through a series of carefully designed simulation experiments. Notably, in the context of PLS1 model, our theoretical result shows that solving the proposed continuous optimization problem provides an exact solution to the best subset solution path problem. This noteworthy result has the potential for further generalization to both PLS2 and PCA models. As part of our ongoing research efforts, we plan to delve deeper into the theoretical underpinnings of our approach and explore its optimally in these broader contexts. 

While the theoretical result shows the optimality of our method, and, the experimental results are promising, solutions obtained may not be guaranteed to be optimal because of the use of the gradient descent algorithm, which can converge to a suboptimal point depending on the initial point, learning rate, and the value of the penalty parameter $\lambda$. We believe future research on improving the gradient based method for this task or application of other alternative continuous optimization methods will overcome these challenges.

We note that the methods proposed by \cite{bertsimas2022solving} (for sparse PCA), \cite{dey2022using} (for sparse PCA) and \cite{watanabe2023branch} (for sparse canonical correlation analysis) are closely related to our work. However, unlike our continuous optimization, they rely on either integer programming or mixed-integer optimization techniques. Similar to these alternative approaches, our unified approach, without any integer constraints, opens the door for future research to explore a wide array of readily available continuous optimization methods, offering opportunities to enhance the already impressive performance of our methodology.

Our framework can be extended in several ways. Utilizing information about the data's structure, such as a group structure (e.g., genes within the same gene pathway sharing similar biological functions), one can aim to construct components based on relevant groups. By enforcing that the variables $t_j$ are equal for the variables $j$ belonging to the same group, one can design BSS-PLS or BSS-PCA to achieve group selection. Additionally, we can consider avoiding the deflation step by employing a block approach, similar to the approach in \cite{journee2010generalized,chavent2017group}.
Finally, in our current PLS framework, we focus on subset selection for the X part, corresponding to a regression setting. One may be interested in obtaining sparse components for both the X and Y parts in the case of a PLS model in a canonical mode, which is closely related to CCA models \cite{watanabe2023branch}.

\medskip

\noindent \textbf{Acknowledgement}\\
\noindent Samuel Muller was supported by the Australian Research Council (DP210100521).




\bibliographystyle{abbrv} 
 \bibliography{biblio}

\begin{thebibliography}{10}

\bibitem{bertsimas2022solving}
D.~Bertsimas and R.~Cory-Wright.
\newblock Solving large-scale sparse pca to certifiable (near) optimality.
\newblock {\em The Journal of Machine Learning Research}, 23(1):566--600, 2022.

\bibitem{BKM16}
D.~Bertsimas, A.~King, and R.~Mazumder.
\newblock {Best subset selection via a modern optimization lens}.
\newblock {\em The Annals of Statistics}, 44(2):813 -- 852, 2016.

\bibitem{broc2021penalized}
C.~Broc, T.~Truong, and B.~Liquet.
\newblock Penalized partial least squares for pleiotropy.
\newblock {\em BMC Bioinformatics}, 22:1--31, 2021.

\bibitem{chavent2017group}
M.~Chavent and G.~Chavent.
\newblock Group-sparse block pca and explained variance.
\newblock {\em arXiv preprint arXiv}, 1705, 2017.

\bibitem{chen2022hybrid}
X.~Chen, W.~Cao, C.~Gan, and M.~Wu.
\newblock A hybrid partial least squares regression-based real time pore
  pressure estimation method for complex geological drilling process.
\newblock {\em Journal of Petroleum Science and Engineering}, 210:109771, 2022.

\bibitem{Chun2010}
H.~Chun and S.~Kele{\c{s}}.
\newblock {Sparse partial least squares regression for simultaneous dimension
  reduction and variable selection}.
\newblock {\em Journal of the Royal Statistical Society: Series B (Statistical
  Methodology)}, 72(1):3--25, 2010.

\bibitem{Chung2010}
D.~Chung and S.~Kele\c{s}.
\newblock {Sparse Partial Least Squares Classification for High Dimensional
  Data}.
\newblock {\em Statistical Applications in Genetics and Molecular Biology},
  9(1):17, 2010.

\bibitem{dey2022using}
S.~S. Dey, R.~Mazumder, and G.~Wang.
\newblock Using l1-relaxation and integer programming to obtain dual bounds for
  sparse pca.
\newblock {\em Operations Research}, 70(3):1914--1932, 2022.

\bibitem{leaps2000}
G.~M. Furnival and R.~W. Wilson.
\newblock Regressions by leaps and bounds.
\newblock {\em Technometrics}, 42:69--79, 2000.

\bibitem{hastie2020best}
T.~Hastie, R.~Tibshirani, and R.~Tibshirani.
\newblock Best subset, forward stepwise or lasso? analysis and recommendations
  based on extensive comparisons.
\newblock {\em Statistical Science}, 35(4):579--592, 2020.

\bibitem{HazimehMazumder2020}
H.~Hazimeh and R.~Mazumder.
\newblock Fast best subset selection: coordinate descent and local
  combinatorial optimization algorithms.
\newblock {\em Operations Research}, 68(5):1517--1537, 2020.

\bibitem{heinig2010trans}
M.~Heinig, E.~Petretto, C.~Wallace, L.~Bottolo, M.~Rotival, H.~Lu, Y.~Li,
  R.~Sarwar, S.~R. Langley, A.~Bauerfeind, et~al.
\newblock A trans-acting locus regulates an anti-viral expression network and
  type 1 diabetes risk.
\newblock {\em Nature}, 467(7314):460--464, 2010.

\bibitem{HockingLeslie1967}
R.~R. Hocking and R.~N. Leslie.
\newblock Selection of the best subset in regression analysis.
\newblock {\em Technometrics}, 9:531--540, 1967.

\bibitem{hui2017joint}
F.~K. Hui, S.~M{\"u}ller, and A.~Welsh.
\newblock Joint selection in mixed models using regularized pql.
\newblock {\em Journal of the American Statistical Association},
  112(519):1323--1333, 2017.

\bibitem{Ji2011}
G.~Ji, Z.~Yang, and W.~You.
\newblock Pls-based gene selection and identification of tumor-specific genes.
\newblock {\em IEEE Transactions on Systems, Man, and Cybernetics, Part C
  (Applications and Reviews)}, 41(6):830--841, 2011.

\bibitem{Jolliffe2005}
I.~Jolliffe.
\newblock {\em Principal Component Analysis}.
\newblock John Wiley \& Sons, Ltd, 2005.

\bibitem{journee2010generalized}
M.~Journ{\'e}e, Y.~Nesterov, P.~Richt{\'a}rik, and R.~Sepulchre.
\newblock Generalized power method for sparse principal component analysis.
\newblock {\em Journal of Machine Learning Research}, 11(2), 2010.

\bibitem{khatri2021review}
P.~Khatri, K.~K. Gupta, and R.~K. Gupta.
\newblock A review of partial least squares modeling (plsm) for water quality
  analysis.
\newblock {\em Modeling Earth Systems and Environment}, 7(2):703--714, 2021.

\bibitem{de2019pls}
P.~L. Lafaye~de Micheaux, B.~Liquet, and M.~Sutton.
\newblock Pls for big data: a unified parallel algorithm for regularised group
  pls.
\newblock {\em Statistics Surveys}, 13:119--149, 2019.

\bibitem{LeCao2011}
K.~L{\^e}~Cao, S.~Boitard, and P.~Besse.
\newblock Sparse pls discriminant analysis: biologically relevant feature
  selection and graphical displays for multiclass problems.
\newblock {\em BMC bioinformatics}, page 253, 2011.

\bibitem{LeCao2008a}
K.-A. L\^{e}~Cao, D.~Rossouw, C.~Robert-Grani\'{e}, and P.~Besse.
\newblock {Sparse PLS: Variable Selection when Integrating Omics data}.
\newblock {\em Statistical Application and Molecular Biology}, 7((1):37), 2008.

\bibitem{le2021multivariate}
K.-A. L{\^e}~Cao and Z.~Welham.
\newblock {\em Multivariate Data Integration Using R: Methods and Applications
  with the mixOmics Package}.
\newblock Chapman and Hall/CRC, 2021.

\bibitem{Lin2014b}
D.~Lin, H.~Cao, V.~D. Calhoun, and Y.-P. Wang.
\newblock Sparse models for correlative and integrative analysis of imaging and
  genetic data.
\newblock {\em Journal of Neuroscience Methods}, 237:69 -- 78, 2014.

\bibitem{liquet2016r2guess}
B.~Liquet, L.~Bottolo, G.~Campanella, S.~Richardson, and M.~Chadeau-Hyam.
\newblock R2guess: a graphics processing unit-based r package for bayesian
  variable selection regression of multivariate responses.
\newblock {\em Journal of Statistical Software}, 69(2), 2016.

\bibitem{R2GUESS}
B.~Liquet and M.~Chadeau-Hyam.
\newblock {\em R2GUESS: Wrapper Functions for GUESS.}, 2014.
\newblock R package version 1.4.

\bibitem{Liquet2016}
B.~Liquet, P.~Lafaye~de Micheaux, B.~Hejblum, and R.~Thi\'{e}baut.
\newblock Group and sparse group partial least square approaches applied in
  genomics context.
\newblock {\em Bioinformatics}, 32:35--42, 2016.

\bibitem{liquet2017bayesian}
B.~Liquet, K.~Mengersen, A.~Pettitt, and M.~Sutton.
\newblock Bayesian variable selection regression of multivariate responses for
  group data.
\newblock {\em Bayesian Analysis}, 12(4):1039--1067, 2017.

\bibitem{mehmood2016diversity}
T.~Mehmood and B.~Ahmed.
\newblock The diversity in the applications of partial least squares: an
  overview.
\newblock {\em Journal of Chemometrics}, 30(1):4--17, 2016.

\bibitem{COMBSS22}
S.~Moka, B.~Liquet, H.~Zhu, and S.~Muller.
\newblock {COMBSS: Best Subset Selection via Continuous Optimization}.
\newblock {\em arXiv, doi: 10.48550/ARXIV.2205.02617}, 2022.

\bibitem{muller2010}
S.~Muller and A.~H. Welsh.
\newblock On model selection curves.
\newblock {\em International Statistical Review}, 78(2):240--256, 2010.

\bibitem{Pedretto}
E.~Petretto, L.~Bottolo, S.~R. Langley, M.~Heinig, C.~McDermott-Roe, R.~Sarwar,
  M.~Pravenec, N.~Hübner, T.~J. Aitman, S.~A. Cook, and S.~Richardson.
\newblock New insights into the genetic control of gene expression using a
  bayesian multi-tissue approach.
\newblock {\em PLOS Computational Biology}, 6(4):1--13, 04 2010.

\bibitem{mixOmics}
F.~Rohart, B.~Gautier, A.~Singh, and K.-A. Le~Cao.
\newblock mixomics: An r package for 'omics feature selection and multiple data
  integration.
\newblock {\em PLoS Computational Biology}, 13(11):e1005752, 2017.

\bibitem{Shen2008}
H.~Shen and J.~Z. Huang.
\newblock Sparse principal component analysis via regularized low rank matrix
  approximation.
\newblock {\em Journal of Multivariate Analysis}, 99(6):1015 -- 1034, 2008.

\bibitem{sutton2018sparse}
M.~Sutton, R.~Thi{\'e}baut, and B.~Liquet.
\newblock Sparse partial least squares with group and subgroup structure.
\newblock {\em Statistics in Medicine}, 37(23):3338--3356, 2018.

\bibitem{YM20}
Y.~Takano and R.~Miyashiro.
\newblock Best subset selection via cross-validation criterion.
\newblock {\em TOP}, 28:475--488, 2020.

\bibitem{tarr2018mplot}
G.~Tarr, S.~M{\"u}ller, and A.~H. Welsh.
\newblock mplot: An r package for graphical model stability and variable
  selection procedures.
\newblock {\em Journal of Statistical Software}, 83:1--28, 2018.

\bibitem{tenenhaus1998}
M.~Tenenhaus.
\newblock {\em La r\'egression PLS: Th\'eorie et Pratique}.
\newblock Paris: Technip, 1998.

\bibitem{Tibshirani1994}
R.~Tibshirani.
\newblock Regression shrinkage and selection via the lasso.
\newblock {\em Journal of the Royal Statistical Society, Series B},
  58:267--288, 1994.

\bibitem{tu2011new}
Y.-K. Tu, G.~Davey~Smith, and M.~S. Gilthorpe.
\newblock A new approach to age-period-cohort analysis using partial least
  squares regression: the trend in blood pressure in the glasgow alumni cohort.
\newblock {\em PloS one}, 6(4):e19401, 2012.

\bibitem{Wang2016}
T.~Wang, Q.~Berthet, and R.~J. Samworth.
\newblock {Statistical and computational trade-offs in estimation of sparse
  principal components}.
\newblock {\em The Annals of Statistics}, 44(5):1896 -- 1930, 2016.

\bibitem{wang2020partial}
Y.~Wang, J.~G. Ibrahim, and H.~Zhu.
\newblock Partial least squares for functional joint models with applications
  to the alzheimer's disease neuroimaging initiative study.
\newblock {\em Biometrics}, 76(4):1109--1119, 2020.

\bibitem{watanabe2023branch}
A.~Watanabe, R.~Tamura, Y.~Takano, and R.~Miyashiro.
\newblock Branch-and-bound algorithm for optimal sparse canonical correlation
  analysis.
\newblock {\em Expert Systems with Applications}, 217:119530, 2023.

\bibitem{Witten2009}
D.~M. Witten, R.~Tibshirani, and T.~Hastie.
\newblock A penalized matrix decomposition, with applications to sparse
  principal components and canonical correlation analysis.
\newblock {\em Biostatistics}, 10(3):515--534, 2009.

\bibitem{Wold1966}
H.~Wold.
\newblock Estimation of principal components and related models by iterative
  least squares.
\newblock In {\em Multivariate Analysis}, pages 391--420, Dayton, Ohio, June
  1966. Academic Press, New York, Wiley.

\bibitem{zou2006sparse}
H.~Zou, T.~Hastie, and R.~Tibshirani.
\newblock Sparse principal component analysis.
\newblock {\em Journal of Computational and Graphical Statistics},
  15(2):265--286, 2006.

\end{thebibliography}


\begin{thebibliography}{1}

\bibitem{de2019pls}
P.~L. Lafaye~de Micheaux, B.~Liquet, and M.~Sutton.
\newblock Pls for big data: a unified parallel algorithm for regularised group
  pls.
\newblock {\em Statistics Surveys}, 13:119--149, 2019.

\bibitem{rosipal2005overview}
R.~Rosipal and N.~Kr{\"a}mer.
\newblock Overview and recent advances in partial least squares.
\newblock In {\em International Statistical and Optimization Perspectives
  Workshop" Subspace, Latent Structure and Feature Selection"}, pages 34--51.
  Springer, 2005.

\bibitem{sutton2018sparse}
M.~Sutton, R.~Thi{\'e}baut, and B.~Liquet.
\newblock Sparse partial least squares with group and subgroup structure.
\newblock {\em Statistics in Medicine}, 37(23):3338--3356, 2018.

\end{thebibliography}

\end{document}


\begin{center}
\Large{Supplementary Material for \\Best Subset Solution Path for Linear Dimension Reduction Models using Continuous Optimization\\by\\
Liquet Benoit and Moka Sarat and Muller Samuel}
\end{center}

\label{firstpage}


%




%

\newpage

\appendix

\section*{Appendix A: Proof of Theorem~3.1}
For the reader's convenience, we recall both the discrete and continuous constrained optimization problems for the PLS1 model. In particular, the exact best subset selection problem is stated as 
\begin{align}
\label{eqn:f-lambda-temp}
\min_{\m s \in \{0, 1\}^p} \lt[ - \frac{\| X_{\m [\m s]}^\top \m y\|}{n}\rt], \quad \text{subject to}\quad \sum_{j=1}^p s_j  \leq k,
\tag{A1}
\end{align}
which is equivalent to  
\begin{equation}
\label{PLS1S-temp}
\min_{\m s \in \{0, 1\}^p} \lt[ - \frac{\| X_{\m [\m s]}^\top \m y\|^2}{n^2}\rt], \quad \text{subject to}\quad \sum_{j=1}^p s_j  \leq k.
\tag{A2}
\end{equation}
Also, recall the Boolean relaxation of \eqref{PLS1S-temp} which is 
\begin{align}
\label{PLS1t-2-temp}
\min_{\m t \in [0, 1]^p} \lt[ -\frac{\|X_{\m t}^\top \m y\|^2}{n^2}\rt], \quad \text{subject to}\quad \sum_{j=1}^pt_j \leq k. 
\tag{A3}
\end{align}
In our method, instead of solving \eqref{PLS1t-2-temp}, we solve 
\begin{equation}
\label{PLS1-temp}
\min_{\m t \in [0, 1]^p} f_{\lambda}^{\plsone}(\m t), 
\tag{A4}
\end{equation}
where 
$f_{\lambda}^{\plsone}(\m t) = - \frac{\|X_{\m t}^\top \m y\|^2}{n^2} + \lambda  \sum_{j=1}^pt_j.$
Now recall Theorem~3.1 using the above equation numbers.
\begin{customthm}{3.1}
\label{thm:equi-s-t}
We have the following equivalence between the optimization problems {\normalfont (A1), (A3)}, and {\normalfont (A4)}.
\begin{itemize}
    \item[] (i) The optimal solutions of the minimization problems defined by {\normalfont (A1)} and {\normalfont (A3)}  are identical. 
    \item[] (ii) For every $k =1, \dots, p$,  there exists $\lambda$ such that an optimal solution of {\normalfont (A3)} is an optimal solution of {\normalfont (A4)}. 
\end{itemize}
\end{customthm}

A key result for proving the theorem is the following lemma which establishes the concavity of the objective function $f_{\lambda}^{\plsone}(\m t)$.
\begin{customlemma}{A1}
\label{lem:concave}
For every $\lambda$, the function $f_{\lambda}^{\plsone}(\m t)$ defined by \eqref{eqn:f-lambda-temp} is a concave function on $\reals_+^p$.
\end{customlemma}
\begin{proof}
We can compute the Hessian of $f_{\lambda}^{\plsone}(\m t)$ easily by differentiating the gradient of $f^{\plsone}_0(\m t)$ which is given by
\[
\nabla f^{\plsone}_0(\m t)  = - \frac{2}{n^2}\lt(\m t \odot X^\top \m y \odot X^\top \m y \rt).
\] 
In particular, we can obtain that the Hessian of $f_{\lambda}^{\plsone}(\m t)$ at a point $\m t$ is a diagonal matrix with the diagonal being $-  \frac{2}{n^2}\lt(X^\top \m y \odot X^\top \m y \rt)$, which indicates that the Hessian is negative definite. Thus, $f_{\lambda}^{\plsone}(\m t)$ is a concave function over $[0, 1]^p$ for all $\lambda$.
\end{proof}
A consequence of Lemma~\ref{lem:concave} is that $f_{\lambda}^{\plsone}(\m t)$ is concave on the hypercube $[0,1]^p$.  

\begin{proof}[\em Proof of Theorem~3.1 (i)] Concavity of $f^{\plsone}_0(\m t)$ on the hypercube implies that the solution of (A3) is achieved at a corner point of the hypercube. 
Specifically, observe from the Boolean relaxation that the objective function $f^{\plsone}_0(\m t) = -\frac{\|X_{\m t}^\top \m y\|^2}{n^2}$ in \eqref{PLS1t-2-temp} is equal to the objective function of the exact optimization  \eqref{PLS1S-temp} at the corner points $\m t = \m s \in \{0, 1\}^p$. 
Thus, we complete the proof if we show that a solution point of \eqref{PLS1t-2-temp} lies at a corner on the hypercube $[0, 1]^p$. Towards this, we recall from Lemma~\ref{lem:concave} that $f_0(\m t)$ is a concave function. This implies, for any two points $\m t, \m t' \in [0,1]^p$ and $a \in [0, 1]$, 
\begin{align*}
f^{\plsone}_0(a \m t + (1 - a) \m t') \geq a f^{\plsone}_0(\m t) + (1 - a) f^{\plsone}_0(\m t') \geq \min \{f^{\plsone}_0(\m t), f^{\plsone}_0(\m t')\}. 
\end{align*} 
This indicates that on a line segment joining $\m t$ and $\m t'$, the function $f_0(\m t)$ achieves its minimum at an end point, i.e.\ either at $\m t$ or at $\m t'$. This holds true even when $\m t$ and $\m t'$ are corner points on the hypercube. 
Thus, we conclude that both the problems \eqref{PLS1S-temp} and \eqref{PLS1t-2-temp} have the same solutions. 
\end{proof}

\begin{proof}[\em Proof of Theorem~3.1 (ii)]
Due to the formulation of the exact optimization, we know that without loss of generality the inequality in the constraint $ \sum_{j =1}^p s_j \leq k$ of \eqref{PLS1S-temp} can be replaced with equality. Similarly, the inequality in the constraint $\sum_{j=1}^pt_j \leq k$ of \eqref{PLS1t-2-temp} can be replaced with an equality. To see this, we observe from the gradient expression of $f^{\plsone}_0(\m t)$ that the $j$th element of the gradient of $\nabla f^{\plsone}_0(\m t)$ is $- c_j t_j$ for some non-negative constant~$c_j$. That means, $f^{\plsone}_0(\m t) $ is decreasing in each coordinate. As a consequence, for every $\m t' \in [0, 1]^p$ such that $\sum_{j = 1}^p t'_j < k$, there exists a point $\m t'' \in [0, 1]^p$ on the hyperplane 
$\sum_{j = 1}^p t_j = k$ such that $ f^{\plsone}_0(\m t') \geq  f^{\plsone}_0(\m t'')$, and thus, the inequality in the constraint of (A3) can be replaced with an equality. In a specific case where $\sum_{j = 1}^p t_j' = k-1$ and $\sum_{j = 1}^p t''_j = k$, we get
\[
f^{\plsone}_0(\m t') \geq f^{\plsone}_0(\m t'').
\]
Furthermore, we can conclude that there exists a sequence of corner points $\m s^{(0)}, \m s^{(1)}, \dots, \m s^{(p)}$ such that $\m s^{(k)}$ is an optimal solution of  (A3) and 
\begin{align}
\label{eqn:dec-seq}
f^{\plsone}_0(\m s^{(0)}) \geq f^{\plsone}_0(\m s^{(1)}) \geq \cdots \geq f^{\plsone}_0(\m s^{(p)}).
\tag{A5}
\end{align}
An interesting property of this decreasing sequence is that the increments are also decreasing, that is, 
\begin{align}
\label{eqn:incr-decr}
f^{\plsone}_0(\m s^{(k-1)}) - f^{\plsone}_0(\m s^{(k)}) \geq f^{\plsone}_0(\m s^{(k)}) - f^{\plsone}_0(\m s^{(k+1)}).
\tag{A6}
\end{align}
To see this,  
note that the number of ones in $\m s^{(k)}$ is  exactly equal to $k$. Furthermore, we can find an index $i \in \{1, \dots, p\}$ such that $s^{(k-1)}_i = 0$ and $s^{(k +1)}_i = 1$. By defining $\m z = (X^\top \m y)/n$, for this index $i$, we can write
\[
f^{\plsone}_0(\m s^{(k-1)} + \m e_i) = - \sum_{j = 1} s^{(k-1)}_j z_j^2 - z_i^2 = f^{\plsone}_0(\m s^{(k-1)}) - z_i^2,
\]
and similarly, 
\[
f^{\plsone}_0(\m s^{(k+1)} - \m e_i) = - \sum_{j = 1} s^{(k+1)}_j z_j^2 + z_i^2 = f^{\plsone}_0(\m s^{(k+1)}) + z_i^2.
\]
Together,
\begin{align*}
    f^{\plsone}_0(\m s^{(k-1)}) + f^{\plsone}_0(\m s^{(k+1)}) &= f^{\plsone}_0(\m s^{(k-1)} + \m e_i) + f^{\plsone}_0(\m s^{(k+1)} - \m e_i)\\
    &\geq 2 f^{\plsone}_0(\m s^{(k)}),
\end{align*}
where the inequality holds because both $\m s^{(k-1)} + \m e_i$ and $\m s^{(k+1)} - \m e_i$ have exactly $k$ ones and $\m s^{(k)}$ is an optimal point of (A3). By reordering the above inequality, we see \eqref{eqn:incr-decr}.

From using \eqref{eqn:dec-seq} and \eqref{eqn:incr-decr}, we can create a univariate function $h: \reals_+ \to \reals$ such that $h$ is convex non-decreasing and $h(k) = f^{\plsone}_0(\m s^{(k)})$. One such function is a piecewise linear function that simply connects the points $(k, f^{\plsone}_0(\m s^{(k)}))$. 

Given such a convex function, the optimization problem (A3) can be restated as the following convex optimization 
\begin{align*}
    &\min_{r \in \reals_+} h(r)\quad \text{subject to}\quad r = k,
\end{align*}
which satisfies the strong duality condition. Therefore, there exists a constant $\lambda^{(k)} \geq 0$ such that 
$h(k) + \lambda^{(k)} k$ is the minimum of the Lagrangian function $L(r, \lambda) = h(r) + \lambda r$. We complete the proof by noting that the minimum value 
\[
L(k, \lambda^{(k)}) = f_0(\m s^{(k)}) + \lambda^{(k)} \sum_{j = 1}^p s_j^{(k)} = \min_{\m t \in [0, 1]^p, \lambda \geq 0} f^{\plsone}_0(\m t) + \lambda \sum_{j = 1}^p t_j = \min_{\m t \in [0, 1]^p, \lambda \geq 0}f^{\plsone}_\lambda(\m t),
\]
where the second equality follows from the fact that $f^{\plsone}_\lambda(\m t)$ is concave on the hypercube and hence achieving minimum at a corner point.

\end{proof}

\section*{Appendix B: Deflation step for sPCA and sPLS }

In this paper, for all the PCA and PLS models (including sparse versions), we use iterative approaches which start by constructing the first component (resp., the first pair of components) for the PCA framework (resp., for the PLS framework) using the original dataset $X$ (resp., datasets $X$ and $Y$). The first component or the first pair of components is a linear combination of the original data.  The next components or the next pairs of components are obtained using successively deflated versions of $X$ and $Y$ (i.e., after removing the information contained in the previous component or pairs of scores). We present in Table~\ref{deflation}  the deflation step to the most popular version of PCA and PLS. We refer the reader to \cite{de2019pls} for other PLS models.

\begin{table}[H]
\caption{Deflation details for sparse PCA and sparse PLS} \label{deflation}

$
\begin{array}{llll|cc}
\hline \text { Method } & \text { initialise } & \text { Deflation } & \text {Component score} & \boldsymbol{c}_h &\boldsymbol{\xi}_h\\  
\hline 
\multirow{1}{*}{\text { sPCA }} & X_0 = X & X_h=X_{h-1}-\boldsymbol{\xi}_h \boldsymbol{c}_h^\top & 
X\boldsymbol{u}_h & \boldsymbol{c}_h= \frac{X_{h-1}^\top\boldsymbol{\xi}_h}{\boldsymbol{\xi}_h^\top \boldsymbol{\xi}_h}&\boldsymbol{\xi}_h=X_{h-1}\boldsymbol{u}_h \\ \hline \hline

\multirow{1}{*}{\text { sPLS }} & X_0 = X & X_h=X_{h-1}-\boldsymbol{\xi}_h\boldsymbol{ c}_h^\top & \boldsymbol{\xi}_h=
X_{h-1}\boldsymbol{u}_h & \boldsymbol{c}_h= \frac{X_{h-1}^\top\boldsymbol{\xi}_h}{\boldsymbol{\xi}_h^\top \boldsymbol{\xi}_h}& \\
\multirow{1}{*}{\text { regression mode }} & Y_0 = Y & Y_h=Y_{h-1}-\boldsymbol{\xi}_h \boldsymbol{d}_h^\top& \boldsymbol{\psi}_h=
Y_{h-1}\boldsymbol{v}_h & \boldsymbol{d}_h= \frac{Y_{h-1}^\top\boldsymbol{\xi}_h}{\boldsymbol{\xi}_h^\top \boldsymbol{\xi}_h}& \\ 
\multirow{1}{*}{\text { canonical mode }} & Y_0 = Y & Y_h=Y_{h-1}-\boldsymbol{\psi}_h \boldsymbol{e}_h^\top& \boldsymbol{\psi}_h=
Y_{h-1}\boldsymbol{v}_h & \boldsymbol{e}_h= \frac{Y_{h-1}^\top\boldsymbol{\xi}_h}{\boldsymbol{\psi}_h^\top \boldsymbol{\xi}_h}& \\ \hline
\end{array}
$
\footnotesize{Here, $\boldsymbol{u}_h$ and $\boldsymbol{v}_h$ being the left and right singular vectors of the largest singular value of $M_{h-1}=X_{h-1}^\top Y_{h-1}$ for PLS and $M_{h-1}=X^\top X$ for PCA.}
\end{table}\vspace{3cm}

\vspace{-3cm}

\section*{Appendix C: Prediction and Adjusted Weights}

The construction of the components using the deflation step approach (presented in Table~\ref{deflation}) leads to decompositions of the original matrices
$$
\begin{aligned}
& X=T C^{\boldsymbol{\top}}+E_X, \\
& Y=S D^{\boldsymbol{\top}}+E_Y,
\end{aligned}
$$
where $T=\left(\boldsymbol{\xi}_1, \cdots, \boldsymbol{\xi}_H\right) \in \mathbb{R}^{n \times H}$ and $S=\left( \boldsymbol{\psi}_1, \cdots, \boldsymbol{\psi}_H\right) \in \mathbb{R}^{n \times H}$ are matrices of estimated latent features called X-scores and Y-scores, $C=\left(\mathbf{c}_1, \cdots, \mathbf{c}_H\right) \in \mathbb{R}^{p \times H}$ and $D=\left(\mathbf{d}_1, \cdots, \mathbf{d}_H\right) \in \mathbb{R}^{q \times H}$ are matrices of X-loadings and Y-loadings, and  $E_X=\left(\mathbf{e}_{1 X}, \cdots, \mathbf{e}_{p X}\right) \in \mathbb{R}^{n\times p}$ and $E_Y=\left(\mathbf{e}_{1 Y}, \cdots, \mathbf{e}_{q Y}\right) \in \mathbb{R}^{n \times q}$ are the residual matrices. The elements of matrices $T$, $S$, $C$ and $D$ are presented in Table~\ref{deflation}.\\

In a regression setting, we have a so-called ``inner relationship'', which provides a link between the latent variables $S$ and $T$ through the relationship $S=T B$ (see \cite{sutton2018sparse}), where $B=\operatorname{diag}\left(b_1, \cdots\right.$, $\left.b_H\right)$. We can use the inner relationship and the relationship $T=X U\left(C^\top U\right)^{-1}$, where $U=\left(\boldsymbol{u}_1, \cdots, \boldsymbol{u}_H\right) \in \mathbb{R}^{p \times H}$ is the matrix of X-weights, to reparameterize the PLS model in terms of the original variables
$$
Y=X \hat{\beta}_{PLS}+E_Y,
$$
where $\hat{\beta}_{PLS}=U\left(C^{\top} U\right)^{-1}B D^{\top}=U\left(C^{\top}U\right)^{-1} T^{\top} Y$ (see page 40 of \cite{rosipal2005overview}). 

For PLS-regression, the matrix of adjusted weights is  $W=\left[\boldsymbol{w}_1, \ldots, \boldsymbol{w}_H\right]$ with $\boldsymbol{w}_1=\boldsymbol{u}_1$ and 
\[
\boldsymbol{w}_h=\prod_{j=1}^{h-1}\left(\boldsymbol{I}-\boldsymbol{u}_j \boldsymbol{c}_j^{\boldsymbol{\top}}\right) \boldsymbol{u}_h =\prod_{j=1}^{h-1}\left(\boldsymbol{I} -\boldsymbol{u}_j\left(\boldsymbol{\xi}_j^{\top} \boldsymbol{\xi}_j\right)^{-1} \boldsymbol{\xi}_j^{\top} X\right) \boldsymbol{u}_h.
\] 
Then, we have $X \boldsymbol{w}_h=X_{h-1} \boldsymbol{u}_h$ and $T=X W$. We can also write $W=U\left(C^\top U\right)^{-1}$.

\section*{Appendix D: Additional results from the simulation study}

\begin{table}[H]
\centering
\caption{Mean square error in prediction on a test set over 100 runs according to different noise levels where $p=15$, $q=10$, $n=100$, and $\gamma=5$.} \label{tabnoise}
\begin{tabular}{ccccc|cccc}
  \hline
& \multicolumn{4}{c}{BSS-PLS} &  \multicolumn{4}{c}{Sparse PLS}\\
\cmidrule{2-5}\cmidrule{6-9}
Subset size &$\sigma =  1.5$ & $\sigma =  3$ & $\sigma =  6$& $\sigma =  8$&$\sigma =  1.5$ & $\sigma =  3$ & $\sigma =  6$& $\sigma =  8$\\
\hline
1 & 149.24 & 167.04 & 202.11 & 232.78 & 149.24 & 167.07 & 202.15 & 232.80 \\ 
  2 & 137.70 & 155.36 & 195.48 & 228.29 & 141.36 & 160.43 & 197.65 & 229.92 \\ 
  3 & 131.14 & 147.45 & 189.76 & 225.03 & 135.76 & 153.25 & 193.96 & 227.56 \\ 
  4 & 126.67 & 141.07 & 185.13 & 222.10 & 131.33 & 147.73 & 190.54 & 225.05 \\ 
  5 & 123.64 & 136.33 & 182.53 & 220.25 & 128.04 & 143.02 & 187.34 & 223.05 \\ 
  6 & 121.26 & 132.48 & 180.60 & 218.93 & 125.42 & 138.90 & 184.78 & 221.60 \\ 
  7 & 119.12 & 129.49 & 178.52 & 217.80 & 123.30 & 135.42 & 182.48 & 220.18 \\ 
  8 & 117.34 & 126.21 & 177.03 & 216.91 & 120.49 & 131.41 & 180.66 & 218.97 \\ 
  9 & 115.62 & 123.84 & 175.68 & 215.85 & 118.46 & 127.70 & 179.02 & 218.10 \\ 
  10 & 114.21 & 122.26 & 175.04 & 215.52 & 114.46 & 124.76 & 177.55 & 217.16 \\ 
  11 & 114.63 & 121.99 & 174.52 & 214.91 & 114.52 & 123.26 & 176.39 & 216.28 \\ 
  12 & 114.79 & 122.05 & 174.22 & 214.65 & 114.59 & 122.64 & 175.44 & 215.60 \\ 
  13 & 114.88 & 122.17 & 173.95 & 214.53 & 114.68 & 122.38 & 174.70 & 215.03 \\ 
  14 & 114.93 & 122.25 & 173.69 & 214.37 & 114.81 & 122.25 & 174.12 & 214.66 \\ 
  15 & 114.94 & 122.28 & 173.69 & 214.34 & 114.94 & 122.28 & 173.69 & 214.34 \\ 
   \hline
\end{tabular}
\end{table}

\begin{table}[H]
\centering
\caption{Sensitivity (sens) and Specificity (spe)  over 100 runs according to different noise levels where $p=15$, $q=10$, $n=100$, and $\gamma=5$.} 
\begin{tabular}{ccccccccc|cccccccc}
  \hline
& \multicolumn{8}{c}{BSS-PLS} &  \multicolumn{8}{c}{Sparse PLS}\\
\cmidrule{2-8}\cmidrule{9-17}  
 Subset & \multicolumn{2}{c}{$\sigma =  1.5$} &\multicolumn{2}{c}{$\sigma =  3$} &\multicolumn{2}{c}{$\sigma =  6$} &\multicolumn{2}{c|}{$\sigma =  8$} &\multicolumn{2}{c}{$\sigma =  1.5$} &\multicolumn{2}{c}{$\sigma =  3$} &\multicolumn{2}{c}{$\sigma =  6$} &\multicolumn{2}{c}{$\sigma =  8$} \\
\cmidrule{2-17}  
size&sens&spe&sens&spe&sens&spe&sens&spe&sens&spe&sens&spe&sens&spe&sens&spe\\ \hline
1 & 0.10 & 1.00 & 0.10 & 1.00 & 0.10 & 1.00 & 0.09 & 0.99 & 0.10 & 1.00 & 0.10 & 1.00 & 0.10 & 1.00 & 0.09 & 0.99 \\ 
  2 & 0.20 & 1.00 & 0.20 & 1.00 & 0.19 & 0.98 & 0.17 & 0.94 & 0.20 & 1.00 & 0.20 & 1.00 & 0.19 & 0.97 & 0.17 & 0.95 \\ 
  3 & 0.30 & 1.00 & 0.30 & 1.00 & 0.28 & 0.96 & 0.25 & 0.90 & 0.30 & 1.00 & 0.30 & 1.00 & 0.28 & 0.96 & 0.25 & 0.89 \\ 
  4 & 0.40 & 1.00 & 0.40 & 1.00 & 0.37 & 0.94 & 0.33 & 0.86 & 0.40 & 1.00 & 0.40 & 1.00 & 0.37 & 0.94 & 0.33 & 0.86 \\ 
  5 & 0.50 & 1.00 & 0.50 & 1.00 & 0.45 & 0.91 & 0.41 & 0.82 & 0.50 & 1.00 & 0.50 & 1.00 & 0.45 & 0.90 & 0.41 & 0.81 \\ 
  6 & 0.60 & 1.00 & 0.60 & 0.99 & 0.53 & 0.86 & 0.48 & 0.77 & 0.60 & 1.00 & 0.60 & 0.99 & 0.53 & 0.86 & 0.48 & 0.76 \\ 
  7 & 0.70 & 1.00 & 0.69 & 0.99 & 0.61 & 0.82 & 0.55 & 0.69 & 0.70 & 1.00 & 0.69 & 0.99 & 0.61 & 0.82 & 0.55 & 0.70 \\ 
  8 & 0.80 & 1.00 & 0.79 & 0.98 & 0.68 & 0.76 & 0.61 & 0.63 & 0.80 & 1.00 & 0.79 & 0.98 & 0.68 & 0.76 & 0.61 & 0.62 \\ 
  9 & 0.90 & 1.00 & 0.88 & 0.96 & 0.74 & 0.69 & 0.67 & 0.54 & 0.90 & 1.00 & 0.88 & 0.96 & 0.74 & 0.68 & 0.67 & 0.54 \\ 
  10 & 1.00 & 1.00 & 0.96 & 0.91 & 0.79 & 0.59 & 0.72 & 0.44 & 1.00 & 1.00 & 0.96 & 0.91 & 0.79 & 0.59 & 0.73 & 0.45 \\ 
  11 & 1.00 & 0.80 & 0.99 & 0.77 & 0.83 & 0.47 & 0.78 & 0.35 & 1.00 & 0.80 & 0.99 & 0.77 & 0.83 & 0.47 & 0.78 & 0.36 \\ 
  12 & 1.00 & 0.60 & 0.99 & 0.59 & 0.88 & 0.35 & 0.84 & 0.27 & 1.00 & 0.60 & 0.99 & 0.59 & 0.88 & 0.36 & 0.84 & 0.27 \\ 
  13 & 1.00 & 0.40 & 1.00 & 0.39 & 0.92 & 0.24 & 0.90 & 0.19 & 1.00 & 0.40 & 1.00 & 0.39 & 0.92 & 0.24 & 0.89 & 0.19 \\ 
  14 & 1.00 & 0.20 & 1.00 & 0.20 & 0.96 & 0.13 & 0.95 & 0.09 & 1.00 & 0.20 & 1.00 & 0.20 & 0.97 & 0.13 & 0.95 & 0.10 \\ 
  15 & 1.00 & 0.00 & 1.00 & 0.00 & 1.00 & 0.00 & 1.00 & 0.00 & 1.00 & 0.00 & 1.00 & 0.00 & 1.00 & 0.00 & 1.00 & 0.00 \\ 
   \hline
\end{tabular}
\end{table}

\begin{table}[h]
\centering
\caption{F1-score for  BSS-PLS and sparse PLS for different subset sizes  over 100 runs for varying noise levels. Here, $p=15$, $q=10$, $n=100$, and $\gamma=5$.} \label{tabnoiseF1}
\begin{tabular}{ccccc|cccc}
  \hline
& \multicolumn{4}{c}{BSS-PLS} &  \multicolumn{4}{c}{Sparse PLS}\\
\cmidrule{2-5}\cmidrule{6-9}
Subset size &$\sigma =  1.5$ & $\sigma =  3$ & $\sigma =  6$& $\sigma =  8$&$\sigma =  1.5$ & $\sigma =  3$ & $\sigma =  6$& $\sigma =  8$\\
  \hline
 \hline
1 & 0.18 & 0.18 & 0.18 & 0.17 & 0.18 & 0.18 & 0.18 & 0.17 \\ 
  2 & 0.33 & 0.33 & 0.31 & 0.29 & 0.33 & 0.33 & 0.31 & 0.29 \\ 
  3 & 0.46 & 0.46 & 0.43 & 0.38 & 0.46 & 0.46 & 0.43 & 0.38 \\ 
  4 & 0.57 & 0.57 & 0.53 & 0.47 & 0.57 & 0.57 & 0.53 & 0.47 \\ 
  5 & 0.67 & 0.66 & 0.61 & 0.54 & 0.67 & 0.66 & 0.60 & 0.54 \\ 
  6 & 0.75 & 0.75 & 0.66 & 0.60 & 0.75 & 0.75 & 0.66 & 0.60 \\ 
  7 & 0.82 & 0.82 & 0.72 & 0.64 & 0.82 & 0.82 & 0.72 & 0.65 \\ 
  8 & 0.89 & 0.88 & 0.76 & 0.68 & 0.89 & 0.88 & 0.76 & 0.68 \\ 
  9 & 0.95 & 0.93 & 0.78 & 0.71 & 0.95 & 0.93 & 0.78 & 0.71 \\ 
  10 & 1.00 & 0.96 & 0.79 & 0.72 & 1.00 & 0.96 & 0.79 & 0.73 \\ 
  11 & 0.95 & 0.94 & 0.79 & 0.74 & 0.95 & 0.94 & 0.79 & 0.74 \\ 
  12 & 0.91 & 0.90 & 0.80 & 0.76 & 0.91 & 0.90 & 0.80 & 0.76 \\ 
  13 & 0.87 & 0.87 & 0.80 & 0.78 & 0.87 & 0.87 & 0.80 & 0.78 \\ 
  14 & 0.83 & 0.83 & 0.80 & 0.79 & 0.83 & 0.83 & 0.80 & 0.79 \\ 
  15 & 0.80 & 0.80 & 0.80 & 0.80 & 0.80 & 0.80 & 0.80 & 0.80 \\ 
   \hline
\end{tabular}
\end{table}

\begin{table}[H]
\centering
\caption{Number of times  BSS-PLS and sparse PLS retrieve the exact (``optimal") best subset of different sizes  over 100 runs for different sample sizes where $p=15$, $q=10$, $\gamma=5$, and $\sigma=6$.} \label{tabsamplesize}
\begin{tabular}{cccc|ccc}
  \hline
& \multicolumn{3}{c}{BSS-PLS} &   \multicolumn{3}{c}{Sparse PLS}\\
\cmidrule{2-4}\cmidrule{5-7}
Subset size & $n =  100$ & $n =  200$ & $n=  500$&$n =  100$ & $n =  200$ & $n=  500$\\ 
  \hline
1 & 100 & 100 & 100 & 95 & 95 & 98 \\ 
  2 & 96 & 96 & 96 & 92 & 91 & 93 \\ 
  3 & 97 & 99 & 99 & 89 & 93 & 94 \\ 
  4 & 98 & 97 & 98 & 85 & 96 & 95 \\ 
  5 & 95 & 97 & 100 & 83 & 93 & 91 \\ 
  6 & 96 & 98 & 99 & 89 & 95 & 97 \\ 
  7 & 95 & 99 & 98 & 89 & 95 & 98 \\ 
  8 & 98 & 99 & 100 & 90 & 92 & 98 \\ 
  9 & 97 & 98 & 100 & 93 & 94 & 98 \\ 
  10 & 99 & 100 & 100 & 96 & 100 & 97 \\ 
  11 & 100 & 100 & 100 & 94 & 96 & 100 \\ 
  12 & 100 & 100 & 100 & 95 & 99 & 99 \\ 
  13 & 100 & 100 & 100 & 94 & 100 & 99 \\ 
  14 & 100 & 100 & 100 & 99 & 100 & 100 \\ 
  15 & 100 & 100 & 100 & 100 & 100 & 100 \\ 
   \hline
\end{tabular}
\end{table}

\begin{table}[H]
\centering
\caption{Mean square error in prediction estimated on a test set over 100 runs for different sample sizes where $p=15$, $q=10$, $\gamma=5$, and $\sigma=6$.} \label{tabsamplesize}
\begin{tabular}{cccc|ccc}
  \hline
& \multicolumn{3}{c}{BSS-PLS} &   \multicolumn{3}{c}{Sparse PLS}\\
\cmidrule{2-4}\cmidrule{5-7}
Subset size & $n =  100$ & $n =  200$ & $n=  500$&$n =  100$ & $n =  200$ & $n=  500$\\ 
 \hline
1 & 202.11 & 208.36 & 201.69 & 202.15 & 208.29 & 201.65 \\ 
  2 & 195.48 & 201.51 & 195.32 & 197.65 & 203.65 & 197.33 \\ 
  3 & 189.76 & 195.75 & 189.77 & 193.96 & 199.50 & 193.25 \\ 
  4 & 185.13 & 191.59 & 185.16 & 190.54 & 195.58 & 189.87 \\ 
  5 & 182.53 & 187.50 & 181.18 & 187.34 & 191.98 & 186.36 \\ 
  6 & 180.60 & 184.26 & 177.45 & 184.78 & 189.28 & 183.16 \\ 
  7 & 178.52 & 181.74 & 174.23 & 182.48 & 186.45 & 180.07 \\ 
  8 & 177.03 & 179.66 & 171.51 & 180.66 & 183.88 & 176.40 \\ 
  9 & 175.68 & 178.13 & 169.37 & 179.02 & 181.79 & 173.29 \\ 
  10 & 175.04 & 176.93 & 167.70 & 177.55 & 179.87 & 169.82 \\ 
  11 & 174.52 & 176.52 & 167.42 & 176.39 & 178.29 & 168.52 \\ 
  12 & 174.22 & 176.16 & 167.53 & 175.44 & 177.41 & 168.04 \\ 
  13 & 173.95 & 175.91 & 167.52 & 174.70 & 176.64 & 167.67 \\ 
  14 & 173.69 & 175.89 & 167.52 & 174.12 & 176.11 & 167.54 \\ 
  15 & 173.69 & 175.89 & 167.52 & 173.69 & 175.89 & 167.52 \\  
   \hline
\end{tabular}
\end{table}

\begin{table}[H]
\centering
\caption{Sensitivity (sens) and specificity (spe)  over 100 runs for different sample sizes where $p=15$, $q=10$, $n=100$, and $\gamma=5$.} 
\begin{tabular}{ccccccc|cccccc}
  \hline
& \multicolumn{6}{c}{BSS-PLS} &  \multicolumn{6}{c}{Sparse PLS}\\
\cmidrule{2-7}\cmidrule{8-11}  
 Subset & \multicolumn{2}{c}{$n=  100$} &\multicolumn{2}{c}{$n =  200$} &\multicolumn{2}{c}{$n=500$} & \multicolumn{2}{c}{$n=  100$} &\multicolumn{2}{c}{$n =  200$} &\multicolumn{2}{c}{$n=500$} \\
\cmidrule{2-13}  
size&sens&spe&sens&spe&sens&spe&sens&spe&sens&spe&sens&spe \\ \hline
\hline
1 & 0.10 & 1.00 & 0.10 & 1.00 & 0.10 & 1.00 & 0.10 & 1.00 & 0.10 & 1.00 & 0.10 & 1.00 \\ 
  2 & 0.19 & 0.98 & 0.20 & 1.00 & 0.20 & 1.00 & 0.19 & 0.97 & 0.20 & 1.00 & 0.20 & 1.00 \\ 
  3 & 0.28 & 0.96 & 0.30 & 0.99 & 0.30 & 1.00 & 0.28 & 0.96 & 0.30 & 0.99 & 0.30 & 1.00 \\ 
  4 & 0.37 & 0.94 & 0.39 & 0.98 & 0.40 & 1.00 & 0.37 & 0.94 & 0.39 & 0.99 & 0.40 & 1.00 \\ 
  5 & 0.45 & 0.91 & 0.49 & 0.97 & 0.50 & 1.00 & 0.45 & 0.90 & 0.48 & 0.97 & 0.50 & 1.00 \\ 
  6 & 0.53 & 0.86 & 0.58 & 0.95 & 0.60 & 1.00 & 0.53 & 0.86 & 0.57 & 0.95 & 0.60 & 1.00 \\ 
  7 & 0.61 & 0.82 & 0.66 & 0.92 & 0.70 & 1.00 & 0.61 & 0.82 & 0.66 & 0.92 & 0.70 & 1.00 \\ 
  8 & 0.68 & 0.76 & 0.74 & 0.87 & 0.79 & 0.99 & 0.68 & 0.76 & 0.74 & 0.87 & 0.79 & 0.99 \\ 
  9 & 0.74 & 0.69 & 0.80 & 0.81 & 0.89 & 0.98 & 0.74 & 0.68 & 0.81 & 0.81 & 0.89 & 0.98 \\ 
  10 & 0.79 & 0.59 & 0.86 & 0.73 & 0.96 & 0.93 & 0.79 & 0.59 & 0.86 & 0.73 & 0.96 & 0.93 \\ 
  11 & 0.83 & 0.47 & 0.91 & 0.61 & 0.99 & 0.78 & 0.83 & 0.47 & 0.91 & 0.62 & 0.99 & 0.78 \\ 
  12 & 0.88 & 0.35 & 0.94 & 0.48 & 0.99 & 0.59 & 0.88 & 0.36 & 0.94 & 0.48 & 0.99 & 0.59 \\ 
  13 & 0.92 & 0.24 & 0.97 & 0.34 & 1.00 & 0.40 & 0.92 & 0.24 & 0.97 & 0.34 & 1.00 & 0.40 \\ 
  14 & 0.96 & 0.13 & 0.99 & 0.18 & 1.00 & 0.20 & 0.97 & 0.13 & 0.99 & 0.18 & 1.00 & 0.20 \\ 
  15 & 1.00 & 0.00 & 1.00 & 0.00 & 1.00 & 0.00 & 1.00 & 0.00 & 1.00 & 0.00 & 1.00 & 0.00 \\ 
 
   \hline
\end{tabular}
\end{table}   

\begin{table}[H]
\centering
\caption{F1-score for different subset sizes  over 100 runs for different sample sizes where $p=15$, $q=10$, $\gamma=5$, and $\sigma=6$.} \label{tabsamplesize}
\begin{tabular}{cccc|ccc}
  \hline
& \multicolumn{3}{c}{BSS-PLS} &   \multicolumn{3}{c}{Sparse PLS}\\
\cmidrule{2-4}\cmidrule{5-7}
Subset size & $n =  100$ & $n =  200$ & $n=  500$&$n =  100$ & $n =  200$ & $n=  500$\\ 
  \hline
1 & 0.18 & 0.18 & 0.18 & 0.18 & 0.18 & 0.18 \\ 
  2 & 0.31 & 0.33 & 0.33 & 0.31 & 0.33 & 0.33 \\ 
  3 & 0.43 & 0.46 & 0.46 & 0.43 & 0.46 & 0.46 \\ 
  4 & 0.53 & 0.56 & 0.57 & 0.53 & 0.56 & 0.57 \\ 
  5 & 0.61 & 0.65 & 0.67 & 0.60 & 0.65 & 0.67 \\ 
  6 & 0.66 & 0.72 & 0.75 & 0.66 & 0.72 & 0.75 \\ 
  7 & 0.72 & 0.78 & 0.82 & 0.72 & 0.78 & 0.82 \\ 
  8 & 0.76 & 0.82 & 0.88 & 0.76 & 0.82 & 0.88 \\ 
  9 & 0.78 & 0.85 & 0.93 & 0.78 & 0.85 & 0.94 \\ 
  10 & 0.79 & 0.86 & 0.96 & 0.79 & 0.86 & 0.96 \\ 
  11 & 0.79 & 0.86 & 0.94 & 0.79 & 0.86 & 0.94 \\ 
  12 & 0.80 & 0.86 & 0.90 & 0.80 & 0.86 & 0.90 \\ 
  13 & 0.80 & 0.84 & 0.87 & 0.80 & 0.84 & 0.87 \\ 
  14 & 0.80 & 0.82 & 0.83 & 0.80 & 0.82 & 0.83 \\ 
  15 & 0.80 & 0.80 & 0.80 & 0.80 & 0.80 & 0.80 \\ 
 \hline
\end{tabular}
\end{table}

\begin{table}[H]
\centering
\caption{Number of times  BSS-PLS and sparse PLS retrieve the exact (``optimal") best subset for different subset sizes  over 100 runs according to the sparsity of the true signal. Here, $p=15$, $q=10$, $n=100$, and $\sigma=5$.} 
\begin{tabular}{ccccc|cccc}
  \hline
& \multicolumn{4}{c}{BSS-PLS} &  \multicolumn{4}{c}{Sparse PLS}\\
\cmidrule{2-5}\cmidrule{6-9}
Subset size &$\gamma =  3$ & $\gamma =  7$& $\gamma =  11$& $\gamma =  13$&$\gamma =  3$ & $\gamma =  7$& $\gamma =  11$& $\gamma =  13$\\
  \hline
 \hline
1 & 100 & 100 & 100 & 99 & 93 & 95 & 92 & 91 \\ 
  2 & 97 & 98 & 93 & 96 & 85 & 89 & 90 & 90 \\ 
  3 & 92 & 96 & 92 & 96 & 88 & 90 & 91 & 86 \\ 
  4 & 95 & 97 & 90 & 96 & 93 & 86 & 86 & 79 \\ 
  5 & 91 & 91 & 91 & 98 & 88 & 84 & 79 & 85 \\ 
  6 & 91 & 94 & 98 & 95 & 87 & 85 & 83 & 87 \\ 
  7 & 96 & 98 & 98 & 94 & 89 & 84 & 87 & 87 \\ 
  8 & 96 & 97 & 100 & 97 & 94 & 87 & 86 & 74 \\ 
  9 & 99 & 99 & 99 & 97 & 95 & 91 & 86 & 85 \\ 
  10 & 98 & 100 & 99 & 98 & 90 & 89 & 85 & 86 \\ 
  11 & 100 & 100 & 100 & 100 & 92 & 92 & 87 & 87 \\ 
  12 & 100 & 100 & 100 & 99 & 94 & 89 & 90 & 89 \\ 
  13 & 100 & 100 & 100 & 100 & 99 & 95 & 97 & 94 \\ 
  14 & 100 & 100 & 100 & 100 & 99 & 95 & 98 & 98 \\ 
  15 & 100 & 100 & 100 & 100 & 100 & 100 & 100 & 100 \\ 
\hline
 \end{tabular}
\end{table}

\begin{table}[H]
\centering
\caption{Mean square error in prediction on a test set over 100 runs for varying sparsity levels. Here, $p=15$, $q=10$, $n=100$, and $\sigma=5$.} \label{tabnoise}
\begin{tabular}{ccccc|cccc}
  \hline
& \multicolumn{4}{c}{BSS-PLS} &  \multicolumn{4}{c}{Sparse PLS}\\
\cmidrule{2-5}\cmidrule{6-9}
Subset size &$\gamma =  3$ & $\gamma =  7$& $\gamma =  9$& $\gamma =  11$&$\gamma =  3$ & $\gamma =  7$& $\gamma =  9$& $\gamma =  11$\\
  \hline
1 & 201.88 & 202.83 & 203.61 & 205.31 & 201.93 & 203.03 & 203.61 & 204.91 \\ 
  2 & 194.93 & 196.62 & 198.16 & 201.92 & 197.36 & 198.83 & 199.99 & 202.55 \\ 
  3 & 189.18 & 192.36 & 195.01 & 199.87 & 193.05 & 195.39 & 197.03 & 200.72 \\ 
  4 & 183.86 & 189.40 & 192.83 & 198.44 & 189.29 & 192.72 & 194.83 & 199.42 \\ 
  5 & 180.12 & 186.87 & 191.71 & 197.65 & 186.11 & 190.33 & 193.52 & 198.26 \\ 
  6 & 177.01 & 185.29 & 190.05 & 196.90 & 182.65 & 188.37 & 191.93 & 197.68 \\ 
  7 & 174.82 & 183.94 & 189.54 & 196.35 & 180.26 & 186.62 & 190.88 & 197.03 \\ 
  8 & 173.18 & 182.81 & 188.79 & 196.29 & 178.18 & 185.35 & 190.01 & 196.60 \\ 
  9 & 171.54 & 182.06 & 188.36 & 195.96 & 175.88 & 184.14 & 189.34 & 196.24 \\ 
  10 & 170.06 & 181.20 & 188.26 & 195.72 & 173.67 & 183.04 & 189.03 & 196.00 \\ 
  11 & 168.93 & 181.00 & 188.02 & 195.59 & 171.99 & 182.07 & 188.52 & 195.77 \\ 
  12 & 168.43 & 180.72 & 187.88 & 195.61 & 170.58 & 181.57 & 188.18 & 195.60 \\ 
  13 & 168.14 & 180.52 & 187.69 & 195.57 & 169.43 & 181.04 & 187.86 & 195.46 \\ 
  14 & 167.75 & 180.36 & 187.68 & 195.54 & 168.39 & 180.63 & 187.67 & 195.46 \\ 
  15 & 167.64 & 180.33 & 187.65 & 195.52 & 167.64 & 180.33 & 187.65 & 195.52 \\ 
   \hline
\end{tabular}
\end{table}

\begin{table}[H]
\centering
\caption{Sensitivity (sens) and specificity (spe)  over 100 runs for varying sparsity levels. Here, $p=15$, $q=10$, $n=100$, and $\sigma=5$.} 
\begin{tabular}{ccccccccc|cccccccc}
  \hline
& \multicolumn{8}{c}{BSS-PLS} &  \multicolumn{8}{c}{Sparse PLS}\\
\cmidrule{2-8}\cmidrule{9-17}  
 Subset & \multicolumn{2}{c}{$\gamma =  3$} &\multicolumn{2}{c}{$\gamma =  7$} &\multicolumn{2}{c}{$\gamma =  9$} &\multicolumn{2}{c|}{$\gamma =  11$} &\multicolumn{2}{c}{$\gamma =  3$} &\multicolumn{2}{c}{$\gamma =  7$} &\multicolumn{2}{c}{$\gamma =  9$} &\multicolumn{2}{c}{$\gamma =  11$} \\
\cmidrule{2-17}  
size&sens&spe&sens&spe&sens&spe&sens&spe&sens&spe&sens&spe&sens&spe&sens&spe\\ \hline
\hline
1 & 0.08 & 1.00 & 0.12 & 0.99 & 0.15 & 0.99 & 0.15 & 0.99 & 0.08 & 1.00 & 0.12 & 0.99 & 0.15 & 0.99 & 0.18 & 0.98 \\ 
  2 & 0.16 & 0.99 & 0.22 & 0.97 & 0.27 & 0.96 & 0.27 & 0.96 & 0.16 & 0.98 & 0.22 & 0.97 & 0.27 & 0.96 & 0.33 & 0.94 \\ 
  3 & 0.24 & 0.97 & 0.32 & 0.94 & 0.38 & 0.92 & 0.38 & 0.92 & 0.24 & 0.97 & 0.32 & 0.94 & 0.39 & 0.92 & 0.44 & 0.89 \\ 
  4 & 0.32 & 0.96 & 0.42 & 0.91 & 0.49 & 0.88 & 0.49 & 0.88 & 0.32 & 0.96 & 0.42 & 0.91 & 0.50 & 0.89 & 0.53 & 0.83 \\ 
  5 & 0.40 & 0.93 & 0.51 & 0.87 & 0.57 & 0.82 & 0.57 & 0.82 & 0.40 & 0.94 & 0.51 & 0.86 & 0.57 & 0.83 & 0.62 & 0.77 \\ 
  6 & 0.48 & 0.90 & 0.59 & 0.82 & 0.66 & 0.77 & 0.66 & 0.77 & 0.47 & 0.90 & 0.59 & 0.81 & 0.65 & 0.77 & 0.69 & 0.71 \\ 
  7 & 0.55 & 0.86 & 0.66 & 0.75 & 0.71 & 0.70 & 0.71 & 0.70 & 0.55 & 0.86 & 0.66 & 0.75 & 0.70 & 0.69 & 0.74 & 0.63 \\ 
  8 & 0.62 & 0.80 & 0.72 & 0.68 & 0.78 & 0.63 & 0.78 & 0.63 & 0.62 & 0.80 & 0.72 & 0.68 & 0.76 & 0.62 & 0.78 & 0.56 \\ 
  9 & 0.69 & 0.74 & 0.77 & 0.60 & 0.81 & 0.54 & 0.81 & 0.54 & 0.68 & 0.74 & 0.77 & 0.60 & 0.81 & 0.54 & 0.82 & 0.48 \\ 
  10 & 0.75 & 0.67 & 0.83 & 0.52 & 0.85 & 0.46 & 0.85 & 0.46 & 0.75 & 0.66 & 0.83 & 0.52 & 0.84 & 0.45 & 0.86 & 0.40 \\ 
  11 & 0.80 & 0.55 & 0.86 & 0.41 & 0.88 & 0.37 & 0.88 & 0.37 & 0.81 & 0.56 & 0.86 & 0.41 & 0.88 & 0.37 & 0.90 & 0.33 \\ 
  12 & 0.86 & 0.44 & 0.90 & 0.31 & 0.92 & 0.28 & 0.92 & 0.28 & 0.86 & 0.44 & 0.90 & 0.31 & 0.92 & 0.28 & 0.93 & 0.25 \\ 
  13 & 0.91 & 0.29 & 0.94 & 0.22 & 0.95 & 0.19 & 0.95 & 0.19 & 0.91 & 0.30 & 0.94 & 0.21 & 0.95 & 0.19 & 0.95 & 0.17 \\ 
  14 & 0.96 & 0.16 & 0.97 & 0.11 & 0.97 & 0.09 & 0.97 & 0.09 & 0.96 & 0.16 & 0.97 & 0.11 & 0.97 & 0.09 & 0.97 & 0.08 \\ 
  15 & 1.00 & 0.00 & 1.00 & 0.00 & 1.00 & 0.00 & 1.00 & 0.00 & 1.00 & 0.00 & 1.00 & 0.00 & 1.00 & 0.00 & 1.00 & 0.00 \\
  
   \hline
\end{tabular}
\end{table}

\begin{table}[H]
\centering
\caption{F1-score for different subset sizes  over 100 runs according to the sparsity of the true signal. Here, $p=15$, $q=10$, $n=100$, and $\sigma=5$.} 
\begin{tabular}{ccccc|cccc}
  \hline
& \multicolumn{4}{c}{BSS-PLS} &  \multicolumn{4}{c}{Sparse PLS}\\
\cmidrule{2-5}\cmidrule{6-9}
Subset size &$\gamma =  3$ & $\gamma =  7$& $\gamma =  9$& $\gamma =  11$&$\gamma =  3$ & $\gamma =  7$& $\gamma =  9$& $\gamma =  11$\\
  \hline
 \hline
1 & 0.15 & 0.21 & 0.25 & 0.29 & 0.15 & 0.21 & 0.25 & 0.30 \\ 
  2 & 0.28 & 0.36 & 0.41 & 0.44 & 0.28 & 0.35 & 0.40 & 0.44 \\ 
  3 & 0.39 & 0.47 & 0.51 & 0.50 & 0.39 & 0.47 & 0.52 & 0.51 \\ 
  4 & 0.48 & 0.56 & 0.59 & 0.54 & 0.48 & 0.56 & 0.59 & 0.53 \\ 
  5 & 0.56 & 0.63 & 0.62 & 0.56 & 0.57 & 0.62 & 0.63 & 0.56 \\ 
  6 & 0.63 & 0.67 & 0.66 & 0.56 & 0.63 & 0.67 & 0.65 & 0.55 \\ 
  7 & 0.69 & 0.70 & 0.66 & 0.54 & 0.69 & 0.70 & 0.65 & 0.54 \\ 
  8 & 0.74 & 0.72 & 0.67 & 0.53 & 0.74 & 0.72 & 0.66 & 0.52 \\ 
  9 & 0.78 & 0.73 & 0.65 & 0.51 & 0.78 & 0.73 & 0.65 & 0.50 \\ 
  10 & 0.82 & 0.74 & 0.64 & 0.50 & 0.82 & 0.74 & 0.63 & 0.49 \\ 
  11 & 0.84 & 0.72 & 0.62 & 0.48 & 0.84 & 0.72 & 0.62 & 0.48 \\ 
  12 & 0.86 & 0.72 & 0.62 & 0.47 & 0.86 & 0.72 & 0.61 & 0.47 \\ 
  13 & 0.87 & 0.72 & 0.60 & 0.45 & 0.87 & 0.71 & 0.60 & 0.45 \\ 
  14 & 0.88 & 0.71 & 0.58 & 0.43 & 0.88 & 0.70 & 0.58 & 0.43 \\ 
  15 & 0.89 & 0.70 & 0.57 & 0.42 & 0.89 & 0.70 & 0.57 & 0.42 \\ 
 \hline
\end{tabular}
\end{table}

\begin{table}[H]
\centering
\caption{Result over 100 runs of MSEP, Sensitivity, Specificity and F1-score for BSS-PLS and Sparse PLS obtained from the subset of size $10$ with $q=10$, $n=100$, and $\sigma=5$.} 
\begin{tabular}{c|cccc|cccc}
  \hline
& \multicolumn{4}{c}{BSS-PLS} &  \multicolumn{4}{c}{Sparse PLS}\\
  \hline
$p$&MSEP&Sensitivity&Specificity&F1-score&MSEP&Sensitivity&Specificity&F1-score\\ \hline

50 & 168.147 & 0.591 & 0.898 & 0.591 & 170.440 & 0.586 & 0.896 &0.586 \\ 
  100 & 171.452 & 0.499 & 0.944 &0.499 & 174.006 & 0.487 & 0.943 & 0.487 \\ 
  200 & 179.016 & 0.397 & 0.968 & 0.397 & 180.329 & 0.393 & 0.968 & 0.393 \\ 
  500 & 186.232 & 0.275 & 0.985 & 0.275 & 186.542 & 0.269 & 0.985 & 0.269 \\ \hline
\end{tabular}
\end{table}

\begin{table}[H]
\centering
\caption{Model accuracy over 100 runs for the univariate response case.}
\begin{tabular}{cccc|cccc|cccc|cccc}
  \hline
\multicolumn{4}{c}{Setting}&\multicolumn{4}{c}{BSS-PLS} &  \multicolumn{4}{c}{Sparse PLS}&\multicolumn{4}{c}{Lasso}\\  \hline
 $n$ &$p$&sparsity&SNR&MSEP&sens&spe&F1&MSEP&sens&spe&F1&MSEP&sens&spe&F1 \\ \hline
400 & 40 & 10 & 10 & 510.22 & 0.99 & 0.82 & 0.97 & 510.46 & 1.00 & 0.48 & 0.92 & 510.63 & 0.81 & 0.28 & 0.78 \\ 
  400 & 40 & 10 & 15 & 651.88 & 0.98 & 0.78 & 0.96 & 652.15 & 1.00 & 0.42 & 0.92 & 652.44 & 0.73 & 0.32 & 0.73 \\ 
  400 & 40 & 30 & 10 & 511.48 & 0.99 & 0.97 & 0.96 & 511.75 & 1.00 & 0.81 & 0.83 & 511.88 & 0.96 & 0.29 & 0.51 \\ 
  400 & 40 & 30 & 15 & 652.99 & 0.99 & 0.94 & 0.94 & 653.30 & 1.00 & 0.75 & 0.79 & 653.63 & 0.93 & 0.32 & 0.51 \\ 
  40 & 80 & 20 & 10 & 941.55 & 0.88 & 0.74 & 0.87 & 943.26 & 0.97 & 0.49 & 0.90 & 935.17 & 0.22 & 0.77 & 0.33 \\ 
  40 & 80 & 20 & 15 & 1975.67 & 0.70 & 0.68 & 0.69 & 1977.20 & 0.75 & 0.53 & 0.70 & 1961.89 & 0.22 & 0.77 & 0.31 \\ 
  40 & 80 & 40 & 10 & 940.82 & 0.87 & 0.83 & 0.84 & 943.64 & 0.93 & 0.58 & 0.79 & 937.08 & 0.24 & 0.78 & 0.31 \\ 
  40 & 80 & 40 & 20 & 1976.26 & 0.73 & 0.75 & 0.68 & 1979.84 & 0.79 & 0.55 & 0.66 & 1965.83 & 0.22 & 0.78 & 0.28 \\ \hline
  \end{tabular}
\end{table}

\newpage
\section*{Appendix E: Illustration of Best Subset Selection for PCA}


\begin{figure}[H]
\begin{center}
\centerline{\includegraphics[width=.9\textwidth]{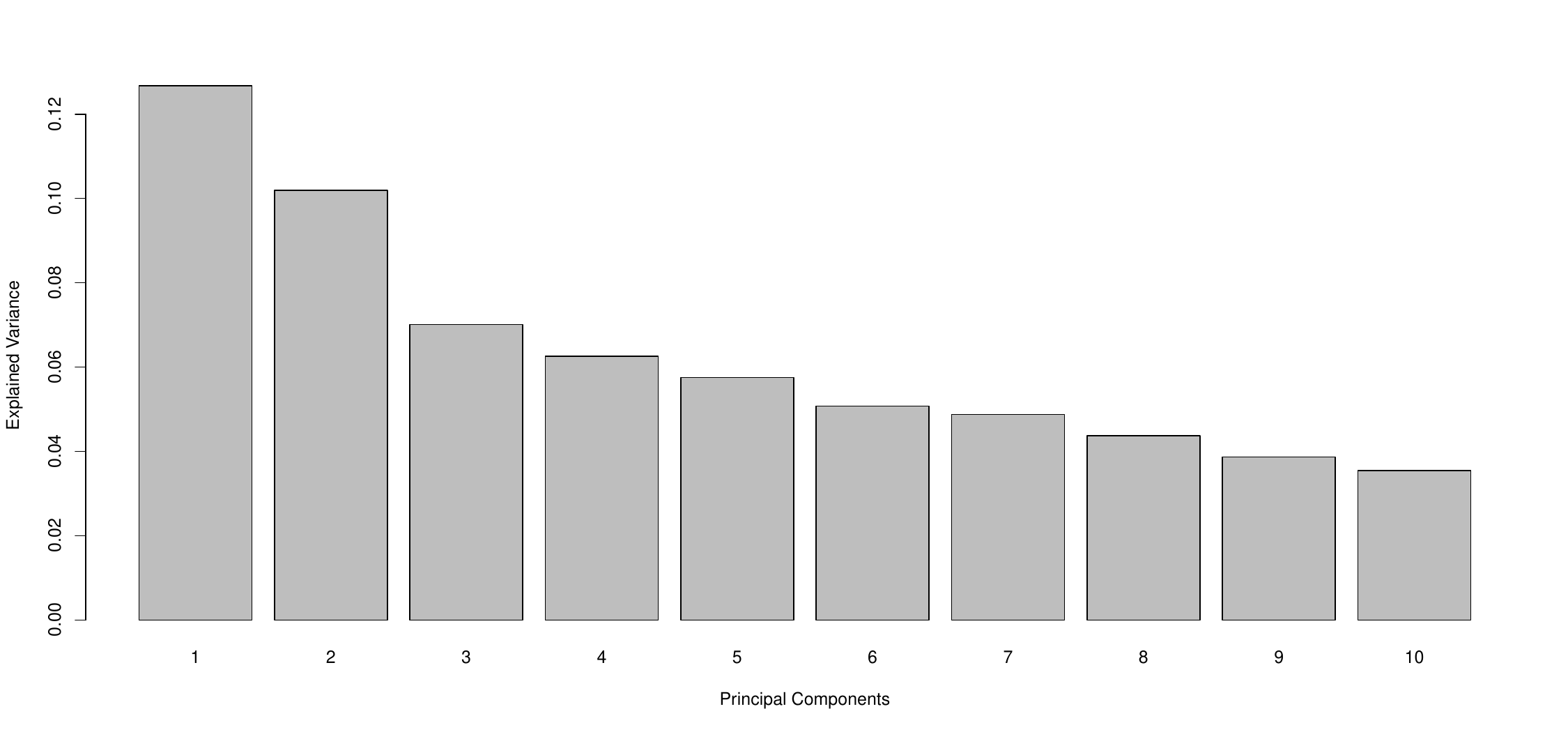}}
\end{center}
\caption{Scree plot for the PCA model on the \texttt{multidrug} dataset.\label{scree_plot}}
\end{figure}

\begin{figure}[H]
\begin{center}
\centerline{\includegraphics[width=.9\textwidth]{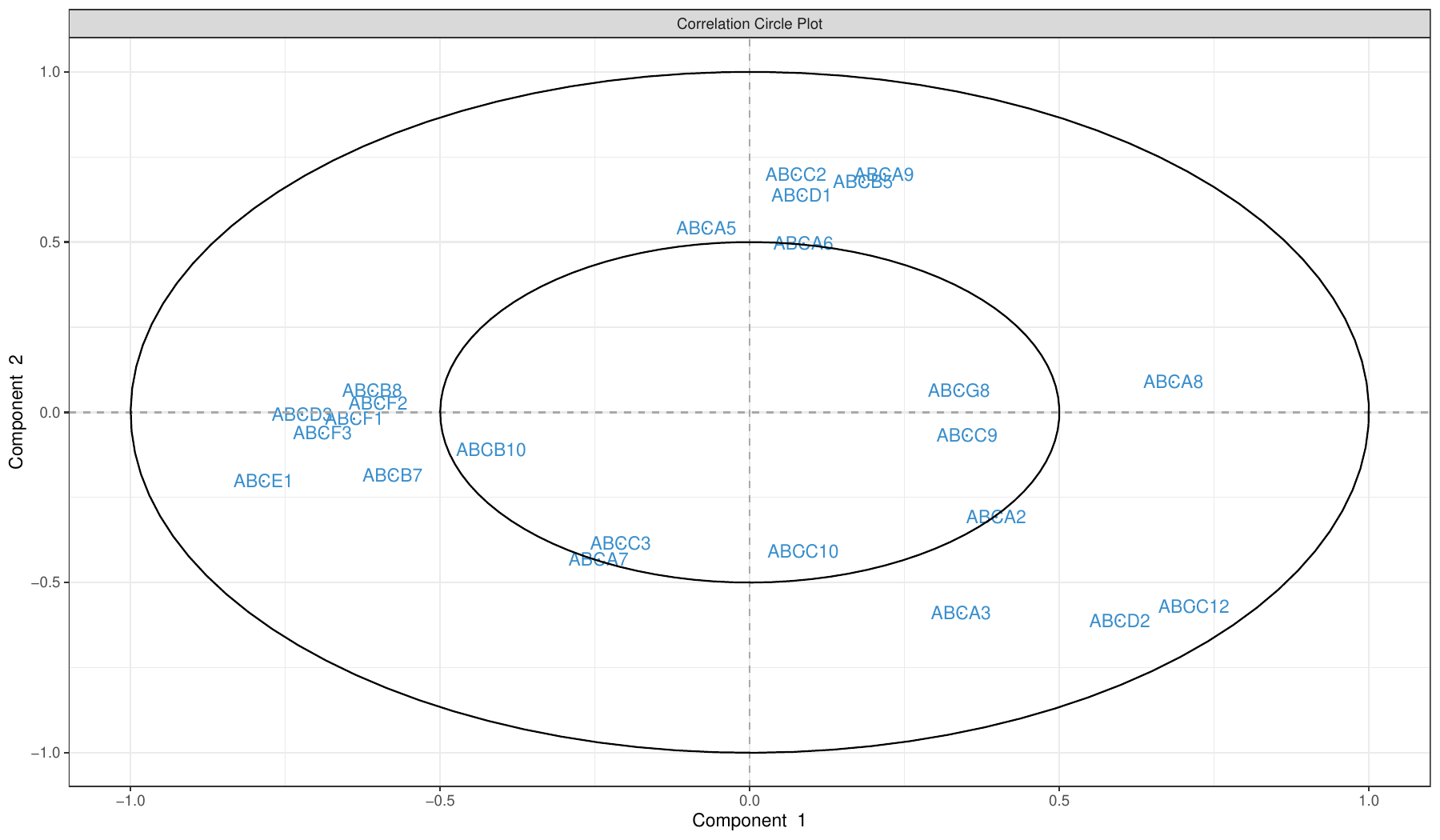}}
\end{center}
\caption{Correlation circle plot from BSS-PCA performed on the \texttt{multidrug} dataset.}
\end{figure}

\newpage

\begin{table}[H]
\centering
\caption{Best Subset results for component 1.} \label{tab.res.pca.1}
\begin{tabular}{ccccc}
  \hline
Dimension & $-\lambda_1/n$& PEV(sparse) & CPEV & cor(sPC,\, PC) \\ 
  \hline
1 & -0.01639 & 2.08 & 2.08 & -0.03 \\ 
  2 & -0.03031 & 3.85 & 3.85 & 0.41 \\ 
  3 & -0.03593 & 4.57 & 4.57 & 0.82 \\ 
  4 & -0.04360 & 5.54 & 5.54 & 0.84 \\ 
  5 & -0.04979 & 6.33 & 6.33 & 0.86 \\ 
  6 & -0.05462 & 6.94 & 6.94 & 0.87 \\ 
  7 & -0.05906 & 7.51 & 7.51 & 0.88 \\ 
  8 & -0.06488 & 8.25 & 8.25 & 0.88 \\ 
  9 & -0.07025 & 8.93 & 8.93 & 0.85 \\ 
  10 & -0.07418 & 9.43 & 9.43 & 0.87 \\ 
  11 & -0.07690 & 9.78 & 9.78 & 0.86 \\ 
  12 & -0.07903 & 10.05 & 10.05 & 0.87 \\ 
  13 & -0.08096 & 10.29 & 10.29 & 0.86 \\ 
  14 & -0.08274 & 10.52 & 10.52 & 0.86 \\ 
  15 & -0.08356 & 10.62 & 10.62 & 0.85 \\ 
  16 & -0.08473 & 10.77 & 10.77 & 0.88 \\ 
  17 & -0.08515 & 10.82 & 10.82 & 0.90 \\ 
  18 & -0.08662 & 11.01 & 11.01 & 0.94 \\ 
  19 & -0.08803 & 11.19 & 11.19 & 0.95 \\ 
  20 & -0.08910 & 11.33 & 11.33 & 0.94 \\ 
   \hline
\end{tabular}
\end{table}

\begin{figure}[H]
\centerline{
 \includegraphics[scale=0.45]{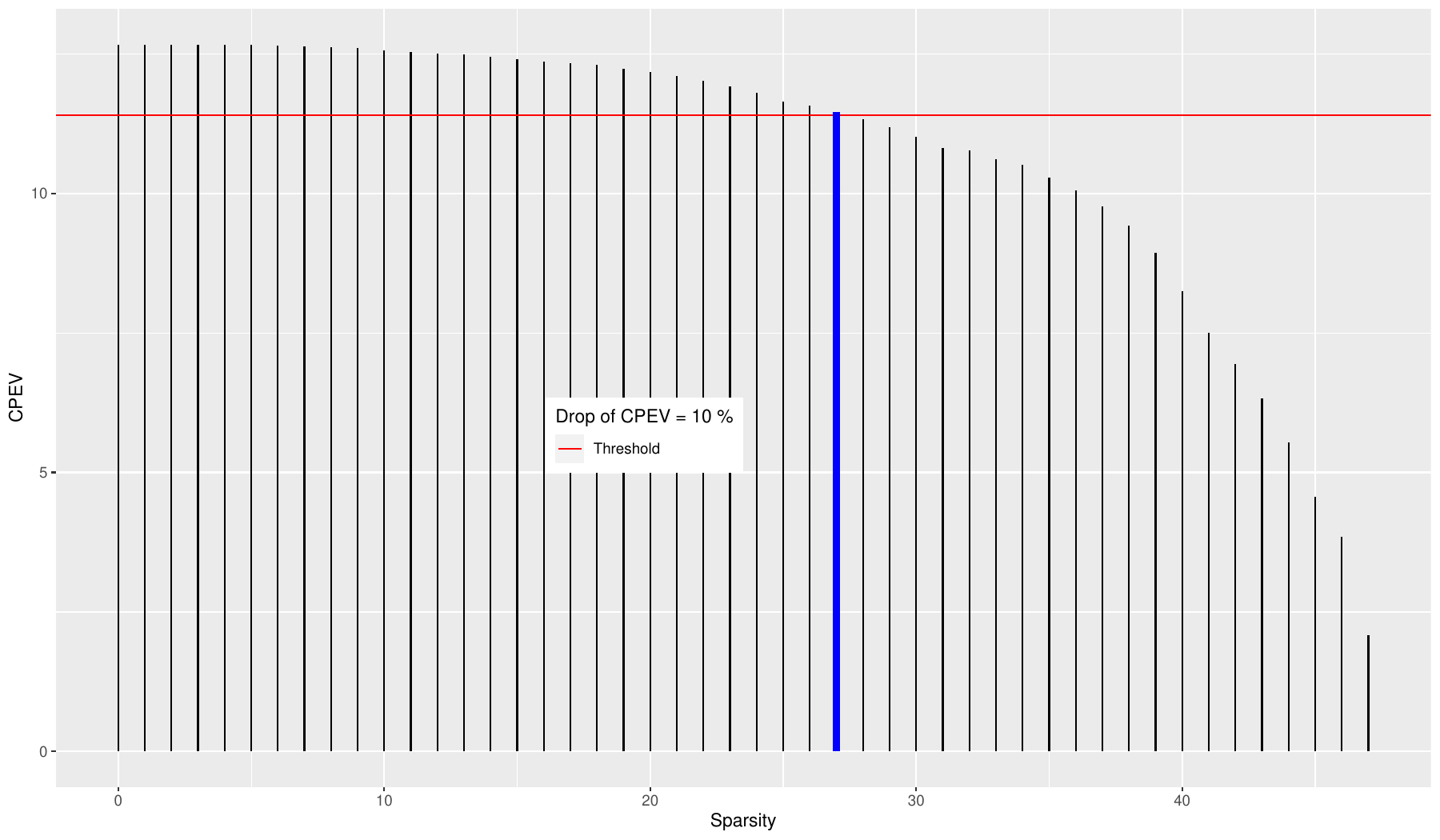}}
 \caption{CPEV as a function of the sparsity ($p$ - size of the subset) for the first component. Here, the blue vertical line indicates the largest value of the sparsity such that  CPEV does not exceed a drop of 10\%.} \label{drop-CPEV1}
\end{figure}

{\small
\begin{table}[H]
\centering
\caption{Best Subset results for component 2.} \label{tab.res.pca.2}
\begin{tabular}{ccccc}
  \hline
Dimension & $-\lambda_1/n$& PEV(sparse) & CPEV & cor(sPC,PC) \\ 
  \hline
1 & -0.01631 & 2.08 & 12.91 & 0.46 \\ 
  2 & -0.02641 & 3.42 & 14.24 & -0.14 \\ 
  3 & -0.03285 & 4.33 & 15.05 & 0.65 \\ 
  4 & -0.03840 & 4.90 & 15.71 & 0.74 \\ 
  5 & -0.04436 & 5.67 & 16.47 & 0.78 \\ 
  6 & -0.04732 & 6.02 & 16.86 & 0.85 \\ 
  7 & -0.05068 & 6.57 & 17.40 & 0.89 \\ 
  8 & -0.05368 & 6.97 & 17.78 & 0.90 \\ 
  9 & -0.05685 & 7.37 & 18.21 & 0.91 \\ 
  10 & -0.05945 & 7.63 & 18.55 & 0.90 \\ 
  11 & -0.06169 & 7.88 & 18.87 & 0.90 \\ 
  12 & -0.06372 & 8.16 & 19.12 & 0.90 \\ 
  13 & -0.06554 & 8.37 & 19.36 & 0.90 \\ 
  14 & -0.06697 & 8.61 & 19.62 & -0.91 \\ 
  15 & -0.06834 & 8.71 & 19.67 & -0.89 \\ 
  16 & -0.06964 & 8.87 & 19.84 & -0.89 \\ 
  17 & -0.07065 & 8.99 & 19.98 & -0.89 \\ 
  18 & -0.07148 & 9.10 & 20.08 & -0.89 \\ 
  19 & -0.07232 & 9.20 & 20.19 & -0.90 \\ 
  20 & -0.07343 & 9.35 & 20.32 & -0.91 \\ 
   \hline
\end{tabular}
\end{table}
}
\begin{figure}[H]

\centerline{
 \includegraphics[scale=0.45]{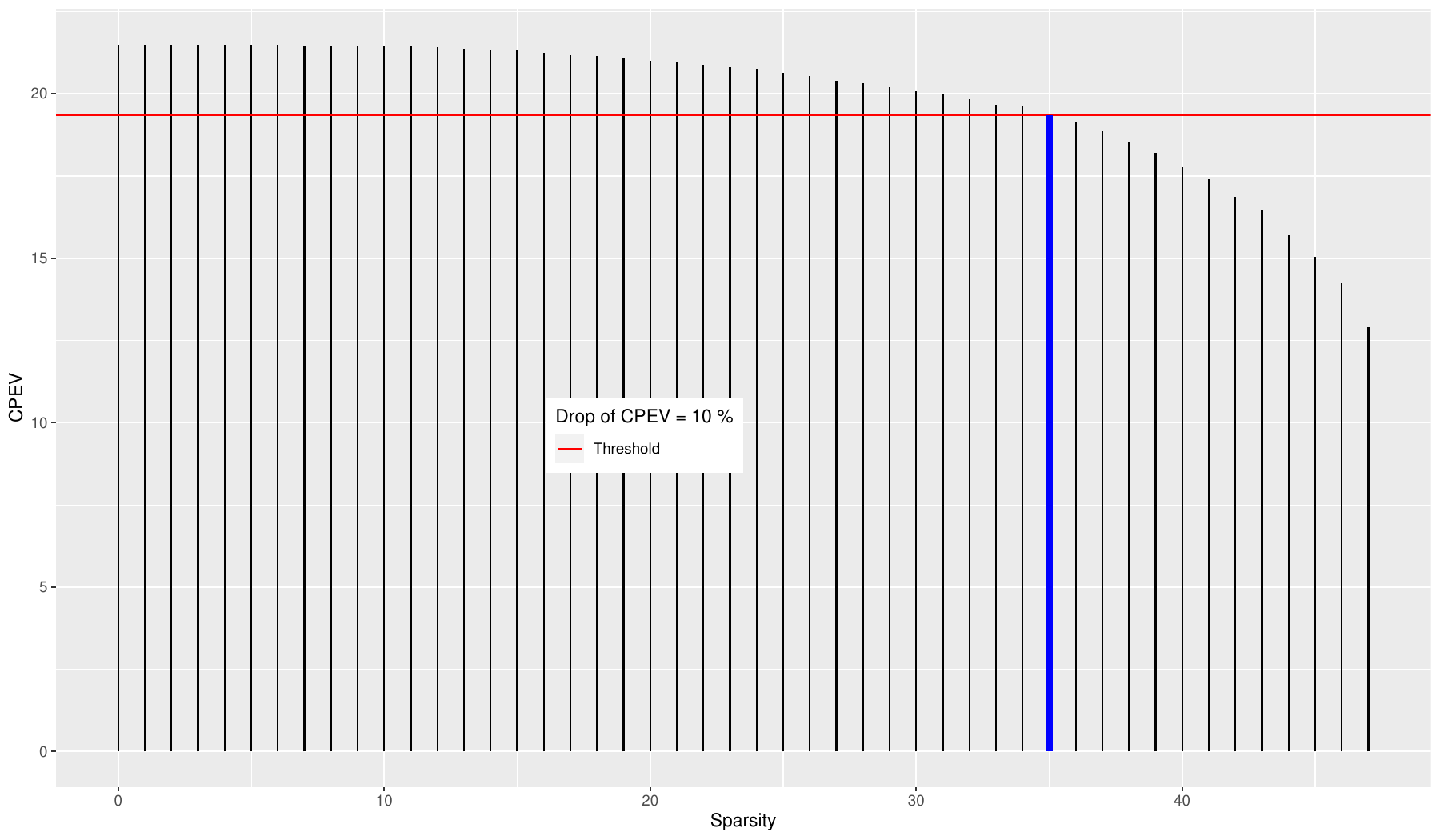}}
 \caption{CPEV as a function of the sparsity ($p$ - size of the subset) for the second component. Here, the blue vertical line indicates the largest value of the sparsity such  that CPEV does not exceed a drop of 10\%.} \label{drop-CPEV2}
\end{figure}

\begin{table}[H]
\centering
\caption{Best Subset results for component 3.} \label{tab.res.pca.3}
\begin{tabular}{ccccc}
  \hline
Dimension & $-\lambda_1/n$& PEV(sparse) & CPEV & cor(sPC,PC) \\ 
  \hline
1 & -0.01617 & 2.08 & 21.20 & -0.01 \\ 
  2 & -0.02622 & 3.42 & 22.53 & -0.60 \\ 
  3 & -0.03016 & 3.92 & 23.04 & 0.67 \\ 
  4 & -0.03355 & 4.35 & 23.47 & -0.72 \\ 
  5 & -0.03572 & 4.68 & 23.79 & -0.76 \\ 
  6 & -0.03814 & 5.01 & 24.13 & -0.75 \\ 
  7 & -0.03961 & 5.17 & 24.28 & -0.79 \\ 
  8 & -0.04163 & 5.45 & 24.57 & -0.81 \\ 
  9 & -0.04302 & 5.72 & 24.74 & -0.81 \\ 
  10 & -0.04391 & 5.77 & 24.82 & -0.84 \\ 
  11 & -0.04501 & 5.75 & 24.87 & -0.94 \\ 
  12 & -0.04641 & 5.95 & 25.06 & -0.95 \\ 
  13 & -0.04751 & 6.13 & 25.21 & 0.96 \\ 
  14 & -0.04875 & 6.26 & 25.36 & 0.96 \\ 
  15 & -0.04999 & 6.43 & 25.51 & 0.96 \\ 
  16 & -0.05088 & 6.57 & 25.65 & -0.96 \\ 
  17 & -0.05138 & 6.58 & 25.70 & -0.97 \\ 
  18 & -0.05184 & 6.62 & 25.75 & -0.97 \\ 
  19 & -0.05221 & 6.67 & 25.81 & -0.97 \\ 
  20 & -0.05264 & 6.71 & 25.86 & -0.98 \\ 
   \hline
\end{tabular}
\end{table}

\begin{figure}[H]
\centerline{
 \includegraphics[scale=0.45]{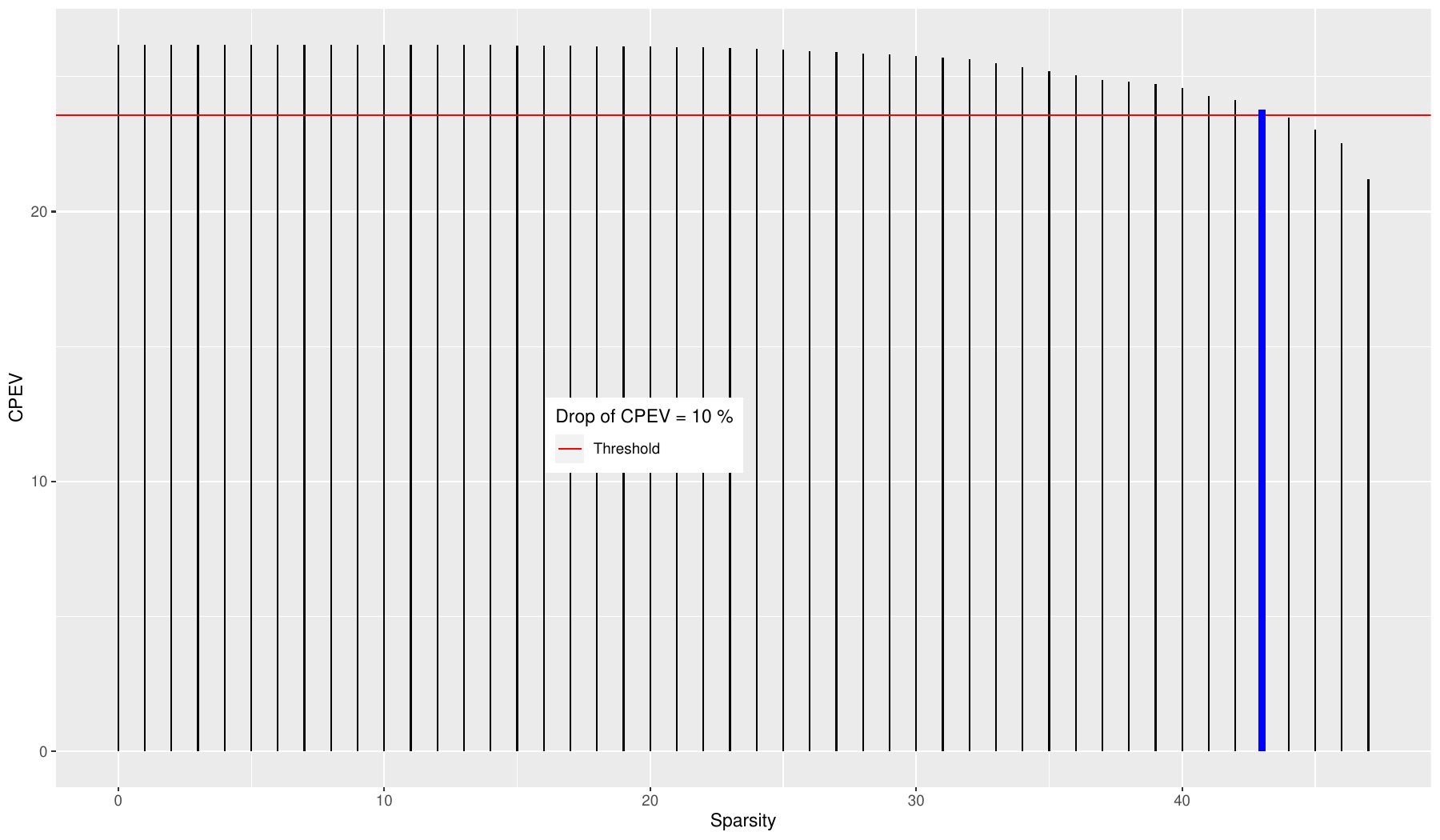}}
 \caption{CPEV as a function of the sparsity ($p$ - size of the subset) for the third component. Here, the blue vertical line indicates the largest value of the sparsity such that CPEV does not exceed a drop of 10\%.} \label{drop-CPEV3}
\end{figure}

\newpage
\section*{Appendix F: Illustration of Best Subset for PLS2 model}

\begin{figure}[H]
\centerline{
 \includegraphics[scale=0.45]{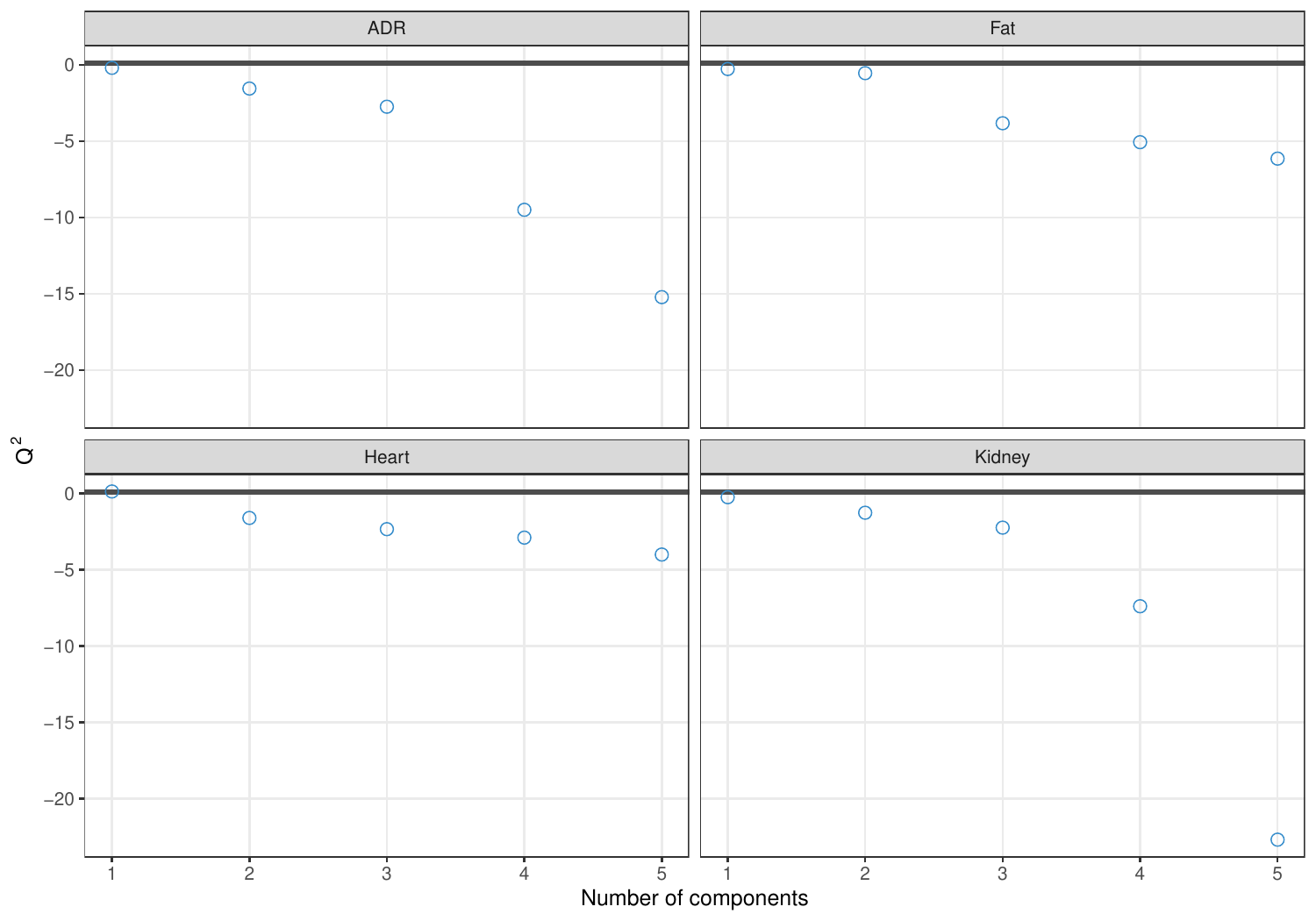}}
\caption{Marginal contribution of components using the $Q^2$ criterion per response variable.} \label{Q2.ind}
\end{figure}

\begin{figure}[H]
\centerline{
 \includegraphics[scale=0.45]{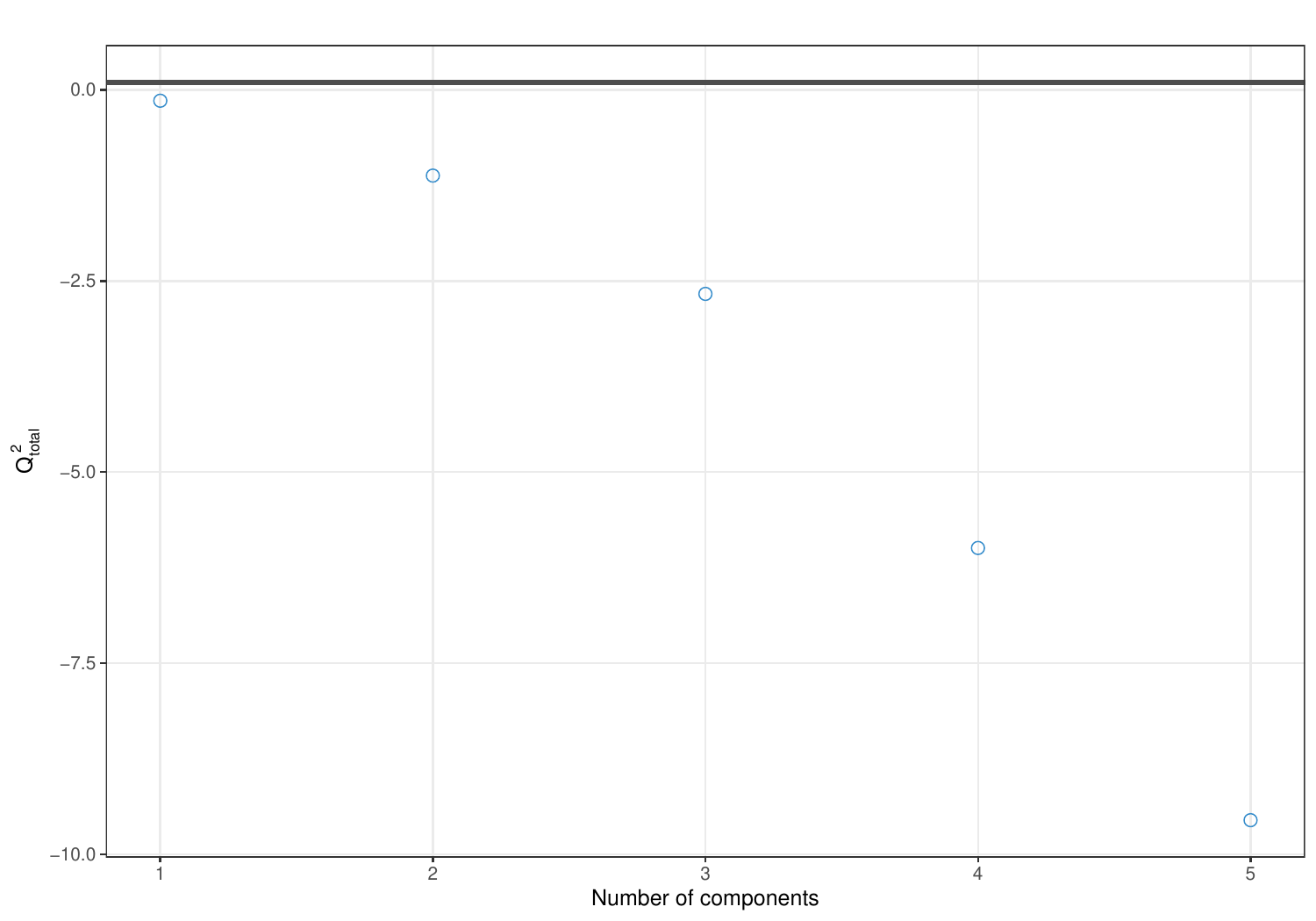}}
 \caption{Aggregation of the performance across all responses variables using the total $Q^2$.} \label{Q2.total}
\end{figure}

\begin{figure}[H]
\centerline{
 \includegraphics[scale=0.45]{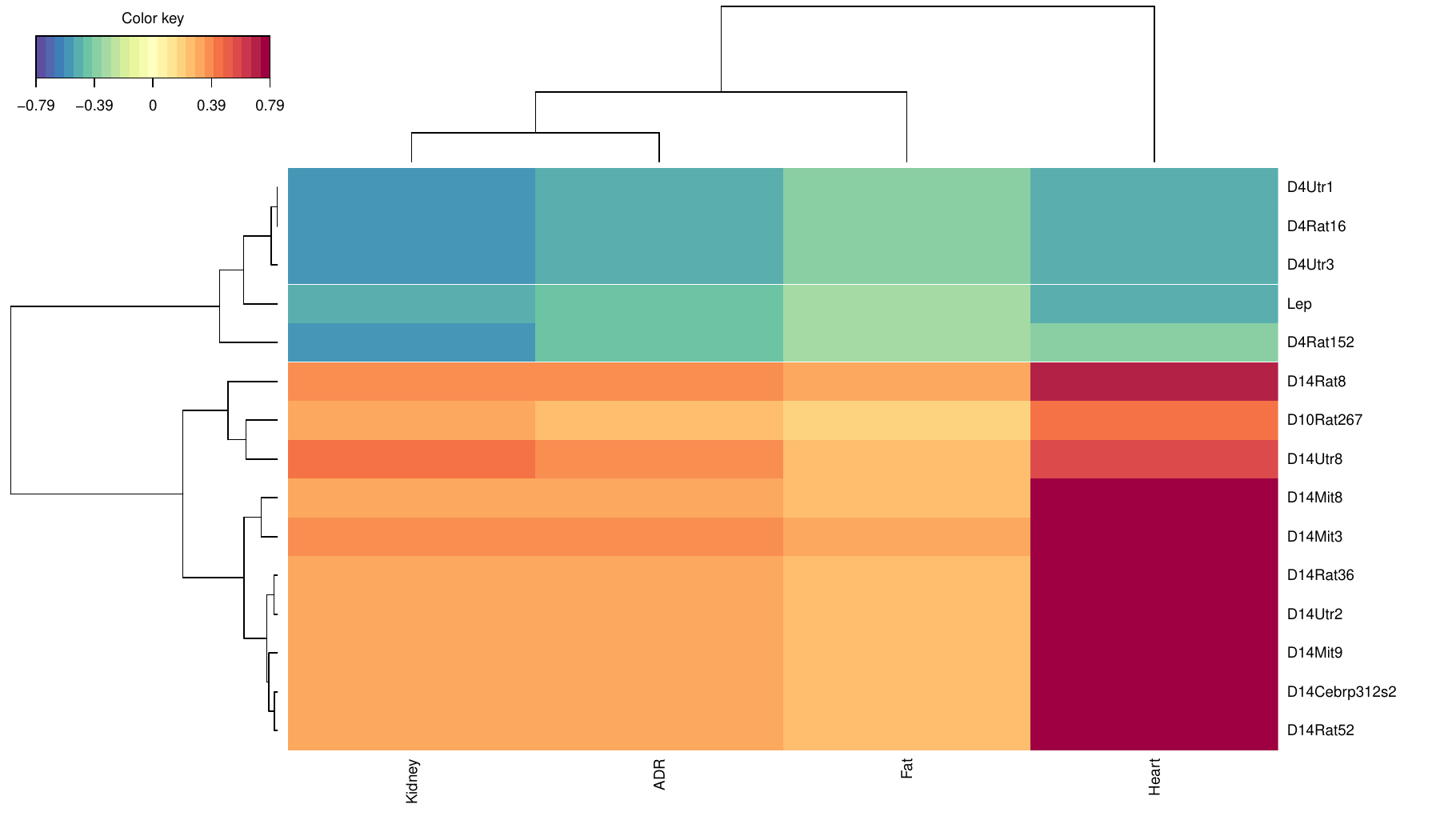}}
 \caption{Clustered image maps from BSS-PLS2 results.} \label{CIM}
\end{figure}

\bibliographystyle{abbrv}

 \bibliography{biblio}

\label{lastpage}